\documentclass[conference]{IEEEtran}
\usepackage{tikz}
\usepackage{amsmath}
\usepackage{color}
\usepackage{amsfonts}
\usepackage{amsthm}
\usepackage{amssymb}
\usepackage{extarrows}
\usepackage{multirow}
\usepackage{graphicx}
\usepackage{fancyhdr}
\usepackage{arydshln}
\usepackage{color}
\usepackage{filecontents}
\usepackage{epstopdf}
\usepackage[justification=centering]{caption}
\allowdisplaybreaks[4]

\definecolor{MyColor}{RGB}{182,222,192}

\ifCLASSINFOpdf
\else
\fi
\begin{document}
%
\title{PriSrv: Privacy-Enhanced and Highly Usable Service Discovery in Wireless Communications}

\author{\IEEEauthorblockN{Yang Yang$^1$, Robert H. Deng$^1$, Guomin Yang$^1$, Yingjiu Li$^2$, HweeHwa Pang$^1$, \\
	  Minming Huang$^1$, Rui Shi$^3$, Jian Weng$^4$}
\IEEEauthorblockA{1. School of Computing and Information Systems,
	Singapore Management University, Singapore
	\\
    \texttt{\{yyang,robertdeng, gmyang,hhpang,mmhuang\}@smu.edu.sg}\\
2. Department of Computer Science, University of Oregon,
USA (\texttt{yingjiul@uoregon.edu})\\
3. Beijing Electronic Science and Technology Institute, Beijing, China (\texttt{ruishi$\_$mail@126.com})\\
4. College of Information Science and Technology, Jinan
University, Guangzhou, China (\texttt{cryptjweng@gmail.com})
}
}


%


\IEEEoverridecommandlockouts
\makeatletter\def\@IEEEpubidpullup{3\baselineskip}\makeatother
\IEEEpubid{\parbox{\columnwidth}{
    This is the full version of the research work published in \textit{Network and Distributed System Security Symposium (NDSS) 2024}.\\
    https://dx.doi.org/10.14722/ndss.2024.24174\\
}
\hspace{\columnsep}\makebox[\columnwidth]{}}

\maketitle

\begin{abstract}
	
Service discovery is essential in wireless communications.
However, existing service discovery protocols provide no or
very limited privacy protection for service providers and clients, and
they often leak sensitive information (e.g., service type, client’s identity and mobility pattern), which leads to various network-based attacks (e.g., spoofing, man-in-the-middle, identification and tracking). In this
paper, we propose a private service discovery protocol, called
PriSrv, which allows a service provider and a client to respectively
specify a fine-grained authentication policy that the other party must
satisfy before a connection is established. PriSrv consists of
a private service broadcast phase and an anonymous mutual
authentication phase with bilateral control, where the private information of both parties is hidden beyond the fact
that a mutual match to the respective authentication policy
occurred. As a core component of PriSrv, we introduce the
notion of anonymous credential-based matchmaking encryption (ACME), which exerts dual-layer matching in one step to
simultaneously achieve bilateral  flexible policy control, selective
attribute disclosure and multi-show unlinkability. As a building block of ACME, we design a fast anonymous credential (FAC) scheme to provide constant size credentials and efficient show/verification mechanisms, which is suitable for privacy-enhanced and highly usable service
discovery in wireless networks. 

We present a concrete PriSrv protocol that is interoperable with popular wireless communication protocols, such as Wi-Fi Extensible Authentication Protocol (EAP), mDNS, BLE and Airdrop, to
offer privacy-enhanced protection. We present formal security proof of our protocol and evaluate its performance on
multiple hardware platforms: desktop, laptop, mobile phone
and Raspberry Pi. PriSrv accomplishes private discovery
and secure connection in less than 0.973 s on the first three
platforms, and in less than 2.712 s on Raspberry Pi 4B. We
also implement PriSrv into IEEE 802.1X in the real
network to demonstrate its practicality.

\end{abstract}


%

\section{Introduction}

Service discovery (SD) protocols, such as Wi-Fi \cite{WiFi}, AirDrop \cite{AirDrop}, and BLE \cite{BLE}, are essential components of networking systems that enable devices and services to dynamically discover and communicate with each other in a network environment. They facilitate the automatic detection and advertisement of available services, making it easier for devices to locate and 
interact with desired resources. 
However, there is a lack of highly usable approaches to sufficiently protect identification and private information in protocol executions, especially for privacy-concerned parties. A survey \cite{konings2013device} showed that about 90$\%$ users considered the exposure of device names from wireless network services as a privacy risk, as such exposure may lead to adversarial inference of users’ private information such as mobility patterns, profiles, and locations \cite{wu2016privacy,zhou2019discovering,stute2019billion,stute2021disrupting}. For instance, in public Wi-Fi, ISP could easily identify a person via the announced device names \cite{cassola2015authenticating}. In IoT networks, an attacker may infer a user's regular routine by collecting the service data from user's smart devices \cite{fawaz2016protecting}.  Several vulnerabilities spanning from Wi-Fi, BLE to Apple
Wireless Direct Link (AWDL) are discovered in recently years which lead to tracking, DoS, and MitM attacks on iOS and macOS \cite{stute2021disrupting}.

On the other hand, users prefer high usability in accessing wireless network services, which include no pre-registered pairing, no third-party dependence for service discovery, and low computation and communication overheads. A major barrier in increasing user satisfaction for accessing wireless network services is the technical difficulty of elevating privacy protection  without sacrificing high usability in wireless network protocols. Existing privacy-aware wireless network protocols and other related works fail to overcome this barrier as they either leak private information \cite{na2021wi,bai2016staying,bai2017apple} or violate high usability requirements in protocol executions \cite{XuSurvey20}.  

Our objective is to develop a privacy-enhanced and highly usable service discovery protocol between wireless network service access point (service provider) and client to enable them to discover each other within range and establish a secure communication channel only if they meet each other's connection requirements. The challenges to achieve this objective are three folds: (1) ensure that services are only discoverable by an authorized set of clients; (2) enable clients to filter out unauthorized services without heavy computation; (3) allow both service provider and client to specify policies the other party must satisfy
in order for their private information to be revealed. In certain service discovery protocols, such as AirDrop and BLE, both service provider and client are wireless devices, which necessitates reciprocal privacy protection.

We propose a dual-layer architecture to solve this problem, which includes an outer layer and an inner layer.
In the outer layer, each service provider or client is associated with a set of public attributes (such as domain name) that can be revealed to everyone and a public authentication policy, which are used for fast bilateral policy matching without decryption. 
Specifically, a service provider broadcasts a ciphertext encrypted by its policy and public attributes. A client first checks whether its public attributes match with the service provider's public policy, and vice versa, which filters the mismatch services accordingly. 
If and only if their public attributes satisfy the bilateral policy, the client can decrypt the ciphertext.
In the inner layer, each party is associated with a set of private attributes (such as  device ID) that are only revealed to the intended peers. 
Only when the decryption of the outer layer ciphertext is successful, the client can recover the private attributes of the service provider, which allows the client to authenticate the service provider by verifying the authenticity of the latter's attributes, including both public attributes and private attributes. The service provider authenticates the client using the same mechanism. Then, 
they establish a session key using a secure key agreement protocol to enable secure communication between them. By applying the above dual-layer architecture, PriSrv builds a private-enhanced service discovery protocol with high usability.

\subsection{Privacy Enhancement and High Usability Requirements}
\label{Subsec:Goal}

To mitigate the leakage of any private information in service discovery, SD protocols should meet the following privacy enhancement requirements. 

\textit{1. Private Service Broadcast}. Service contents broadcasted by service providers must be both confidential and unforgeable, preventing unintended clients from learning service content and enabling the detection of bogus service providers broadcasting fraudulent services.

\textit{2. Mutual Authentication}. Service providers and clients authenticate each other in a secure manner to ensure that the private information of both parties will not be leaked to any unauthenticated entity.

\textit{3. Bilateral Anonymity}. Both service providers and clients remain anonymous to a third-party during protocol execution, and no third-party can identify the private information of the involved parties.

\textit{4. Bilateral Flexible Policy Control}. Both service providers and clients can specify fine-grained access policies for authorized peers and simultaneously check the satisfaction of policies from both sides, which guarantees that private information of both sides are only exposed to their authorized peers.

\textit{5. Selective Attribute Disclosure}. 
It refers to the ability of an entity (either service provider or client) to choose which specific attributes they disclose to the other, while keeping other attributes undisclosed. It allows each entity to share only the necessary and relevant information while maintaining control over their private information.

\textit{6. Multi-Show Unlinkability}. 
It allows a user to prove possession of a credential or  attributes without revealing their identity or linking their actions across multiple sessions.


In addition to these privacy enhancement requirements, SD protocols are expected to meet the following high usability requirements.

\textit{1. No Pre-registered Pairing}. Clients are not required to subscribe to or share a secret key with any service providers beforehand. It allows clients to discover and connect to service providers seamlessly without any manual setup or configuration.

\textit{2. No Third-party Dependency during Service Discovery Process}. Service discovery should not depend on any external services such as a third-party server or a directory provider during protocol execution. Protocols relying on external servers presume a reliable Internet connection for mobile devices. However, this presumption may not hold in wireless communications (e.g., BLE communications).

\textit{3. No In-advance Identity Issuance}. Users are not required to register to a third-party to obtain identity certification documents, such as certificates, credentials, etc. In-advance identity issuance has less impact on the usability of service discovery process since it occurs only once before the execution of SD protocol. We note that PriSrv requires in-advance identity issuance.



\subsection{Contributions}
\label{Subsec:Contrib}

We propose PriSrv, a service discovery protocol, to meet both privacy enhancement and high usability requirements. 
The main contributions of this work are summarized as follows.

$\bullet$ \textbf{A New Privacy-Enhanced Service Discovery Protocol with High Usability}. PriSrv is the first privacy-enhanced and highly usable service discovery protocol that can be integrated into a wide range of wireless applications.

$\bullet$ \textbf{Anonymous Credential-based Matchmaking Encryption (ACME)}. We propose a novel cryptographic primitive called anonymous credential-based matchmaking encryption (ACME). ACME supports bilateral fine-grained policies and selective attribute disclosure for private mutual authentication in service discovery. ACME outperforms the matchmaking encryption (ME) in CRYPTO'19 \cite{ateniese2019match} in terms of functionality and efficiency. This is a contribution of independent interest for the advancement of matchmaking encryption.

$\bullet$ \textbf{Fast Anonymous Credential}. As a building block of ACME, we propose a fast anonymous credential (FAC) scheme to support anonymous authentication with selective attribute disclosure and multi-show unlinkability.
A comprehensive comparison with existing anonymous credential schemes demonstrates its superior efficiency
for credential showing and verification with constant and small credential.

$\bullet$ \textbf{Interoperability with Existing Protocols}. To demonstrate interoperability, we present concrete methods for integrating PriSrv with mainstream service discovery protocols including  Extensible Authentication Protocol (EAP), mDNS, BLE and AirDrop. Through experimentation, we show the applicability and effectiveness of PriSrv in real-world
scenarios.

$\bullet$ \textbf{Formal Security Proofs}. We provide formal security proofs for the security and privacy properties of PriSrv in a security model that captures various attack vectors, such as intercepting, tampering with channel messages, replaying, injecting
data packets, and interleaving messages among different sessions in realistic settings.

$\bullet$ \textbf{Deployment on Multiple Platforms in Real Networks}: The performance of PriSrv is evaluated on multiple hardware
platforms, including desktop, laptop, mobile phone and Raspberry Pi, in the Wi-Fi WPA-Enterprise framework. Our experiments demonstrate the efficiency of PriSrv across different
platforms. The private service broadcast phase in PriSrv takes less than 0.483
seconds, and the anonymous mutual authentication phase takes less
than 0.973 seconds on the first three devices.
The delays stay well below 1 second, which humans perceive as an “immediate response” \cite{heinrich2021privatedrop,Stuart1991Information}. While on Raspberry Pi, the delays are 1.189 and 2.712 seconds for private broadcast and mutual authenticationon, respectively, which demonstrates additional costs on IoT devices.

\renewcommand{\arraystretch}{1.4}
\begin{table*}[thbp]\centering 	\small
	\setlength{\tabcolsep}{1.2mm}
	\begin{tabular}{|c|c|c|c|c|c|c|c|c|c|c|c|c|c|c|c|c|}		
		\hline
		\multirow{3}{*}{\textbf{SD Protocols}} & \multicolumn{6}{c|}{\textbf{Privacy Enhancement}} & \multicolumn{3}{c|}{\textbf{High Usability}}\\
		\cline{2-11}		
		& Private & Mutual  & Bilateral & Bilateral Flex. & Sel. Attr. & Multi-Show & No Pre-reg. & No 3rd-party & No In-advance \\
		& Broadcast & Authn. & Anon. & Pol. Ctrl. & Disclosure & Unlinkability & Pairing & Dependence & ID Issuance\\
		\hline
		DNS-SD \cite{DNS-SD}
		& $\times$ & $\times$ & $\times$ & $\times$ & $\times$ & $\times$ & $\surd$ &$\times$ & $\times$ \\	
		mDNS \cite{mDNS}
		& $\times$ & $\times$ & $\times$ & $\times$ & $\times$ & $\times$ & $\surd$ &$\surd$ & $\times$\\
		SSDP \cite{SSDP}
		& $\times$ & $\times$ & $\times$ & $\times$ & $\times$ & $\times$ & $\surd$ &$\surd$ & $\surd$\\
		UPnP \cite{UPnP}
		& $\times$ & $\times$ & $\times$ & $\times$ & $\times$ & $\times$ & $\surd$ &$\surd$ & $\surd$\\
		Wi-Fi \cite{WiFi}  & $\times$ & $\surd$ & $\times$ & $\times$ & $\times$ & $\times$ & $\surd$ &$\surd$ & $\times$ \\
		BLE \cite{BLE}  & $\times$ & $\surd$ & $\times$ & $\times$ & $\times$ & $\times$ & $\surd$ &$\surd$ & $\surd$ \\
		AirDrop \cite{AirDrop}
		& $\times$ & $\surd$ & $\times$ & $\times$ & $\times$ & $\times$ & $\surd$ &$\surd$ & $\times$ \\		
		PrivateDrop \cite{heinrich2021privatedrop}
		& $\times$ & $\surd$ & $\surd$ & $\times$ & $\times$ & $\times$ & $\surd$ &$\surd$ & $\times$ \\
		CBN \cite{cassola2015authenticating} & $\times$ & $\times$ & $\times$ & $\times$ & $\times$ & $\times$ & $\times$ &$\surd$ & $\times$\\		WTSB \cite{wu2016privacy} & $\surd$& $\surd$ & $\surd$ & $\times$ & $\times$ & $\times$&$\surd$ &$\surd$ & $\times$\\
		\textbf{PriSrv} & $\surd$ & $\surd$ & $\surd$ & $\surd$ & $\surd$ & $\surd$ & $\surd$ &$\surd$ & $\times$\\
		\hline
	\end{tabular}
	\caption{Comparison of Service Discovery Protocols}
	\label{Tab:ComparePriSvc}
\end{table*}

\section{Related Work}
\label{Sec:RelatedWork}

A variety of protocols have been developed for service discover in network environments. As shown in Table \ref{Tab:ComparePriSvc}, none of them, except PriSrv, satisfy all privacy enhancement requirements.

In particular, the protocols DNS-SD \cite{DNS-SD}, mDNS \cite{mDNS}, SSDP \cite{SSDP}, UPnP \cite{UPnP} and CBN \cite{cassola2015authenticating} do not meet any privacy enhancement requirement. First, DNS-based Service Discovery (DNS-SD) \cite{DNS-SD} utilizes the Domain Name System (DNS) to enable service discovery. It allows service providers to advertise their services by registering them with a DNS server, and clients can discover these services by querying the DNS server, which is widely used in local networks and the Internet. 
Second, multicast DNS (mDNS) \cite{mDNS} enables service discovery in local networks without the need for a central DNS server, and allows service providers to announce their services using multicast DNS packets, and clients can resolve and discover these services directly.
Third, Simple Service Discovery Protocol (SSDP) \cite{SSDP} is designed based on the Internet protocol suite for advertisement and discovery of network services and presence information. Fourth, Universal Plug and Play (UPnP) \cite{UPnP} permits networked devices, such as personal computers, printers, Internet gateways, Wi-Fi access points and mobile devices to seamlessly discover each other's presence on the network and establish functional network services. Lastly, CBN scheme \cite{cassola2015authenticating} requires clients to subscribe to service providers so that service providers can unilaterally authenticate clients anonymously for service discovery. 

The above SD protocols are vulnerable to man-in-the-middle (MitM) attacks, spoofing attacks and denial-of-service (DoS) attacks due to the lack of proper privacy protection. 
Bai et al. \cite{bai2016staying} launched MitM attacks against mDNS and  illustrated how a malicious device can impersonate a printer by spoofing its mDNS hostname. According to Wang et al. \cite{wang2019looking}, UPnP is vulnerable to DoS attacks: a device receiving a request from a potentially spoofed control point may respond to the supposed requester, unknowingly contributing to the amplification and intensification of the attack.
CBN scheme  \cite{cassola2015authenticating} only requires clients to anonymously authenticate to service providers in a private manner, while the authentication/anonymity of service providers and private broadcast are not supported, making it vulnerable to MitM attacks and spoofing attacks.

Among these protocols, DNS-SD relies on DNS records to advertise and discover services within a network. CBN scheme
relies on a pre-registration pairing mechanism:  service provider maintains a directory to control the access of subscribers while clients are required to register to service providers beforehand, where the size of directory grows linearly with the number of clients.

Although Wi-Fi \cite{WiFi} and BLE \cite{BLE} support mutual authentication, they dissatisfy other privacy enhancement requirements, including private broadcast, bilateral anonymity, bilateral flexible policy control, selective attribute disclosure and multi-show unlinkability. Wi-Fi \cite{WiFi} enables devices to discover and connect to services available on a local-area network.  Bluetooth Low Energy (BLE) \cite{BLE} is designed for low-power devices, such as IoT devices and wearable devices, to advertise their available services, allowing other devices to discover and connect to them for data exchange and interaction. 

A common problem of Wi-Fi and BLE is that the private information of service providers and clients is advertised publicly in wireless network, which may induce user identification, impersonation attacks and spoofing attacks. 
A survey \cite{konings2013device} indicated that 59$\%$ investigated devices periodically announce their owners’ real names for Wi-Fi network, which is deemed as a privacy risky by about 90$\%$
users. A deep-learning-based identification mechanism (with accuracy over 80$\%$) was demonstrated in \cite{yu2020you} to identify mobile devices from broadcast and multicast packets. Na et al. \cite{na2021wi} proposed Wi-attack to leverage the
wide-deployed Wi-Fi devices (such as Wi-Fi APs) to conduct
poisonous impersonation attacks, where the vulnerability is caused by the open nature of these cleartext advertisements.
Similarly, BLE-equipped devices consistently advertise their unique identifiers in cleartext 
\cite{fawaz2016protecting}, making them vulnerable to BLE Spoofing Attacks (BLESA) \cite{wu2020blesa}.

Revealing of device identifiers in Wi-Fi and BLE is a stepping stone toward advanced attacks such as user profiling and tracking \cite{fawaz2016protecting}. Large-scale tracking attack in real-time can be  mounted by deploying multiple low-cost Wi-Fi and BLE nodes throughout an area. 
This allows adversaries to infer
additional user information such as home and work locations,
movement patterns and behavior profiling, which are useful for
targeted tracking  \cite{venkatnarayan2020leveraging}.

AirDrop \cite{AirDrop}, PrivateDrop \cite{heinrich2021privatedrop} and WTSB  \cite{wu2016privacy} employ encryption and authentication mechanisms to protect communications in service discovery. 
AirDrop \cite{AirDrop} is an SD protocol for file-sharing on Apple devices, which utilizes a combination of Wi-Fi and Bluetooth technologies to enable devices in close proximity to discover each other and share files wirelessly. AirDrop and PrivateDrop need to establish TLS connection with client and server certificates for authentication. PrivateDrop realizes \textit{private} mutual authentication for AirDrop by protecting device identifiers in an optimized private set intersection protocol 
\cite{heinrich2021privatedrop}. 
WTSB \cite{wu2016privacy} realizes  private service discovery by leveraging prefix encryption (a variant of identity-based encryption) and standard digital signature-based key exchange protocol. 
WTSB \cite{wu2016privacy} supports private broadcast, mutual authentication and bilateral anonymity.

However, these SD protocols (AirDrop, PrivateDrop \cite{heinrich2021privatedrop} and WTSB \cite{wu2016privacy}) suffer from MitM attacks, DoS attacks, impersonation attacks or user tracking attacks due to the lack of privacy enhanced properties, such as bilateral policy control, selective attribute disclosure and multi-show unlinkability. The attacker is able to link multiple sessions using client and server certificates in AirDrop and PrivateDrop protocols. Stute et al. \cite{stute2019billion} exposed several security and privacy vulnerabilities in Apple Wireless Direct Link (AWDL) ranging from design flaws to implementation bugs leading to (i) MitM attacks enabling stealthy modification of files transmitted via AirDrop, (ii) DoS
attacks disrupting communications, and (iii) privacy leaks enabling
user identification and long-term tracking. Bai et al. \cite{bai2016staying} demonstrated impersonation and spoofing attacks on certain Zeroconf protocols (e.g. AirDrop), which even allows attackers to steal clients' SMS messages, documents, email notifications and photos \cite{bai2017apple}. 

Heinrich et al. \cite{heinrich2021privatedrop} discovered a series of flaws in AirDrop that allow attackers to learn phone numbers and email addresses of both sender and receiver devices.  As stated in the work \cite{heinrich2021privatedrop}, users of PrivateDrop can be tracked via UUIDs
in the TLS certificates used for establishing the protocol communication channels. 
WTBS \cite{wu2016privacy} dissatisfies bilateral flexible policy control: service providers have the ability to specify the type of clients they intend to communicate with, but clients do not have the option to choose the service providers they want to communicate with. 
Furthermore, WTBS  \cite{wu2016privacy} is susceptible to user tracking attack due to the lack of multi-show unlinkability.

AirDrop and PrivateDrop offer mutual authentication, while PrivateDrop provides an additional feature of bilateral anonymity. The fundamental building block of PrivateDrop is a Diffie-Hellman-based Private Set Intersection (PSI) scheme, which exclusively entails exponentiation computations. Moreover, WTBS \cite{wu2016privacy} further enhances privacy by encrypting broadcast messages, achieving private broadcasting in addition to these features. WTSB has the advantage
of high efficiency due to the usage of efficient identity-based prefix encryption scheme. Conversely, PriSrv utilizes both exponentiation and bilinear pairing operations within ACME, and the time consumption increases with the complexity
of access policy. Therefore, PriSrv achieves improved privacy but incurs a higher computational overhead as a trade-off.

After conducting a  comprehensive comparison, it becomes evident that PriSrv stands out as the only SD protocol that successfully meets all the privacy enhancement requirements. As for high usability, PriSrv satisfies no pre-registerd pairing and no third-party dependence during service discovery.
PriSrv does require in-advance identity issuance, but it does not affect the service discovery process.

\section{Preliminaries}
\label{Sec:Preliminaries}

\subsection{Notation and Bilinear Pairing}
\label{Subsec:Notations}

Let $\vec{x}$ denote the full attribute set, 
$\vec{x}^{(in)}$ the private attributes for an inner layer and 
$\vec{x}^{(out)}$ the public attributes for an outer layer, where $\vec{x}^{(in)},\vec{x}^{(out)}\subseteq\vec{x}$.
Let $f:\{0,1\}^n\rightarrow\{0,1\}$ denote the policy; $f(\vec{x})=1$ denote $\vec{x}$ satisfying $f$, and $f(\vec{x})=0$ denote $\vec{x}$ not satisfying $f$.

Let $s\xleftarrow{\$}S$ denote $s$ sampled uniformly at random from a set $S$; $\mathbb{N}$ denote the natural number; $\lambda\in\mathbb{N}$ denote the security parameter; $[n_1,n_2]$ denote  $\{n_1,\cdots,n_2\}$; PPT denote probabilistic polynomial time;  $\mathbb{Z}_p$ represent the group of integers modulo $p$, and $\mathbb{Z}_p^*=\mathbb{Z}_p\backslash\{0\}$. We use lower case boldface to denote (column) vectors and upper case boldface to denote matrices. Denote a bilinear group with Type-3 pairings as $\mathcal{BG}=(G_1,G_2,G_T,e,p)$, where there is no efficiently computable isomophism between $G_1$ and $G_2$.
Let $g_1\in G_1$, $g_2\in G_2$ and $g_T=e(g_1,g_2)\in G_T$ be the respective generators.
For a matrix $\textbf{A}$ over $\mathbb{Z}_p$, define $[\textbf{A}]_1:=g_1^{\textbf{A}}$, $[\textbf{A}]_2:=g_2^{\textbf{A}}$, $[\textbf{A}]_T:=g_T^{\textbf{A}}$, where exponentiation is carried out component-wise.


\subsection{Assumptions}
\label{Subsec:Assumptions}

$\quad$\textit{Definition 3.1.} (Discrete Logarithm (DL) Assumption). Let $g$ be a generator of a cyclic group $G$. 
DL assumption holds if for all PPT adversary $\mathcal{A}$, the advantage function $\textsf{Adv}_{\mathcal{A}}^{\text{DL}}(\lambda):=Pr[\mathcal{A}(g,g^a)=a]$
is negligible, where $a \xleftarrow{\$}\mathbb{Z}_p^*$.

\textit{Definition 3.2.} (Decisional Diffie-Hellman (DDH) Assumption). Let $g$ be a generator of $G$ and $\mathcal{T}=(g,g^a,g^b)\in G^3$, where $a,b\xleftarrow{\$}\mathbb{Z}_p^*$. DDH assumption holds if for all PPT adversary $\mathcal{A}$, the advantage 
$|Pr[\mathcal{A}(\mathcal{T},g^{ab})=1]-Pr[\mathcal{A}(\mathcal{T},g^{c})=1]|$
is negligible, where $c\xleftarrow{\$}\mathbb{Z}_p^*$.

\textit{Definition 3.3.} (Matrix DDH (MDDH$_{k}$) Assumption) \cite{kowalczyk2019compact}. Let $\ell>k\geq1$, $d\geq 1$.  MDDH$_{k}$ assumption holds if for all PPT adversary $\mathcal{A}$, the advantage $\textsf{Adv}_{\mathcal{A}}^{\text{MDDH}_{k}}(\lambda):=$
$|Pr[\mathcal{A}([\textbf{M}]_1,[\textbf{MS}]_1)=1]-Pr[\mathcal{A}([\textbf{M}]_1,[\textbf{U}]_1)=1]|$
is negligible, where $\textbf{M}\xleftarrow{\$}\mathbb{Z}_p^{\ell\times k}$, $\textbf{S}\xleftarrow{\$}\mathbb{Z}_p^{k\times d}$ and $\textbf{U}\xleftarrow{\$}\mathbb{Z}_p^{\ell\times d}$.

\subsection{Linear Secret Sharing for Monotone Boolean Formulae}
\label{SubSec:LSS}

The information-theoretic linear secret sharing for monotone Boolean formulae \cite{kowalczyk2019compact,katsumata2020compact} is described below.

$\underline{\textsf{share}(f,\mu)}$.
Input: A formula $f:\{0,1\}^n\rightarrow\{0,1\}$ of size $m$ (i.e., the number of edges in $f$ is $m$), and a secret $\mu\in\mathbb{Z}_p$.
1) For each non-output wire $j=1,\cdots,m-1$, select  $\hat{\mu}_j\xleftarrow{\$}\mathbb{Z}_p$. For the output wire, set $\hat{\mu}_m:=\mu$.
2) For each outgoing wire $j$ from input node $i$, add $\mu_j:=\hat{\mu}_j$ to the output set of shares and set $\rho(j):=i$.
3) For each AND gate $g$ with input wires $a$, $b$ and output wire $c$, add $\mu_{c_a}:=\hat{\mu}_c+\hat{\mu}_a+\hat{\mu}_b\in\mathbb{Z}_p$ to the output set of shares and set $\rho(c):=0$.
4) For each OR gate $g$ with input wires $a$, $b$ and output wire $c$, add $\mu_{c_a}:=\hat{\mu}_c+\hat{\mu}_a\in\mathbb{Z}_p$ and
$\mu_{c_b}:=\hat{\mu}_c+\hat{\mu}_b\in\mathbb{Z}_p$ to the output set of shares and set $\rho(c_a):=0$ and $\rho(c_b):=0$.
5) Output $(\{\mu_j\}_{j\in[\hat{m}]},\rho)$.

$\underline{\textsf{reconstruct}(f,x,\{\mu_j\}_{\rho(j)=0\vee x_{\rho(j)}=1})}$.
Input: A formula $f$, $\vec{x}\in\{0,1\}^n$, and $\{\mu_j\}_{\rho(j)=0\vee x_{\rho(j)}=1}$.
From the leaves of the formula to the root, calculate the output wire value $\hat{\mu}_c$ at each node.
1) Given $\hat{\mu}_a,\hat{\mu}_b$ associated with the input wires $a$ and $b$ of an AND gate,  compute $\hat{\mu}_c=\mu_c-\hat{\mu}_a-\hat{\mu}_b$.
2) Given $\hat{\mu}_a$ (or $\hat{\mu}_b$) associated with the input wires $a$ (or $b$) of an OR gate,  compute $\hat{\mu}_c=\mu_{c_a}-\hat{\mu}_a$ (or $\hat{\mu}_c=\mu_{c_b}-\hat{\mu}_b$).
3) Output $\mu=\hat{\mu}_m$.

\subsection{$\textsf{NC}^1$ Circuit and Monotone Formulae}
\label{Subsec:NC1}

We define $\textsf{NC}^1$ circuit and monotone Boolean formulae following Kowalczyk's \cite{kowalczyk2019compact} and Katsumata's \cite{katsumata2020compact} works.
A monotone Boolean formula $f:\{0,1\}^n\rightarrow\{0,1\}$ is specified by a directed acyclic graph (DAG) with three kinds of nodes: input gate nodes, gate nodes and a single output node. Input nodes have in-degree 0 and out-degree 1, AND/OR nodes have in-degree (fan-in) 2 and out-degree (fan-out) 1, and the output node has in-degree 1 and out-degree 0. We number the edges (wires) $1,2,\cdots,m$, and each gate node is defined by a tuple $(g,a_g,b_g,c_g)$, where $g:\{0,1\}^2\rightarrow\{0,1\}$ is either AND or OR, $a_g,b_g$ are the incoming wires, $c_g$ is the outgoing wire and $a_g,b_g< c_g$. The size $m$ of a formula is the number of edges in the underlying DAG and the depth $d$ of a formula is the length of the longest path from the output node. Lemma 2.1 in Katsumata's work \cite{katsumata2020compact} states the well-known equivalence between the monotone Boolean formulae and $\textsf{NC}^1$ circuits.

\section{PriSrv's Overview}
\label{Sec:PriSrvOverview}


First, we present a technical overview of PriSrv. Next, we
provide an example to illustrate how PriSrv is used. Finally,
we highlight how PriSrv meets all privacy-enhancement and
high usability requirements.

\begin{figure}
	\begin{center}
		\includegraphics[width=3.5in]{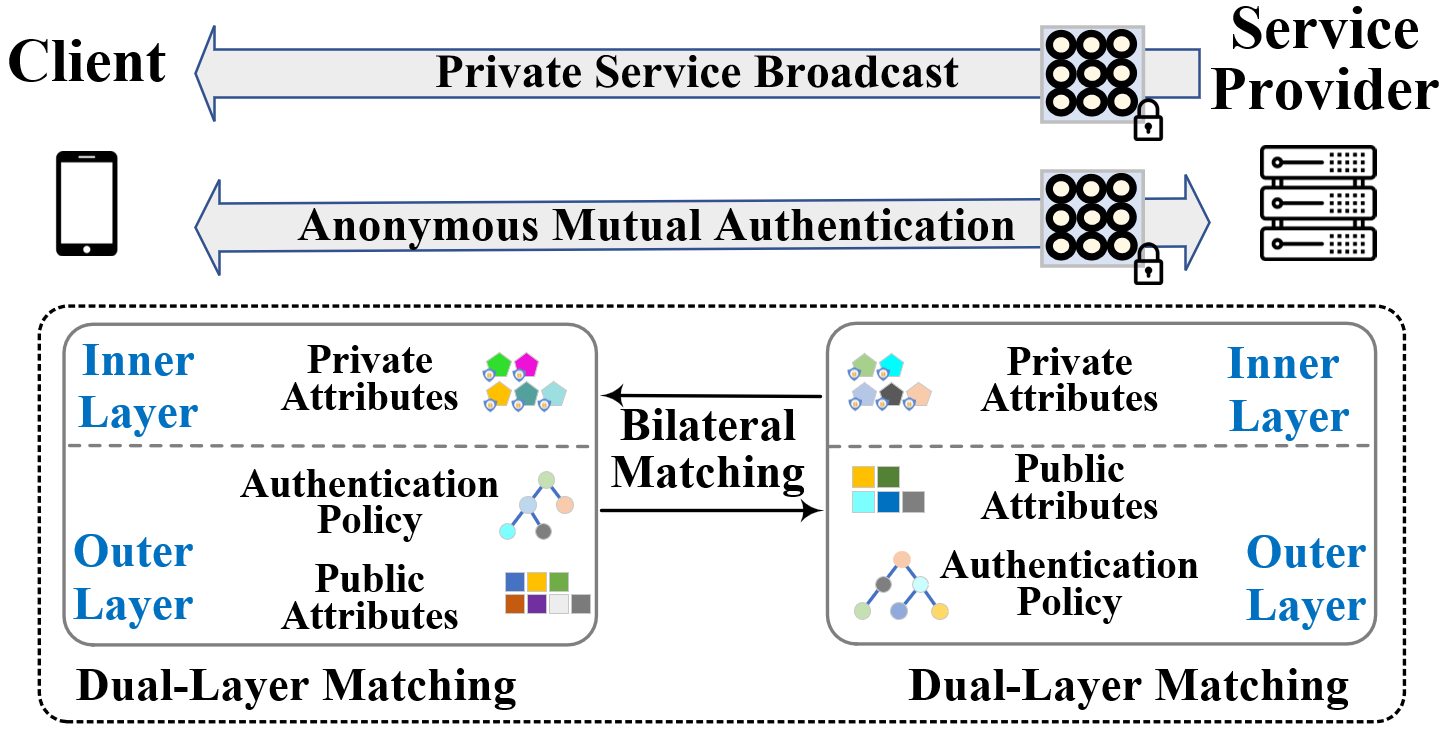}
		\caption{Overview of PriSrv}
		\label{Fig:Overview}
	\end{center}
\end{figure}

\noindent\textbf{Technical overview.}
At a high level, PriSrv is a private service discovery protocol that ensures services are only discoverable by an authorized set of clients. PriSrv consists of a private service broadcast phase and an anonymous mutual authentication phase as shown in Fig. \ref{Fig:Overview}.

PriSrv's design incorporates a novel crypto-enforced  construction that enables both service providers and clients to express flexible access control policies and disclose partial attributes. To meet the privacy enhancement and high usability requirements outlinted in $\S$\ref{Subsec:Goal}, we design a dual-layer matching mechanism: an outer layer defines bilateral public authorization policies for filtering unauthorized service providers and clients based on their public attributes; an inner layer performs mutual authentication based on the selectively disclosed private attributes. We design a new cryptographic primitive, named anonymous credential-based matchmaking encryption (ACME), to realize such a dual-layer design in PriSrv.

Anonymous credential (AC) realizes attribute based anonymous authentication with selective attribute disclosure, making it a potential tool for ACME construction. Existing AC schemes suffer from either large credential sizes or cumbersome show and verification mechanisms \cite{ringers2017efficient,camenisch2015composable,fuchsbauer2019structure}, rendering them unsuitable for privacy-enhanced and highly usable service discovery in wireless networks. We design a new AC scheme, named fast anonymous credential (FAC) as a building block of ACME.

To realize bilateral policy control in ACME, one promising technology is the Matchmaking Encryption (ME) proposed by Ateniese et al. in CRYPTO'19 \cite{ateniese2019match}. In ME, sender ($\textsf{snd}$) and receiver ($\textsf{rcv}$) possess a set of attributes $\vec{x}_\textsf{snd}$ and $\vec{x}_\textsf{rcv}$, respectively. The sender is able to specify an authorization policy $f_{\textsf{snd}}$ for the receiver's attributes $\vec{x}_{\textsf{rcv}}$ to satisfy, and vice versa. ME enables both participants to specify fine-grained policies for encrypted data, which satisfies our need for bilateral policy control. 
Nonetheless, the ME in \cite{ateniese2019match} has three limitations: (1) the conception of ME to support expressive  policies relies on heavy cryptographic tools, including Functional Encryption (FE) and general Zero-Knowledge Proof (ZKP), whose known instantiations are still far from practical; (2) ME does not support selective attribute disclosure; (3) concrete instantiations of ME \cite{ateniese2019match, chen2022identity} only support identity-based equality matching. It remains an open problem to develop an efficient ME that supports fine-grained policy based fuzzy matching \cite{ateniese2019match}.
We develop ACME to solve this open problem and overcome the above limitations. 
We further develop PriSrv based on ACME to meet both privacy enhancement and high usability requirements.

\noindent\textbf{Example.} We provide a smart office example to exemplify the use of bilateral policy control and selective attribute disclosure in PriSrv. Consider a screen mirroring service provided by a smart TV, which only allows authorized devices to connect to it. On the other hand, a client device should only project its screen to an authorized screen mirroring service device to prevent any leakage of private information. The \textit{service type} in this scenario is the screen mirroring service, and the \textit{service parameters} include resolution, refresh rate, etc. The smart TV is associated with a set of attributes: $\vec{x}_s$=(device type, vendor, model, OS, domain name, device name, location, IP address, security domain), where the first five are public attributes and the rest are private. The mirroring service provider may select a set of public attributes  $\vec{x}^{(out)}_s$=(device type, vendor, domain name) to be used in the outer layer, and a set of private attributes $\vec{x}^{(in)}_s$=(IP address) to be used in the inner layer. The client device is associated with another set of  attributes: $\vec{x}_c$=(device type, model, OS, department, device name, classified device, IP address, security domain), where the first four are public attributes and the rest are private.
The client selects a set of public attributes $\vec{x}^{(out)}_c$=(device type, OS, department) for outer layer matching, and a set of private attributes $\vec{x}^{(in)}_c$=(classified device, security domain) for inner layer authentication.

To realize bilateral control, the service provider (i.e., the smart TV) sets a service policy as 
\begin{eqnarray*}
	f_s=\big(&&\text{Device Type = ``Smart phone} \vee \text{Laptop"}\\
	&\bigwedge& \text{OS = ``Android} \vee \text{iOS} \vee \text{Windows"}\\
	&\bigwedge& \text{Department = ``A} \vee \text{B"}\big).
\end{eqnarray*}
The client device specifies a connection policy as
\begin{eqnarray*}
	f_c=\big(&&\text{Device Type = ``TV" }\bigwedge \text{Vendor = ``C} \vee \text{D"}\\
	&\bigwedge& \text{Domain Name = ``*.XYZ.COM"}\big).
\end{eqnarray*}

The screen mirroring service can be discovered by the client if and only if $f_s(\vec{x}^{(out)}_c)=1\wedge f_c(\vec{x}^{(out)}_s)=1$, which indicates that the public attributes of the service provider (and the client, respectively) satisfy the policy of its peer. The private attributes selected by smart TV and client device are used for mutual authentication.

\noindent\textbf{How PriSrv Meets Requirements.} PriSrv meets both privacy enhancement and high usability requirements as outlined in $\S$\ref{Subsec:Goal}.

- \textit{Private Service Broadcast $\&$ Mutual Authentication}. The messages broadcasted by service providers are encrypted using ACME such that only intended clients can obtain the decrypted information. Both service providers and clients authenticate each other's private attributes before establishing a secure communication channel. 

- \textit{Bilateral Anonymity $\&$ Bilateral Flexible Policy Control}. 
Both service providers and clients maintain their anonymity during the discovery process. Bilateral flexible policy control is achieved via ACME, as decryption fails if any protocol participant's policy is not satisfied by its peers' attributes.

- \textit{Selective Attribute Disclosure $\&$ Multi-Show Unlinkability}. According to the minimum privacy leakage principle, any participant in PriSrv only reveals a subset of its attributes to its peer. Both service provider and client select a subset of their attributes, including public attributes and private attributes to generate their authentication tokens. 
Multi-show unlinkability of PriSrv is inherited from that of FAC, which ensures the unlinkability of multiple instances of authentication tokens generated by the same entity across multiple protocol sessions (even using the same subset of non-unique attributes).

- \textit{No Pre-registered Pairing $\&$ No Third-party Dependency for Service Discovery}. PriSrv protocol execution does not require any service provider to know its clients, or any client to subscribe to its service providers in advance.  
PriSrv operates without relying on any external services during protocol execution.

\textbf{Threat and Attacker Model.} The credential issuer is considered trustworthy to issue and revoke anonymous credentials. Both service providers and clients in the protocol are considered untrustworthy, as they have the potential to launch any passive or active attacks. Specifically, a service provider may attempt to impersonate other providers by broadcasting deceptive messages or to track clients' activities. Likewise, a client may impersonate other clients to obtain unauthorized network access.

Following the Canetti-Krawczyk model for authenticated key-exchange (AKE) in \cite{canetti2001analysis,canetti2002security} and the service discovery model in \cite{wu2016privacy}, the attackers against PriSrv include malicious service providers, clients, and external adversaries.
We aim to comprehensively model the attackers' capabilities in the real world to gain full control over public network communication. This control encompasses actions such as revealing certain protocol secrets, intercepting or tampering with channel messages, replaying, delaying, injecting or dropping data packets, and interleaving messages from different sessions, etc. They are capable to launch various types of attacks, including eavesdropping attacks, spoofing attacks, impersonation attacks, man-in-the-middle attacks, etc. The attackers' goals include: (1) breaking authenticated key-exchange security; and (2) revealing sensitive information pertaining to clients or service providers, enabling attackers to track their activities.

\textbf{Formal Security Definition and Analysis.} The formal security models of private service discovery include service discovery security and bilateral anonymity, which is followed by formal security proofs. The formal security models and proofs are shown in Appendix \ref{Appendix:PriSrv}.

\section{Fast Anonymous Credential}
\label{Sec:FAC}

We propose a fast anonymous credential scheme (FAC) as a component of ACME to enable fast anonymous authentication while maintaining a constant and small credential size. 
FAC supports  re-randomization of credentials to support multi-show unlinkability, and selective attribute disclosure.
We provide the syntax for anonymous credentials and proceed to construct a concrete FAC scheme for mobile devices.

\subsection{Syntax of Anonymous Credential}
\label{Subsec:ACSyntax}

Anonymous credential (AC) is formally defined by the following PPT algorithms \cite{chase2014algebraic,sanders2020efficient}.

$\bullet$ $\textsf{Setup}(1^{\lambda}, 1^{n})\rightarrow \textsf{pp}$: On input a security parameter $\lambda$ and a function parameter $1^n$, it outputs public parameter $\textsf{pp}$, which is an implicit input to all the other algorithms.

$\bullet$ $\textsf{CredKeyGen}(\textsf{pp})\rightarrow(\textsf{pk},\textsf{sk})$: On input $\textsf{pp}$, this algorithm creates credential issuer's public/secret keys $\textsf{pk}$/$\textsf{sk}$, where $\textsf{pk}$ is an implicit input to the algorithms below.

$\bullet$ $\textsf{UserKeyGen}(\textsf{pp})\rightarrow(\textsf{upk},\textsf{usk})$: On input $\textsf{pp}$, the algorithm generates user's public key $\textsf{upk}$ and secret key $\textsf{usk}$.

$\bullet$ $\left<\textsf{Issue.I}(\textsf{sk},\textsf{upk})\rightleftarrows\textsf{Issue.U}(uid,\vec{x},\textsf{usk})\right>\rightarrow \textsf{cred}$. This is an interactive protocol for AC issuance executed between the issuer and a user over a secure channel. The user executes the protocol by inputting a user's identity $uid$, an attribute set $\vec{x}$ and a secret key $\textsf{usk}$. The credential issuer runs the protocol by inputting $\textsf{sk}$ and $\textsf{upk}$. The issuer hands over a credential $\textsf{cred}$ to user via secure channel.

$\bullet$ $\textsf{Show}(uid,\{x_i\}_{i\in\mathcal{I}},\textsf{cred},\textsf{usk},m)\rightarrow \textsf{tok}$: On input $uid$, an attribute subset $\{x_i\}_{i\in\mathcal{I}}\subseteq\vec{x}$ ($\mathcal{I}\subseteq [1,n]$), $\textsf{cred}$, $\textsf{usk}$ and a message $m$, it outputs an authentication token $\textsf{tok}$.

$\bullet$ $\textsf{Verify}(\textsf{tok},m)\rightarrow b\in\{0,1\}$. On input $\textsf{tok}$ and $m$, it outputs $b=1$ if $\textsf{tok}$ is valid; otherwise, it outputs $b=0$.

Following the security definitions in \cite{chase2014algebraic,sanders2020efficient}, the \textit{correctness}, \textit{unforgeability}, \textit{anonymity} and \textit{unlinkability} of AC are defined, which are shown in Appendix \ref{Appendix:FAC}.

\subsection{Construction of FAC}

Our construction of FAC is given below.

$\bullet$ $\textsf{Setup}(1^{\lambda},1^n)\rightarrow \textsf{pp}$: Let $\lambda$ be the security parameter, and $n$ the attribute number in the system. Run $\mathbb{G}=(p,G_1,G_2,G_T,e)\xleftarrow{\$}\mathsf{GGen}(1^{\lambda})$, and output $\textsf{pp}=(g,h,n)$, where $g,h$ are the generators of $G_1$, $G_2$, respectively.

$\bullet$ $\textsf{CredKeyGen}(\textsf{pp})\rightarrow(\textsf{pk},\textsf{sk})$: The issuer samples $\tau,y_i\xleftarrow{\$}\mathbb{Z}_p^*$, computes $W\leftarrow g^{\tau}$, $X_i\leftarrow h^{y_i}$, $Y_i\leftarrow g^{y_i}$ for $i\in[0,n+1]$, and $Z_{i,j}=g^{y_i\cdot y_j}$ for $0\leq i\neq j\leq n+1$. Then, it outputs secret key $\textsf{sk}=(\tau,\{y_i\}_{i\in[0,n+1]})$ and public key $\textsf{pk}\leftarrow(W,\{X_i,Y_{i}\}_{i\in[0,n+1]},\{Z_{i,j}\}_{0\leq i\neq j\leq n+1})$. 

$\bullet$ $\textsf{UserKeyGen}(\textsf{pp})\rightarrow(\textsf{upk},\textsf{usk})$: The user with $uid$ samples $\textsf{usk}\xleftarrow{\$}\mathbb{Z}_p^*$, computes $\textsf{upk}\leftarrow h^{\textsf{usk}}$, and creates a signature proof of knowledge  $\pi_1$ as
$\textsf{SPK}\{\textsf{usk}:\textsf{upk}= h^{\textsf{usk}}\}.$
The issuer registers $\textsf{upk}$ if $\textsf{Verify}_{\textsf{SPK}}(\textsf{upk},\pi_1)=1$ holds. 

$\bullet$ $\left<\textsf{Issue.I}(\textsf{sk},\textsf{upk})\rightleftarrows\textsf{Issue.U}(uid,\vec{x},\textsf{usk})\right>\rightarrow\textsf{cred}$. The secure channel between issuer and user can be established by standard protocols, such as TLS.

(1) User sends $uid$ and attributes $\vec{x}=\{x_i\}_{i\in[1,n]}$ to issuer.

(2) The issuer samples $r\overset{\$}{\leftarrow}\mathbb{Z}_p^*$ to calculate $\textsf{cred}\leftarrow (\sigma_1,\sigma_2)$, where $$\sigma_1\leftarrow h^r,\sigma_2\leftarrow \textsf{upk}^{r\cdot y_0}\cdot h^{r(\tau+\sum_{i=1}^ny_ix_i+y_{n+1}\cdot uid)}.$$

(3) The user accepts the credential $\textsf{cred}$ if the following equation holds $$e(W\cdot Y_0^{\textsf{usk}}\cdot Y_{n+1}^{uid}\prod\nolimits_{i=1}^nY_i^{x_i},\sigma_1)=e(g,\sigma_2).$$

$\bullet$ $\textsf{Show}(uid,\{x_i\}_{i\in\mathcal{I}},\textsf{cred},\textsf{usk},m)\rightarrow \textsf{tok}$: The user generates a token on selected attribute subset $\{x_i\}_{i\in\mathcal{I}}$, $\mathcal{I}\subseteq [1,n]$.
Select $t_1,t_2\overset{\$}{\leftarrow}\mathbb{Z}_p^*$ to compute   $$T_1=g^{t_1}\prod_{j\in[1,n]\backslash\mathcal{I}}Y_j^{x_j},~~T_2=(\prod_{i\in\mathcal{I}'}Y_i)^{t_1}\prod_{i\in\mathcal{I}',j\in[1,n]\backslash\mathcal{I}}Z_{i,j}^{x_j},$$
$\bar{\sigma}_1=\sigma_1^{t_2}$, $\bar{\sigma}_2=\sigma_2^{t_2}\bar{\sigma}_1^{t_1}$, and create $\pi_2$ as
$$\textsf{SPK}\left\{(\textsf{usk},uid):\begin{matrix}\bar{\sigma}_1=\sigma_1^{t_2},\bar{\sigma}_2=\sigma_2^{t_2}\bar{\sigma}_1^{t_1},\sigma_1= h^r,\\
	\sigma_2= (h^{\textsf{usk}})^{r\cdot y_0}h^{r(\tau+\sum\limits_{i=1}^ny_ix_i+y_{n+1}\cdot uid)}
\end{matrix} \right\}(m),$$
where $\mathcal{I}'=\mathcal{I}\cup\{0,n+1\}$.
The token is $$\textsf{tok}\leftarrow(\{x_i\}_{i\in\mathcal{I}},T_1,T_2,\bar{\sigma}_1,\bar{\sigma}_2,\pi_2).$$

$\bullet$ $\textsf{Verify}(\mathsf{tok},m)\rightarrow b\in\{0,1\}$. The algorithm outputs $b=1$ if $\textsf{Verify}_{\textsf{SPK}}(\textsf{tok},m)=1$. Otherwise, it returns $b=0$.

\textbf{Instantiation of \textsf{SPK}}. Following 
the standard Fiat-Shamir paradigm, \textsf{SPK}s in FAC are instantiated as follows. 

\underline{The \textsf{SPK} $\pi_1$}:

\textsf{Prove:} Prover selects $\widetilde{\textsf{usk}}\overset{\$}{\leftarrow}\mathbb{Z}_p^*$ and computes $\gamma\leftarrow h^{\widetilde{\textsf{usk}}}$,
$c\leftarrow H(\textsf{upk},\gamma)$,
$\overline{\textsf{usk}}=\widetilde{\textsf{usk}}-c\cdot \textsf{usk} \mod p$. Return $\pi_1\leftarrow(c,\gamma,\overline{\textsf{usk}})$.

\textsf{Verify:} Given  $\textsf{upk}$ and SPK $\pi_1$, the verifier checks
$c\overset{?}{=}H(\textsf{upk},\gamma)$, $\gamma\overset{?}{=}h^{\overline{\textsf{usk}}}\textsf{upk}^{c}$. It outputs 1 if these equations hold, and 0 otherwise.

\underline{The \textsf{SPK} $\pi_2$}:

\textsf{Prove:} Prover selects $\widetilde{uid},\widetilde{\textsf{usk}}\overset{\$}{\leftarrow}\mathbb{Z}_p^*$ and computes $\Lambda\leftarrow e(Y_0^{\widetilde{\textsf{usk}}}Y_{n+1}^{\widetilde{uid}},\bar{\sigma}_1)$, $c\leftarrow H(m,\{x_i\}_{i\in\mathcal{I}},\Lambda,T_1,T_2,\bar{\sigma}_1,\bar{\sigma}_2)$,
$\overline{uid}\leftarrow \widetilde{uid}-c\cdot uid \mod p$, $\overline{\textsf{usk}}\leftarrow \widetilde{\textsf{usk}}-c\cdot \textsf{usk} \mod p$. Set $\pi_2\leftarrow(c,\overline{uid},\overline{\textsf{usk}},\Lambda)$, and return $\textsf{tok}\leftarrow(\{x_i\}_{i\in\mathcal{I}},T_1,T_2,\bar{\sigma}_1,\bar{\sigma}_2,\pi_2)$.

\textsf{Verify:} Given $\textsf{tok}$, the verifier checks
$c\overset{?}{=}H(m,\{x_i\}_{i\in\mathcal{I}},\Lambda,$ $T_1,T_2,\bar{\sigma}_1,\bar{\sigma}_2)$, $e(Y_0^{\overline{\textsf{usk}}}Y_{n+1}^{\overline{uid}},\bar{\sigma}_1)^{-1}\cdot\Lambda\overset{?}{=}[e(g,\bar{\sigma}_2)\cdot\Gamma]^{c}$, $e(T_1,$ $\prod\nolimits_{i\in\mathcal{I}'}X_i)\overset{?}{=}e(T_2,h)$, where $\Gamma=e(W\cdot T_1\cdot\prod\nolimits_{i\in\mathcal{I}}Y_i^{x_i},\bar{\sigma}_1)^{-1}$. It outputs 1 if these equations hold, and 0 otherwise.

Our fast anonymous credential (FAC) scheme has the following advantages: 1) FAC offers a non-interactive $\textsf{Show}\leftrightarrows \textsf{Verify}$ process, ensuring fast anonymous authentication. 2) FAC generates anonymous credentials of a constant and small size. 3) An authentication token generated in FAC consists of only two group elements. The construction is based on the unlinkable redactable signature (URS) scheme \cite{sanders2020efficient}, which is one of the initial frameworks for generating constant-size redactable signatures on attributes $\vec{x}=(x_1,\cdots,x_n)$. 
FAC generates an anonymous credential $\textsf{cred}$ based on the URS  scheme \cite{sanders2020efficient}. When a request is made to verify the authenticity of a subset of attributes $\{x_i\}_{i\in\mathcal{I}}\subseteq\vec{x}$, the $\textsf{Show}$ algorithm in FAC performs the following steps: it derives an authentication token $\textsf{tok}$ from the anonymous credential $\textsf{cred}$, and then produces a signature proof of knowledge (\textsf{SPK}) for the authentication token. The $\textsf{Verify}$ algorithm in FAC is responsible for checking the validity of $\textsf{tok}$. 
The correctness proof of FAC is shown in Appendix \ref{Appendix:FAC}.

\textsf{Theorem 5.1.}
\label{Theo:FACSec}
The FAC scheme is secure (i.e., achieves unforgeability, anonymity and unlinkability) under the DL and DDH assumptions.

The proof of Theorem 5.1 is shown in Appendix \ref{Appendix:FAC}.

\section{Anonymous credential-based matchmaking encryption (ACME)}
\label{Sec:ACME}

We introduce a new cryptographic primitive named ACME to support several core features in PriSrv protocol, including bilateral policy control, anonymous authentication and selective attribute disclosure. ACME is a variant of ME where the sender and the receiver can use anonymous credentials to prove their attributes without revealing their identities. This is useful because it allows for stronger privacy guarantees and flexible policy enforcement in scenarios such as secure online dating, e-voting, and anonymous whistleblowing, where the parties do not trust each other or third parties. ACME is of independent interests for advancing research on Matchmaking Encryption.

\begin{figure}
	\begin{center}
		\includegraphics[width=3.5in]{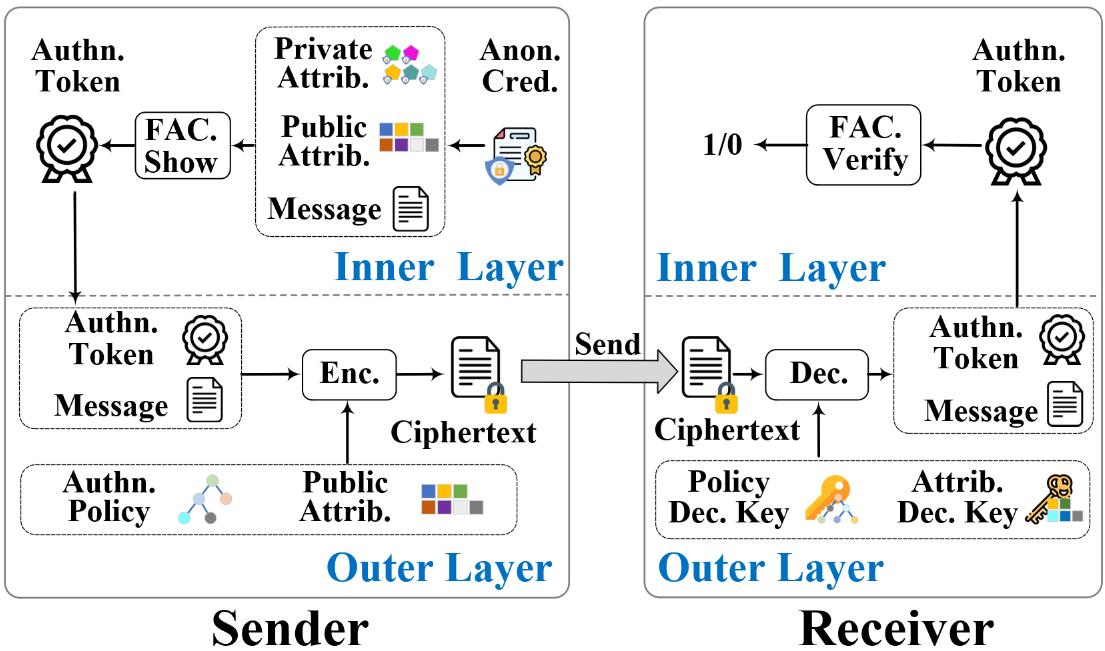}
		\caption{Architecture of ACME}
		\label{Fig:ACME-Overview}
	\end{center}
\end{figure}

\subsection{Design Intuition}

Matchmaking Encryption (ME) is a natural starting point to construct ACME. However, the conception of ME \cite{ateniese2019match} that can simultaneously support expressive policy (e.g., monotone Boolean formulae) and policy hiding is of theoretical interest only since no concrete instantiation has been proposed. 
Although  	identity-based ME schemes supporting equality policies were introduced in \cite{ateniese2019match,chen2022identity}, 
they do not fit for highly-usable service discovery since in general participants of service discovery are unaware of their peers' identities and thus cannot define identity-based equality policies.
Meanwhile, we notice that the original ME schemes \cite{ateniese2019match,chen2022identity} support hidden policies, but they are not ideal for service discovery because such schemes require clients to blindly decrypt every service advertisement, bringing high costs when multiple services are in presence.

To balance fast service discovery and privacy protection, ACME adopts a dual-layer matching design for a sender (\textsf{snd}) to encrypt any message $M$ and send the ciphertext to a receiver (\textsf{rcv}) with bilateral policy control. Sender \textsf{snd} receives an anonymous credential $\textsf{cred}_{\textsf{snd}}$ from a credential issuer for all its attributes $\vec{x}_{\textsf{snd}}$. As shown in Fig. \ref{Fig:ACME-Overview}, ACME consists of an inner layer and an outer layer. 
In the inner layer, 
sender \textsf{snd} generates an authentication token using $\mathcal{FAC}.\textsf{Show}$ from a message $M$ and selected attributes (including public and private attributes) based on the received credential $\textsf{cred}_{\textsf{snd}}$. In the outer layer, sender \textsf{snd} encrypts the authentication token and the message $M$ using an authentication policy $f_{\textsf{snd}}$ (specified by $\textsf{snd}$ for $\textsf{rcv}$) and the sender's selected public attributes. Then, sender $\textsf{snd}$ transmits the ciphertext to receiver $\textsf{rcv}$.

On the receiver side, the ciphertext is decrypted in the outer layer using receiver's policy decryption key and attribute decryption key to recover the authentication token and message $M$. In the inner layer, the authentication token is verified using $\mathcal{FAC}.\textsf{Verify}$ to authenticate the sender's selected private attributes, public attributes and the message $M$. 

\textbf{Impersonation Resistance}. ACME is the core component of PriSrv to prevent impersonation attacks.
As shown in Fig. \ref{Fig:ACME-Overview}, both the public and private attributes are used as inputs for authentication token generation in the inner layer. This design has been purposefully engineered to provide robust protection against impersonation attacks. Although the public attributes used in the service provider's outer layer are public, a malicous service provider (without all the authorized public attributes) is not able to impersonate any legal provider since the forged authentication token cannot pass the verification  by the receiver (using $\mathcal{FAC}.\textsf{Verify}$). On the other hand, if the public attributes used in the outer layer are not unique, PriSrv relies on the inner layer to authenticate both public and private attributes, which rules out any impersonation attack. Meanwhile, an attacker impersonating a legitimate receiver cannot be successful in decryption without a valid secret key.

%

\subsection{Syntax of ACME}
\label{SubSec:ACMEConsSyntax}

Anonymous credential-based matchmaking encryption (ACME) is formally defined below, and the correctness of ACME is defined in Appendix \ref{Appendix:ACME}.

$\bullet$ $\textsf{Setup}(1^{\lambda},1^n)$: On input a security parameter $1^{\lambda}$ and a function parameter $1^n$, this algorithm outputs the master public/secret keys $\textsf{mpk}$/$\textsf{msk}$. Note that $\textsf{mpk}$ is implicit input in all the following algorithms. 

$\bullet$ $\textsf{CredKeyGen}(\textsf{mpk})\rightarrow(\textsf{pk},\textsf{sk})$: On input $\textsf{mpk}$, this algorithm creates credential issuer's public key $\textsf{pk}$ and secret key $\textsf{sk}$. $\textsf{pk}$ is an implicit input to the following algorithms.

$\bullet$ $\textsf{UserKeyGen}(\textsf{mpk})\rightarrow(\textsf{upk},\textsf{usk})$: On input $\textsf{mpk}$, the algorithm generates user's public key $\textsf{upk}$ and secret key $\textsf{usk}$.

$\bullet$ $\left<\textsf{Issue.I}(\textsf{sk},\textsf{upk})\rightleftarrows\textsf{Issue.U}(uid,\vec{x},\textsf{usk})\right>\rightarrow \textsf{cred}$. The issuer inputs $\textsf{sk}$, $\textsf{upk}$ and the user inputs $uid$, $\textsf{usk}$, full attributes $\vec{x}$. The issuer interacts with user to generate a credential $\textsf{cred}$ for the user.

$\bullet$
$\textsf{DKGen}(\mathsf{msk},\vec{x}_{\textsf{rcv}})$: On input $\mathsf{msk}$ and attributes
$\vec{x}_{\textsf{rcv}}$, this algorithm outputs an attribute decryption key $\textsf{DK}_{\vec{x}_{\textsf{rcv}}}$.

$\bullet$
$\textsf{PolGen}(\mathsf{msk},f_{\textsf{rcv}})$: On input $\mathsf{msk}$ and policy $f_{\textsf{rcv}}$, this algorithm outputs a policy decryption key $\textsf{DK}_{f_{\textsf{rcv}}}$. 

$\bullet$ $\mathsf{Enc}(\textsf{cred}_{\textsf{snd}},\vec{x}_{\textsf{snd}},f_{\textsf{snd}},M)$: On input  $\textsf{cred}_{\textsf{snd}}$, full attributes $\vec{x}_{\textsf{snd}}$,
policy $f_{\textsf{snd}}$ and message $M$ as input, the sender selects a set of private attributes $\vec{x}_{\textsf{snd}}^{(in)}$ for an inner layer and a set of public attributes $\vec{x}_{\textsf{snd}}^{(out)}$ for an outer layer from $\vec{x}_{\textsf{snd}}$. It firstly generates a token $\textsf{tok}_{\textsf{snd}}$ for $\vec{x}_{\textsf{snd}}^{(in)}$, $\vec{x}_{\textsf{snd}}^{(out)}$ and message $M$. 
Then, it encrypts $(M,\textsf{tok}_{\textsf{snd}})$ using the public attributes $\vec{x}_{\textsf{snd}}^{(out)}$ and policy $f_{\textsf{snd}}$, and outputs a ciphertext $\textsf{CT}_{\vec{x}_{\textsf{snd}},f_{\textsf{snd}}}$.

$\bullet$ $\mathsf{Dec}(\textsf{DK}_{\vec{x}_{\textsf{rcv}}},\textsf{DK}_{f_{\textsf{rcv}}},\textsf{CT}_{\vec{x}_{\textsf{snd}},f_{\textsf{snd}}})$: On input $\textsf{DK}_{\vec{x}_{\textsf{rcv}}}$,  $\textsf{DK}_{f_{\textsf{rcv}}}$ and $\textsf{CT}_{\vec{x}_{\textsf{snd}},f_{\textsf{snd}}}$, the receiver
recovers $(M,\textsf{tok}_{\textsf{snd}})$ iff $f_{\textsf{snd}}(\vec{x}_{\textsf{rcv}}^{(out)})=1$ and $f_{\textsf{rcv}}(\vec{x}_{\textsf{snd}}^{(out)})=1$; otherwise, it outputs $\bot$. If the above step succeeds,  the receiver verifies  $\textsf{tok}_{\textsf{snd}}$ for $\vec{x}_{\textsf{snd}}^{(in)}\cup\vec{x}_{\textsf{snd}}^{(out)}$ and $M$. It outputs the message $M$ if the token is valid;
otherwise, it outputs $\bot$.

\textbf{Remark.}
In encryption algorithm, the authentication token $\textsf{tok}_{\textsf{snd}}$ is generated for selected public and private attributes rather than just private attributes. The purpose is to authenticate sender's selective attributes in both layers to prevent spoofing attacks. The token also authenticates $M$ to prevent message forgery.

\textsf{Definition 6.1.}
An ACME scheme is secure if it satisfies \textit{privacy}, \textit{authenticity}, \textit{anonymity} and  \textit{unlinkability}. 

The formal definitions of these security properties are provided in Appendix \ref{Appendix:ACME}.


\subsection{Construction of ACME}

FAC in $\S$\ref{Sec:FAC} is leveraged in the inner layer of ACME for authentication.
For outer-layer encryption and bilateral policy control, we resort to attribute-based encryption (ABE) that supports expressive access policies. However, ABE only supports unilateral policy control.
To enable bilateral control, a potential solution is to integrate key policy (KP-)ABE and ciphertext policy (CP-)ABE so that the secret key of CP-ABE (resp. KP-ABE) functions as attribute decryption key (resp. policy decryption key) produced by the $\textsf{DKGen}$ (resp. $\textsf{PolGen}$) algorithm.
Although the idea seems straightforward, there are a few subtleties to be addressed.
Firstly, 
compact ABE schemes are preferred for compact ciphertext size and key size.
The compact KP-ABE and CP-ABE schemes proposed by Kowalczyk et al. \cite{kowalczyk2019compact} in Eurocrypt'19 are natural candidates because they are in the dual form with common parameters and support Boolean formulae (equivalent to $\textsf{NC}^1$ circuits\footnote{In computational complexity theory, $\textsf{NC}^i$ is the class of decision problems decidable by uniform boolean circuits with a polynomial number of gates of at most two inputs and depth $O(\log^i n)$, or the class of decision problems solvable in time $O(\log^i n)$ on a parallel computer with a polynomial number of processors, where $\textsf{NC}$ is short for "Nick Pippenger's Class".}). 
Nonetheless, the decryption process of the dual ABE schemes in  \cite{kowalczyk2019compact}  involves 
a large number of time-consuming pairing operations (depending on the complexity of $\textsf{NC}^1$).
If we construct ACME based on the dual schemes given in \cite{kowalczyk2019compact}, such ACME would incur high computational costs for wireless devices. 
Meanwhile, we notice that for KP-ABE, Katsumata et al. proposed an improved scheme in \cite{katsumata2020compact} with faster decryption, which requires only a constant number of pairing operations. We apply the technique in \cite{katsumata2020compact} to improve CP-ABE scheme in \cite{kowalczyk2019compact} to achieve fast decryption with a constant number of  pairing operations. By integrating the improved CP-ABE with Katsumata's KP-ABE \cite{katsumata2020compact}, which are also in a dual form, we can achieve both fine-grained bilateral policy control and fast decryption.

\textbf{Concrete Construction}. Our ACME scheme for general policies is built from the above fast anonymous credential scheme $\mathcal{FAC}$, a symmetric encryption scheme
$\mathcal{SE}=(\textsf{SGen},\textsf{SEnc},\textsf{SDec})$ with key space $\mathcal{K}$, and a hash function
$H:\{0,1\}^*\rightarrow\mathcal{K}$. 

$\mathsf{Setup}(1^{\lambda},1^{n})$: Run $\mathbb{G}=(p,G_1,G_2,G_T,e)\xleftarrow{\$}\mathsf{GGen}(1^{\lambda})$. Let $g,h$ be the generators of $G_1$, $G_2$, respectively. Run $\mathcal{FAC}.\textsf{Setup}(1^{\lambda}, 1^{n})$ to get $\textsf{pp}$. Sample $\mathbf{A}\xleftarrow{\$}\mathbb{Z}_p^{k\times 2k}$,
$\mathbf{B}\xleftarrow{\$}\mathbb{Z}_p^{k\times k}$, $\textbf{U}_0,\textbf{W}_i\xleftarrow{\$}\mathbb{Z}_p^{2k\times k}$ for $i\in[n]$, $\textbf{v}\xleftarrow{\$}\mathbb{Z}_p^{2k}$, output 
\begin{eqnarray*}
	&\mathsf{msk}=(\textbf{v},\textbf{B},\textbf{U}_0,\textbf{W}_1,\cdots,\textbf{W}_n),\\
	&\mathsf{mpk}=(\textsf{pp},[\textbf{A}]_1,[\textbf{AU}_0]_1,[\textbf{AW}_1]_1,\cdots,[\textbf{AW}_n]_1,e([\textbf{A}]_1,[\textbf{v}]_2)).
\end{eqnarray*}

$\textsf{CredKeyGen}(\textsf{mpk})\rightarrow(\textsf{pk},\textsf{sk})$: This algorithm executes $\mathcal{FAC}.\textsf{CredKeyGen}$ to generate issuer's $\textsf{pk}$ and $\textsf{sk}$.

$\textsf{UserKeyGen}(\textsf{mpk})\rightarrow(\textsf{upk},\textsf{usk})$: This algorithm executes $\mathcal{FAC}.\textsf{UserKeyGen}$ to generate user's $\textsf{upk}$ and $\textsf{usk}$.

$\left<\textsf{Issue.I}(\textsf{sk},\textsf{upk})\rightleftarrows\textsf{Issue.U}(uid,\vec{x},\textsf{usk})\right>\rightarrow \textsf{cred}$. This algorithm executes $\mathcal{FAC}.\textsf{Issue}$ to create user's credential $\textsf{cred}$.

$\mathsf{DKGen}(\mathsf{msk},\vec{x}_{\textsf{rcv}})$: To generate an attribute decryption key for receiver's attributes $\vec{x}_{\textsf{rcv}}$, it samples $\textbf{r}\xleftarrow{\$}\mathbb{Z}_p^{k}$
and outputs $\textsf{DK}_{\vec{x}_{\textsf{rcv}}}=(\textsf{dk}_1,\textsf{dk}_2,\textsf{dk}_{3})$:
$$\textsf{dk}_1=[\textbf{v}+\textbf{U}_0\textbf{B}\textbf{r}]_2,\textsf{dk}_2=[\textbf{B}\textbf{r}]_2,\textsf{dk}_{3}=[\sum\nolimits_{i:x_{r,i}^{(out)}=1}\textbf{W}_{i}\textbf{B}\textbf{r}]_2.$$

$\mathsf{PolGen}(\mathsf{msk},f_{\textsf{rcv}})$: To generate a policy decryption key for receiver's policy $f_{\textsf{rcv}}$, this algorithm samples $(\{\textbf{v}_j\}_{j\in[\hat{m}_r]},\rho_r)\xleftarrow{\$}\textsf{share}(f_{\textsf{rcv}},\textbf{v})$, $\textbf{r}_j\xleftarrow{\$}\mathbb{Z}_p^{k}$
and outputs $\textsf{DK}_{f_{\textsf{rcv}}}=(\{\textsf{dk}_j,\textsf{dk}_{\rho_r(j),j},\{\textsf{dk}_{i,j}\}_{i\in[n]\backslash\{\rho_r(j)\}}\}_{j\in[\hat{m}_r]}):$
$$\textsf{dk}_j=[\textbf{r}_j]_2,\textsf{dk}_{\rho_r(j),j}=[\textbf{v}_j+\textbf{W}_{\rho_r(j)}\textbf{r}_j]_2,\textsf{dk}_{i,j}=[\textbf{W}_{i}\textbf{r}_j]_2,$$
where $\textbf{W}_0=\textbf{0}$, $\hat{m}_r$ is the number of shares for receiver's policy, and $\rho_r$ is a mapping from the indices of the shares to the indices
of receiver's public attributes\footnote{Please refer to the details of linear secret sharing for $\textsf{NC}^1$ in \S 5.1 of \cite{kowalczyk2019compact}.}. For $\rho_r(j)=0$, we have $[n]\backslash\{\rho_r(j)\}=[n]$.

$\mathsf{Enc}(\textsf{cred}_{\textsf{snd}},\vec{x}_{\textsf{snd}}, f_{\textsf{snd}},M)$: The sender selects the private attributes $\vec{x}_{\textsf{snd}}^{(in)}$ for inner layer and public attributes $\vec{x}_{\textsf{snd}}^{(out)}$ for outer layer from $\vec{x}_{\textsf{snd}}$. Then, it runs $\mathcal{FAC}.\textsf{Show}$ to obtain $\textsf{tok}_{\textsf{snd}}$ for  $\vec{x}_{\textsf{snd}}^{(in)}$, $\vec{x}_{\textsf{snd}}^{(out)}$ and $M\in\{0,1\}^*$. Next, it encrypts $(M,\textsf{tok}_{\textsf{snd}})$ using the public attributes $\vec{x}_{\textsf{snd}}^{(out)}$ and policy $f_{\textsf{snd}}$ as follows. 

The sender samples $\widetilde{\textbf{s}},\textbf{s},\textbf{s}_j\xleftarrow{\$}\mathbb{Z}_p^{k}$, $(\{\textbf{u}_j^{\top}\}_{j\in[\hat{m}_s]},\rho_s)\xleftarrow{\$}\textsf{share}(f_{\textsf{snd}},\textbf{s}^{\top}\textbf{A}\textbf{U}_0)$, $K\in G_T$, and compute
$\textsf{CT}_{\vec{x}_{\textsf{snd}}, f_{\textsf{snd}}}=(\textsf{ct}_M,\textsf{ct}_0,\textsf{ct}_1',\textsf{ct}_2',\textsf{ct}_1,\{\widetilde{\textsf{ct}}_{j},\textsf{ct}_{\rho_s(j),j},\{\textsf{ct}_{i,j}\}_{i\in[n]\backslash\{\rho_s(j)\}}\}_{j\in[\hat{m}_s]}):$
\begin{eqnarray*}
	&&\textsf{ct}_M=\mathcal{SE}.\textsf{SEnc}(H(K),(M,\textsf{tok}_{\textsf{snd}})),\\
	&&\textsf{ct}_0=e([\widetilde{\textbf{s}}^{\top}\textbf{A}+\textbf{s}^{\top}\textbf{A}]_1,[\textbf{v}]_2)\cdot K,\\
	&&\textsf{ct}_1'=[\widetilde{\textbf{s}}^{\top}\textbf{A}]_1,~~~\textsf{ct}_2'=\big[\widetilde{\textbf{s}}^{\top}\sum\nolimits_{i:x_{s,i}^{(out)}=1}\textbf{AW}_i\big]_1,\\
	&&\textsf{ct}_1=[\textbf{s}^{\top}\textbf{A}]_1,~~~\widetilde{\textsf{ct}}_{j}=[\textbf{s}_j^{\top}\textbf{A}]_1,\\
	&&\textsf{ct}_{\rho_s(j),j}=\big[\textbf{u}_j^{\top}+\textbf{s}_j^{\top}\textbf{AW}_{\rho_s(j)}\big]_1,~~\textsf{ct}_{i,j}=\big[\textbf{s}_j^{\top}\textbf{AW}_i\big]_1,
\end{eqnarray*}
where $\textbf{W}_0=\textbf{0}$, $x_{s,i}^{(out)}$ is sender's $i$-th public attribute for outer layer, $\hat{m}_s$ is the number of shares for sender's policy, and $\rho_s$ is a mapping from the indices of the shares to the indices
of sender's public attributes.

$\mathsf{Dec}(\textsf{DK}_{\vec{x}_{\textsf{rcv}}},\textsf{DK}_{ f_{\textsf{rcv}}},\textsf{CT}_{\vec{x}_{\textsf{snd}}, f_{\textsf{snd}}})$: The receiver recovers $(M,\textsf{tok}_{\textsf{snd}})$ using $(\textsf{DK}_{\vec{x}_{\textsf{rcv}}},\textsf{DK}_{ f_{\textsf{rcv}}})$ as follows.
It compute $\omega_j,\mu_j$ such that $\textbf{v}=\sum\nolimits_{j\in\mathcal{S}_r}\omega_j\textbf{v}_j$, $\textbf{s}^{\top}\textbf{A}\textbf{U}_0=\sum\nolimits_{j\in\mathcal{S}_s}\mu_j\textbf{u}_j^{\top},$ and calculates
\begin{eqnarray*}
	\small
	K&=&\textsf{ct}_0\cdot \frac{e\big(\textsf{ct}_2',\prod\nolimits_{j\in\mathcal{S}_r}\textsf{dk}_j^{\omega_j}\big)\cdot}{e\big(\textsf{ct}_1',\prod\nolimits_{j\in\mathcal{S}_r}\big(\prod_{i:x_{s,i}^{(out)}=1}\textsf{dk}_{i,j}\big)^{\omega_j}\big)}\\
	&&\cdot \frac{e(\prod\nolimits_{j\in\mathcal{S}_s}(\prod\nolimits_{i:x_{r,i}^{(out)}=1}\textsf{ct}_{i,j})^{\mu_j},\textsf{dk}_2)}{e(\textsf{ct}_1,\textsf{dk}_1)e(\prod\nolimits_{j\in\mathcal{S}_s}\widetilde{\textsf{ct}}_{j}^{\mu_j},\textsf{dk}_3)},
\end{eqnarray*}
\normalsize
where $\mathcal{S}_r=\{j:\rho_r(j)=0\vee x_{s,\rho_r(j)}^{(out)}=1\}$, $\mathcal{S}_s=\{j:\rho_s(j)=0\vee x_{r,\rho_s(j)}^{(out)}=1\}$ and $x_{r,i}^{(out)}$ is receiver's $i$-th public attribute for outer layer.

If  $f_{\textsf{snd}}(\vec{x}_{\textsf{rcv}}^{(out)})=0 \vee f_{\textsf{rcv}}(\vec{x}_{\textsf{snd}}^{(out)})=0$, it outputs $\bot$; otherwise, it recovers
$(M,\textsf{tok}_{\textsf{snd}})\leftarrow\mathcal{SE}.\textsf{SDec}(H(K),\textsf{ct}_M)$. Then, the receiver runs $\mathcal{FAC}.\textsf{Verify}(\textsf{tok}_{\textsf{snd}},M)$ to verify  $\textsf{tok}_{\textsf{snd}}$ for $\vec{x}_{\textsf{snd}}^{(in)}\cup\vec{x}_{\textsf{snd}}^{(out)}$ and $M$. It outputs the message $M$ if the token is valid;
otherwise, it outputs $\bot$. 

The correctness of ACME scheme is analyzed below.

Denote $\mathcal{S}_s=\{j:\rho_s(j)=0\vee x_{r,\rho_s(j)}^{(out)}=1\}$ and $\mathcal{S}_r=\{j:\rho_r(j)=0\vee x_{s,\rho_r(j)}^{(out)}=1\}$.
The correctness of ACME relies on the fact that $\prod\nolimits_{j\in\mathcal{S}_r}\textsf{dk}_j^{\omega_j}=\prod\nolimits_{j\in\mathcal{S}_r}[\textbf{r}_j]_2^{\omega_j}=[\hat{\textbf{r}}]_2$,
\begin{eqnarray*}
	&&\prod\limits_{j\in\mathcal{S}_r}\big(\prod\limits_{i:x_{s,i}^{(out)}=1}\textsf{dk}_{i,j}\big)^{\omega_j}=[\textbf{v}+\sum\limits_{i:x_{s,i}^{(out)}=1}\textbf{W}_{i}\hat{\textbf{r}}]_2,
\end{eqnarray*}
where $\hat{\textbf{r}}=\sum_{j\in\mathcal{S}_r}\omega_j\textbf{r}_j$.
Also we have,
\begin{eqnarray*}
	&&e(\textsf{ct}_1,\textsf{dk}_1)=[\textsf{s}^{\top}\textbf{A}\textbf{v}+\textsf{s}^{\top}\textbf{A}\textbf{U}_0\textbf{B}\textbf{r}]_{T},\\
	&&\prod\nolimits_{j\in\mathcal{S}_s}\widetilde{\textsf{ct}}_{j}^{\mu_j}=\prod\nolimits_{j\in\mathcal{S}_s}[\textbf{s}_j^{\top}\textbf{A}]_1^{\mu_j}=[\hat{\textbf{s}}^{\top}\textbf{A}]_1,\\\\
	&&\prod\limits_{j\in\mathcal{S}_s}\big(\prod\limits_{i:x_{r,i}^{(out)}=1}\textsf{ct}_{i,j}\big)^{\mu_j}=[\textbf{s}^{\top}\textbf{A}\textbf{U}_0+\hat{\textbf{s}}^{\top}\sum\limits_{i:x_{r,i}^{(out)}=1}\textbf{A}\textbf{W}_{i}]_1,
\end{eqnarray*}
where $\hat{\textbf{s}}^{\top}=\sum\nolimits_{j\in\mathcal{S}_s}\mu_j\textbf{s}_j^{\top}$.

Therefore, for all $f_{\textsf{rcv}},\vec{x}_{\textsf{snd}}$ such that $f_{\textsf{rcv}}(\vec{x}_{\textsf{snd}}^{(out)})=1$, we have:
\begin{eqnarray*}
	&& \frac{e\big(\textsf{ct}_2',\prod\nolimits_{j\in\mathcal{S}_r}\textsf{dk}_j^{\omega_j}\big)\cdot}{e\big(\textsf{ct}_1',\prod\nolimits_{j\in\mathcal{S}_r}\big(\prod_{i:x_{s,i}^{(out)}=1}\textsf{dk}_{i,j}\big)^{\omega_j}\big)}\\
	&=& \frac{e\big(\big[\widetilde{\textbf{s}}^{\top}\sum_{i:x_{s,i}^{(out)}=1}\textbf{AW}_i\big]_1,[\hat{\textbf{r}}]_2)} {e\big([\widetilde{\textbf{s}}^{\top}\textbf{A}]_1,[\textbf{v}+\sum_{i:x_{s,i}^{(out)}=1}\textbf{W}_{i}\hat{\textbf{r}}]_2\big)}\\
	&=&\frac{[\widetilde{\textbf{s}}^{\top}\textbf{A}\hat{\textbf{r}}\sum_{i:x_{s,i}^{(out)}=1}\textbf{W}_i\big]_T}
	{[\widetilde{\textbf{s}}^{\top}\textbf{A}\textbf{v}+\widetilde{\textbf{s}}^{\top}\textbf{A}\hat{\textbf{r}}\sum_{i:x_{s,i}^{(out)}=1}\textbf{W}_{i}]_T}=([\widetilde{\textbf{s}}^{\top}\textbf{A}\textbf{v}]_T)^{-1}.
\end{eqnarray*}
where $\hat{\textbf{r}}=\sum_{j\in\mathcal{S}_r}\omega_j\textbf{r}_j$.

For all $f_{\textsf{snd}},\vec{x}_{\textsf{rcv}}$ such that $f_{\textsf{snd}}(\vec{x}_{\textsf{rcv}}^{(out)})=1$, we have:
\begin{eqnarray*}
	&&\frac{e(\prod\nolimits_{j\in\mathcal{S}_s}(\prod\nolimits_{i:x_{r,i}^{(out)}=1}\textsf{ct}_{i,j})^{\mu_j},\textsf{dk}_2)}{e(\textsf{ct}_1,\textsf{dk}_1)\cdot e(\prod\nolimits_{j\in\mathcal{S}_s}\widetilde{\textsf{ct}}_{j}^{\mu_j},\textsf{dk}_3)}\\
	&=&\frac{e([\textbf{s}^{\top}\textbf{A}\textbf{U}_0+\hat{\textbf{s}}^{\top}\sum\nolimits_{i:x_{r,i}^{(out)}=1}\textbf{A}\textbf{W}_{i}]_1,[\textbf{B}\textbf{r}]_2)}{[\textbf{s}^{\top}\textbf{A}\textbf{v}+\textbf{s}^{\top}\textbf{A}\textbf{U}_0\textbf{B}\textbf{r}]_T\cdot e([\hat{\textbf{s}}^{\top}\textbf{A}]_1,[\sum\nolimits_{i:x_{r,i}^{(out)}=1}\textbf{W}_{i}\textbf{B}\textbf{r}]_2)}\\
	&=&\frac{[\textbf{s}^{\top}\textbf{A}\textbf{U}_0\textbf{B}\textbf{r}]_T}{[\textbf{s}^{\top}\textbf{A}\textbf{v}+\textbf{s}^{\top}\textbf{A}\textbf{U}_0\textbf{B}\textbf{r}]_T}=([\textbf{s}^{\top}\textbf{A}\textbf{v}]_T)^{-1}.
\end{eqnarray*}

	\textsf{Theorem 6.2.}
	\label{Theo:ACME}	
	The ACME scheme achieves privacy, authenticity, anonymity and unlinkability if the $MDDH_k$ assumption holds and the underlying FAC is secure.

The proof of Theorem 6.2 is shown in Appendix \ref{Appendix:ACME}.

\subsection{Comparison of ME Schemes}
\label{SubSec:CompareME}

Mathmaking encryption (ME) protects data confidentiality with bilateral control for both senders and receivers in communications.
The existing instantiations of ME include an identity-based scheme (IBME) \cite{ateniese2019match} proposed in CRYPTO'19 and a security enhanced version \cite{chen2022identity} in Asiacrypt'22, but they do not support fine-grained access control.
Table \ref{Tab:CompareME} compares our ACME with IBME \cite{ateniese2019match,chen2022identity}. Since IBME simply sets $\vec{x}_{\textsf{snd}}=\textsf{snd}$, $f_{\textsf{snd}}=\textsf{rcv}$ and $\vec{x}_{\textsf{rcv}}=\textsf{rcv}$, $f_{\textsf{rcv}}=\textsf{snd}$, it requires pre-registration pairing between service providers and clients. On the contrary, ACME relies on bilateral policy matching for service discovery and thus it does not need pre-registration pairing. Furthermore, ACME supports expressive policy (i.e., Boolean formulae equivalent to $\textsf{NC}^1$ circuit), while IBME is constrained to equality policy. On the other hand, the expressive policy in ACME is public to enable fast service discovery, while the equality policy is hidden from the public in IBME.

\renewcommand{\arraystretch}{1.2}
\begin{table}[thbp]\centering 
	\small
	\setlength{\tabcolsep}{0.5mm}
	\begin{tabular}{ccccccc}
		\hline \multirow{2}{*}{Scheme} & \multirow{2}{*}{ME} & \multirow{2}{*}{Anon.} &
		Complex 
		& Selective  & No Pre-reg. & Hidden\\
		&  &  
		& Policy &
		Disclosure & Pairing & Policy\\
		\hline
		IBME \cite{ateniese2019match,chen2022identity} & $\surd$ & $\surd$ & $\times$ &$\times$ & $\times$ & $\surd$\\
		ACME  & $\surd$ & $\surd$ & $\surd$ &$\surd$  &$\surd$& $\times$\\
		\hline
	\end{tabular}
	\caption{Comparison of ME Schemes}		
	\label{Tab:CompareME}	
\end{table}

\section{PriSrv: Privacy-enhanced Fast Service Discovery}
\label{Sec:PriSrv}

\subsection{PriSrv Protocol  and Security}
\label{SubSec:PriSrvProtocol}

\begin{figure*}[htbp]
	\centering
	\small
	\renewcommand{\arraystretch}{1.4}
	\begin{tabular}{|ccc|}
		\hline
		\multicolumn{3}{|c|}{\textbf{Service Broadcast Phase}}\\
		\cdashline{1-3}[1.5pt/2pt]
		\multicolumn{3}{|c|}{Service Provider $S$'s Broadcast: $bid, \textsf{CT}_{B}\leftarrow\mathcal{ACME}.\textsf{Enc}(\textsf{cred}_s,\vec{x}_s, f_s,MSG_B)$}\\
		\multicolumn{3}{|c|}{where $MSG_B=\{bid||Z||Service_{Type}||Service_{Par}||K_c\}$, $z\xleftarrow{\$} \mathbb{Z}_p^*$, $Z\leftarrow h^z\in G_2$, $K_c\leftarrow\mathcal{MAC}.\textsf{KeyGen}(1^{\lambda})$}\\
		\hline\hline
		\multicolumn{3}{|c|}{\textbf{Anonymous Mutual Authentication Phase}}\\
		\cdashline{1-3}[1.5pt/2pt]
		\textbf{Client ($C$)}  && \textbf{Service Provider ($S$)} \\	
		$(\textsf{cred}_c,\textsf{DK}_{\vec{x}_{c}},\textsf{DK}_{ f_c})$ && $(\textsf{cred}_s,\textsf{DK}_{\vec{x}_s},\textsf{DK}_{ f_s})$\\
		\hline
		$MSG_B\leftarrow \mathcal{ACME}.\textsf{Dec}(\textsf{DK}_{\vec{x}_c},\textsf{DK}_{f_c},\textsf{CT}_{B})$&&\\
		$x_1,x_2\xleftarrow{\$} \mathbb{Z}_p^*$, $X_1\leftarrow g^{x_1}\in G_1$, $X_2\leftarrow h^{x_2}\in G_2$  &&\\		
		$\sigma_c\leftarrow\mathcal{MAC}.\textsf{MAC}(K_c,M_c)$&&\\
		
		where $M_c=(``C\rightarrow S",bid,sid,X_1,X_2,Z)$&&\\	
		
		$K_s\leftarrow\mathcal{MAC}.\textsf{KeyGen}(1^{\lambda})$&$\xrightarrow{~~~~~~
			bid,sid,\sigma_c,\textsf{CT}_{c}~~~~~}
		$		
		&$MSG_c\leftarrow \mathcal{ACME}.\textsf{Dec}(\textsf{DK}_{\vec{x}_s},\textsf{DK}_{ f_s},\textsf{CT}_{c})$\\	
		
		$\textsf{CT}_{c}\leftarrow\mathcal{ACME}.\textsf{Enc}(\textsf{cred}_c,\vec{x}_c, f_c,MSG_c)$&&$b_c\leftarrow\mathcal{MAC}.\textsf{Verify}(K_c,M_c,\sigma_c)$\\	
		
		where $MSG_c=(K_s,M_c)$&& If $b_c=0$, abort; otherwise,\\


		&&$y\xleftarrow{\$} \mathbb{Z}_p^*$, $Y\leftarrow g^y\in G_1$\\
		
		$b_s\leftarrow\mathcal{MAC}.\textsf{Verify}(K_s,M_s,\sigma_s)$	
		&&  $\sigma_s\leftarrow\mathcal{MAC}.\textsf{MAC}(K_s,M_s)$\\

		If $b_s=0$, abort; otherwise,
		&$\xleftarrow{~~~~~~~~~~
			M_s,\sigma_s~~~~~~~~~~}$
		&where $M_s=(``S\rightarrow C",bid,sid,X_1,X_2,Y,Z)$\\
		
		$SSK_{c,s}\leftarrow H(Y^{x_1},Z^{x_2})$
		&&$SSK_{c,s}\leftarrow H(X_1^{y},X_2^{z})$\\
		\hline
	\end{tabular}	
	\caption{PriSrv Protocol} 
	\label{Fig:PriSrv} 
\end{figure*}



Fig. \ref{Fig:PriSrv} shows PriSrv, which consists of a service broadcast phase and an  anonymous mutual authentication phase. A unique broadcast identifier $bid$ is assigned to each broadcast cycle; and a unique session identifier $sid$ is assigned to each session. A lifetime should be set for each broadcast cycle (e.g., 30 seconds) by including a timestamp (which can be part of $bid$) and a client verifies the timestamp upon successful decryption to ensure the freshness. Let  $\mathcal{MAC}=(\textsf{Setup},\textsf{KeyGen},\textsf{MAC},\textsf{Verify})$ be a message authentication code (MAC) scheme \cite{chase2014algebraic,chase2020signal,zhang2016practical}, and $H:\{0,1\}^*\rightarrow\mathcal{K}$ be a hash function, where $\mathcal{K}$ is the secret session key space. We assume that 
the generation and dissemination of  anonymous credential, attribute and policy decryption keys to both service provider ($S$) and client ($C$) are performed according to ACME. 

\textit{Service Broadcast Phase}. To initiate a broadcast session with identifier $bid$, $S$ defines a policy $f_s$ to be satisfied by $C$. $S$ selects an ephemeral Diffie-Hellman (DH) exponent $z\xleftarrow{\$} \mathbb{Z}_p^*$ and calculates $Z\leftarrow h^z$. $S$ also runs $K_c\leftarrow\mathcal{MAC}.\textsf{KeyGen}(1^{\lambda})$ to generate an MAC key.
$S$ generates the broadcast message $MSG_B=\{bid||Z||$ $Service_{Type}||Service_{Par}||K_c\}$ including the broadcast identifier, service type and parameters as well as a MAC key for the client. Next, $S$ encrypts $MSG_B$ to a broadcast ciphertext $\textsf{CT}_B=\textsf{CT}_{\vec{x}_s, f_s}\leftarrow\mathcal{ACME}.\textsf{Enc}(\textsf{cred}_s,\vec{x}_s, f_s,MSG_B).$ Then, the broadcast identifier $bid$ and service ciphertext $\textsf{CT}_B$ are announced over the public network.

\textit{Anonymous Mutual Authentication Phase}. To establish a secure session between $C$ and $S$, the anonymous mutual authentication is executed to establish a session key $SSK_{c,s}$. 

(1) To discover the private service, $C$ firstly checks whether  $\vec{x}_c^{(out)}$ satisfies with the anonounced access policy $f_s$ of $S$, i.e. $f_s(\vec{x}_c^{(out)})\overset{?}{=}1$. $C$ quickly filters out mismatched services without decryption when $f_s(\vec{x}_c^{(out)})=0$. Otherwise, $C$ attempts to decrypt $\textsf{CT}_B$ using its attribute and policy decryption keys $(\textsf{DK}_{\vec{x}_c},\textsf{DK}_{f_c})$. If the decryption fails which means $f_{s}(\vec{x}_{c}^{(out)})=0 \vee f_{c}(\vec{x}_{s}^{(out)})=0$, then $C$ aborts. Otherwise, $C$ responds to the broadcast message by executing $\mathcal{ACME}.\textsf{Dec}$ to recover $MSG_B$. 
Next, $C$ selects ephemeral DH exponents $x_1,x_2\xleftarrow{\$} \mathbb{Z}_p^*$ and calculates $X_1=g^{x_1}$, $X_2=h^{x_2}$.
$C$ computes a MAC key $K_s$ and an authentication tag $\sigma_c$ of $M_c=(``C\rightarrow S",bid,sid,X_1,X_2,Z)$ using $K_c$ from $MSG_B$, where $``C\rightarrow S"$ denotes the message direction. Then, $C$ defines a policy $f_c$ to be satisfied by $S$, and selects a set of public attributes and a set of private attributes to be disclosed to $S$. $C$ runs $\mathcal{ACME}.\textsf{Enc}$ to compute $\textsf{CT}_{c}=\textsf{CT}_{\vec{x}_c,f_c}$ and sends it to $S$.

(2) $S$ authenticates $C$'s service access request and computes a secret session key.
$S$ executes $\mathcal{ACME}.\textsf{Dec}$ to recover $MSG_c$. $S$ aborts the protocol if decryption fails.
Next, $S$ verifies $\sigma_c$ and selects DH exponent $y\xleftarrow{\$} \mathbb{Z}_p^*$ to calculate $Y\leftarrow g^y$. $S$ sets $M_s=(``S\rightarrow C",bid,sid,X_1,X_2,Y,Z)$  and generates a tag $\sigma_s\leftarrow\mathcal{MAC}.\textsf{MAC}(K_s,M_s)$ using $K_s$ from $MSG_c$.
Then, $S$  computes a secret session key $SSK_{c,s}\leftarrow H(X_1^{y},X_2^{z})$ and sends $(M_s,\sigma_s)$ to $C$.

(3) Receiving the message, $C$ checks the validity of $\sigma_s$. If it is valid, $C$ computes $SSK_{c,s}\leftarrow H(Y^{x_1},Z^{x_2})$ using the secret values $(x_1,x_2)$. 
Therefore, $C$ and $S$ derive the same session key $SSK_{c,s}$ since $X_1^{y}=Y^{x_1}=g^{x_1y}\in G_1$ and $X_2^{z}=Z^{x_2}=h^{x_2z}\in G_2$.

The following theorem shows the security of PriSrv.

\textsf{Theorem 7.1.}
\label{Theo:PriSrvSecurity}
Suppose that the DDH assumption holds, $\mathcal{ACME}$ is secure, $\mathcal{MAC}$ is unforgeable, and $H$ is a random oracle, then  $\textsf{PriSrv}$ is a secure service discovery protocol and satisfies bilateral anonymity.

The proof of Theorem 7.1 is shown in Appendix \ref{Appendix:PriSrv}.


\subsection{PriSrv Credential Management}

Now we discuss credential management, including credential issuance, credential interoperability, and credential revocation. 

\textit{Credential Issuance}. 
PriSrv leverages FAC to implement a digital identity system for service providers and clients, offering the advantages of unforgeability, anonymous authentication, unlinkability, and selective attribute disclosure.
W3C published Decentralized Identifiers (DIDs) \cite{Manu2021DID} and Verifiable Credentials (VC) \cite{Manu2021W3CVC} specifications to regulate verifiable and decentralized digital identities. Decentralized Identity Foundation (DIF) \cite{DIF} developed a set of standards to support a decentralized identity ecosystem \cite{lesavre2019taxonomic}. Technology giants, such as IBM \cite{IBMDID} and Microsoft \cite{MicrosoftDID}, also provide flexible identity governance and administration services for credentials. CanDID proposed in \cite{maram2021candid} allows user's attributes to be verified by issuers or imported from existing authority systems. PriSrv may follow any of these existing DID frameworks to issue credentials.

\textit{Credential Interoperability}. Credentials complying with standard specifications are interoperable across different platforms.
DID \cite{Manu2021DID} and VC \cite{Manu2021W3CVC} have regulated the process for inteoperable usage of credentials, which is also supported by DIF \cite{DIF}. Backed by Microsoft, Google, Yahoo, IBM, VeriSign, PayPal, and Facebook, the OpenID Foundation\footnote{OpenID Foundation: https://openid.net/foundation.} promotes identity management, federation and interoperation, in compliance with the specifications of W3C. PriSrv may follow these specifications to ensure  credential interoperability when deployed in various service discovery settings.

\textit{Credential Revocation}.
Another consideration of PriSrv is to manage revocation of user's credentials whenever it is necessary. Credential revocation has been intensively studied in the last decade: various types of dynamic accumulators (such as RSA-based and bilinear map based) with ZKP are adopted for credential revocation \cite{boneh2019batching,camenisch2009accumulator,baldimtsi2017accumulators}. It can also be achieved by the combination of ElGamal encryption and Schnorr proofs \cite{bogatov2021anonymous}, or $n$-times unlinkable proofs \cite{camenisch2016scalable}.
PriSrv may incorporate the above techniques to realize credential revocation.

\subsection{Interoperability of PriSrv with Existing Protocols}
\label{SubSec:PriSvcCompatibility}

There are two approaches to make PriSrv work on top of/with different layers of different wireless protocols. The first approach is to position PriSrv at the application layer providing application payload to lower layers. If the payload of PriSrv is oversized in lower layers, the lower layer protocols need to perform segmentation on the sender side and assembling on the receiver side without changing the protocol logics. The second approach is to substitute target protocols at lower layers with PriSrv. However, the second approach requests for specific adaptations of the concret protocols. In the following, we give two examples for each approach, including mDNS and BLE for the first approach, and EAP,  AirDrop for the second approach.

\subsubsection{Privacy Enhanced mDNS and BLE}
\label{SubsubSec:PriSvcDNSBLE}

PriSrv can be integrated in the Vanadium\footnote{Vanadium. https://vanadium.github.io/.} framework for developing privacy enhanced mDNS and BLE. Vanadium provides service discovery APIs to broadcast and scan services over widely deployed protocols, such as mDNS \cite{cheshire2013dns,cheshire2013multicast} and BLE \cite{BLE}. 
mDNS can work in conjunction with DNS Service Discovery (DNS-SD), a companion zero-configuration networking technique specified separately in RFC 6763\footnote{https://tools.ietf.org/html/rfc6763.}. DNS-SD extends the functionality of mDNS by adding additional attributes to the service discovery process. Specifically, the TXT (Text) resource record can be used to carry the attributes in the payload, where the maximum size for a single TXT record in DNS is 65535 bytes. The service broadcast of PriSrv is in the form $(bid,\textsf{CT}_B)$, which takes 531996 bytes in communications on BN256 elliptic curve (100-bit security) \cite{NIST-100bit}. Therefore, privacy enhanced mDNS may use 9 TXT records in DNS-SD to transmit the broadcast ciphertext of PriSrv.




On the other hand, the payload of BLE broadcast is constrained to 31 bytes, which is too small for carrying a broadcast ciphertext in PriSrv. To enable privacy enhanced BLE using PriSrv, the BLE Attribute Protocol (ATT) and Attribute Protocol Data Unit (PDU) Segmentation techniques can be leveraged to extend the payload size. If the payload exceeds the standard packet size in BLE, the ATT protocol (which is used for exchanging data between devices) can segment the payload data into multiple Attribute Protocol Data Units (PDUs) and transmit them sequentially. These PDUs can be reassembled on the receiver side to recover the original payload for the ciphertext in PriSrv.

\begin{figure}
	\begin{center}
		\includegraphics[width=3.3in]{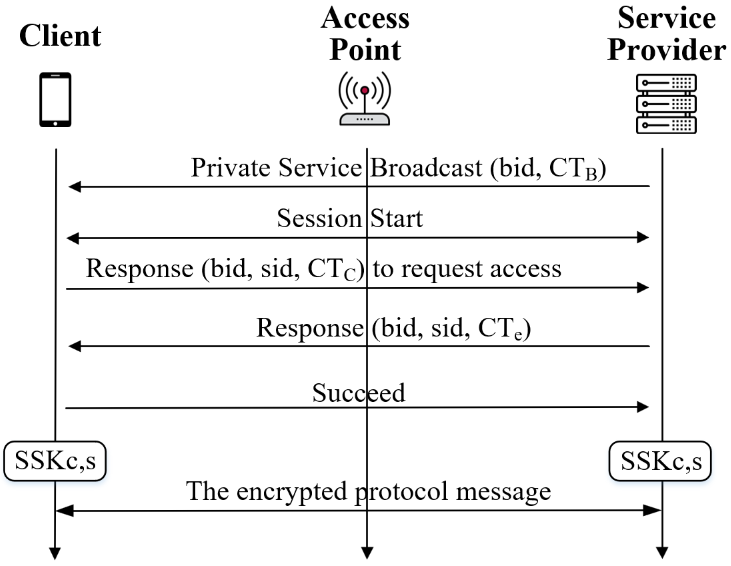}
		\caption{Architecture of Privacy Enahnced EAP}
		\label{Fig:EAP-PS}
	\end{center}
\end{figure}

\subsubsection{Privacy Enhanced EAP}
\label{SubsubSec:PriSvcEAP}

Figure \ref{Fig:EAP-PS} presents the architecture of privacy enhanced EAP using PriSrv, which extends RFC 3748 on Extensible Authentication Protocol (EAP) \cite{aboba2004extensible} to support private service discovery. An access point (AP) is involved in the interactions between client and service provider, which 
acts as a pass-through agent for a backend authentication
server \cite{aboba2004extensible}. The anonymous authentication exchange in privacy enhanced EAP proceeds as follows. (1) At the beginning, the service provider announces private service broadcast information via AP, which contains the broadcast identifier $bid$ and the broadcast ciphertext $\textsf{CT}_B=\textsf{CT}_{\vec{x}_s, f_s}=\mathcal{ACME}.\textsf{Enc}(\cdots,MSG_B)$. This step corresponds to the statement ``the authenticator sends a request to authenticate the peer" in EAP Standard. (2) If the client can decrypt $MSG_B$ from $\textsf{CT}_B$, he/she sends a response packet $(bid,sid,\sigma_c,\textsf{CT}_c)$ as reply to the service provider, where $\textsf{CT}_c=\textsf{CT}_{\vec{x}_c, f_c}=\mathcal{ACME}.\textsf{Enc}(\cdots,MSG_C)$ and $sid$ is a session identifier. (3) Receiving the response, the service provider proceeds to recover $MSG_c$ from $\textsf{CT}_c$. If it succeeds, 
the service provider sends $(M_s,\sigma_s)$ to client, where $M_s$ contains the DH shares for computing a session key and $\sigma_s$ is the corresponding MAC value. After the client verifies $\sigma_s$, it calculates a secret session key $SSK_{c,s}$, and responds with a message ``succeed". (4) Finally, the service provider also computes $SSK_{c,s}$ so that a secure session is established between service provider and client. All subsequent protocol messages are encapsulated in EAPOL frames and re-encapsulated as RADIUS packets on the back-haul. Following \cite{cassola2015authenticating}, the privacy enhanced EAP can be adopted to enhance the privacy of Wi-Fi connections.


\subsubsection{Privacy Enhanced Apple AirDrop}
\label{SubsubSec:PriSvcAirDrop}

AirDrop applies BLE to advertise the hashed identity of a service provider to look for potential clients in their proximity. If a match is confirmed, a TLS handshake is performed to exchange their certificates in cleartext. Both hashed identities and certificates are disclosed to the public, which is subject to  identification and tracking attacks. Following the PrivateDrop mechanism in \cite{heinrich2021privatedrop}, we can improve the privacy of AirDrop by avoiding transmitting private information (such as identifier) of service provider during the advertising phase using BLE, and then encrypt the certificates of both parties using ACME at the beginning of TLS handshake. Apple may take the role of credential issuer in this case to generate necessary secret keys and credentials in addition to their existing iCloud certificates.


\subsection{Limitations of PriSrv}

One limitation of PriSrv lies in its large message size when compared to existing protocols. This large size of the outer discovery broadcast poses a scalability challenge, particularly on slower networks like BLE, resulting in high transmission overhead and reception delays. Moreover, on networks such as Wi-Fi, broadcasts must always be transmitted at the lowest feasible speed, further exacerbating airtime congestion.

The issue of large message sizes also compounds another challenge in wireless networks: packet loss, especially when using opportunistic transmission protocols like mDNS, which relies on UDP. Although UDP packets can theoretically reach sizes of up to 64K, they are fragmented to align with the Maximum Transmission Unit (MTU) of the physical network. Any loss of a single fragment results in the entire packet being discarded. While Wi-Fi incorporates a rudimentary acknowledgment and retry mechanism, this only applies to unicast traffic and can only recover from brief RF disruptions.  Consequently, clients must wait for the broadcast ciphertext in the subsequent round to receive full packets, causing additional delays in reception. How to design efficient privacy-preserving discovery protocols remain an open problem for future research.

PriSrv protects its own payloads for achieving unlinkability at its positioned layer. As for achieving unlinkability at lower layers, the lower layer headers must be protected using specific anti-tracking mechanisms designed at lower layers.
For example, PriSrv can work  with MAC randomization  mechanism at data link layer. Smartphone manufacturers (e.g., Apple iOS) incorporate
MAC randomization for Wi-Fi and AWDL connections to
provide unlinkability at the link layer, but devices can still be
tracked at the layer where PriSrv resides. PriSrv complements the MAC randomization mechanism to realize unlinkability in different layers. Nevertheless, the current MAC address randomization approach (e.g., as implemented in Android
and iOS) only performs randomization once when connecting
to a new network and not with each subsequent connection.
To achieve more robust unlinkability, a more effective MAC
address randomization strategy should be devised to ensure
unlinkability for each individual connection. Achieving unlinkability across multiple layers remains a persistent challenge.

\section{Implementation and Comparison}
\label{Sec:Evaluation}

\label{SubSec:CompareDiscovery}

We benchmark the performance of PriSrv on various hardware platforms, including desktop, laptop, smartphone, and Raspberry Pi as shown  in Table \ref{Tab:EvaSys}. 
Three asymmetric elliptic curves are selected from the MIRACL library \cite{MIRACL} for evaluation, including MNT159 (80-bit security), MNT201 (90-bit security), and BN256 (100-bit security) \cite{NIST-100bit}. We use AES-CTR with 100-bit keys to instantiate the $\textsf{SEnc}$/$\textsf{SDec}$ algorithms in PriSrv, using SHA-256 as the hash function, and use $\textsf{MAC}_{\textsf{GGM}}$ \cite{chase2014algebraic} as suf-cma secure MAC. The source code of our experiments is written in C/C++ and publicly available on GitHub\footnote{Source Code: https://github.com/prisrv.}. For each test case, we report the average over 20 executions.

\renewcommand{\arraystretch}{1.2}
\begin{table}[thbp]\centering 
	\small
	\setlength{\tabcolsep}{0.1mm}
	\begin{tabular}{ccc}
		\hline No. & Type & Hardware Platforms \\
		\hline
		1 & Desktop & Intel® Core™ i9-7920X CPU @ 2.9GHz$\times$12, 16GB \\
		2 & Laptop & Intel® Core™ i5-10210U CPU @ 1.6GHz$\times$4, 8GB \\
		3 & Phone & ARM Cortex @2.84GHz+3$\times$2.4GHz, 4GB \\
		4 & Raspberry Pi & ARM Cortex @1.5GHz$\times$4, 2GB \\
		\hline
	\end{tabular}	
	\caption{Hardware Platforms for Experiments}
	\label{Tab:EvaSys}	
\end{table}

\subsection{Evaluation of FAC}
\label{SubSec:CompareAC}

In Table \ref{Tab:CompareAC}, we compare FAC with typical anonymous credential schemes.
FAC constructs a constant-size anonymous credential. With FAC, a verifier only needs to conduct $k$ operations to check the proof of $k$ attributes, which is an  up-to-date optimal solution. 
The $O(1)$ communication complexity in  \cite{camenisch2015composable} for its $\textsf{Show}$ algorithm (i.e., $|\textsf{Show}|$) is composed of about 100 group elements. 
Since the scheme in \cite{camenisch2015composable} is the only one to achieve UC security in Table \ref{Tab:CompareAC}, these overheads can be seen as a tradoff between efficiency and security.
Compared with  \cite{fuchsbauer2019structure}, our credential only consists of 2 elements in $G_2$, which is approximately $2\times$ more efficient than that of \cite{fuchsbauer2019structure} (i.e., $3|G_1|+|G_2|+2|\mathbb{Z}_p|$).
To show a credential in FAC, a user transmits 2 elements in $G_1$, 2 in $G_2$, 1 in $G_T$ and three scalar elements, which is smaller than 8 elements in $G_1$, 1 in $G_2$ and two scalar elements for \cite{fuchsbauer2019structure}.

\renewcommand{\arraystretch}{1.2}
\begin{table}[thbp]\centering 
	\setlength{\tabcolsep}{0.9mm}
	\small
	\begin{tabular}{ccccccccc}		
		\hline 
		Ref. & 
		$\textsf{Issue}$ &
		$|\textsf{cred}|$ & $|\textsf{Show}|$ & $\textsf{Show}$ & $\textsf{Verify}$ \\
		\hline
		\cite{camenisch2001efficient} & $O(1)$ & $2|QR_N|+|\ell_N|$ & $O(k)$ &  $O(k)$ &  $O(k)$ \\
		\cite{paquin2011u} & $O(1)$ & $|G_1|+2|Z_q|$ & $O(n)$ &  $O(n)$ &  $O(n)$ \\
		\cite{ringers2017efficient} &  $O(n)$ & $(2n+4)(|G_1|+|Z_q|)$ & $O(n)$ &  $O(n)$ &  $O(n)$ \\
		\cite{camenisch2015composable} & $O(1)$ & $6|G_1|+2|G_2|+|Z_p|$ & $O(1)$ &  $O(n-k)$ &  $O(k)$ \\
		\cite{fuchsbauer2019structure} & $O(1)$ &  $3|G_1|+|G_2|+2|Z_p|$ & $O(1)$ &  $O(n-k)$ &  $O(k)$ \\		
		FAC & $O(1)$ & $2|G_2|$& $O(1)$ & $O(n-k)$ &  $O(k)$  \\
		\hline
	\end{tabular}
	\caption{Comparison of Anonymous Credential Schemes}	
	\label{Tab:CompareAC}
	\vspace{-1mm}
	\begin{flushleft}
		$|\textsf{Show}|$ indicates the communication cost for showing $k$ attributes. $\textsf{Show}$ and $\textsf{Verify}$ represent the computational costs. $QR_N$ represents the group of quadratic residues modulo a composite $N$, and $\ell_N$ is an RSA moduli defined in \cite{camenisch2001efficient}.
	\end{flushleft}
\end{table}

\begin{table}[thbp]\centering 
	\small
	\setlength{\tabcolsep}{2.2mm}
	\begin{tabular}{cccccc}
		\hline Ref. &  $\boldsymbol{|\textsf{Cred}|}$  & $\boldsymbol{\textsf{Issue}}$  & $\boldsymbol{\textsf{Show}}$  &
		$\boldsymbol{\textsf{Verify}}$ \\
		\hline
		Idemix \cite{camenisch2001efficient} &0.671 & 76.437		&283.245	&210.783		\\
		UProve \cite{paquin2011u} &0.768 & 37.422	&12.264	&33.231	 \\
		\cite{camenisch2015composable} &1.352 & 389.513	&657.024	&253.453	\\
		\cite{fuchsbauer2019structure} &0.736 & 371.126	& 87.625	&284.719 \\
		FAC  &0.544	 &39.387	&28.302	&65.819 \\
		\hline
	\end{tabular}
	\caption{Performance of AC (ms/KB) (BN256)}
	\label{Tab:PerFACCompare}	
\end{table}

Table \ref{Tab:PerFACCompare} compares the performance of FAC with Idemix, UProve and the schemes in \cite{camenisch2015composable,fuchsbauer2019structure} on desktop. The parameters for FAC are $n=10$ and $|\mathcal{I}|=4$. UProve incurs a low cost without providing multi-show unlinkability, while the other schemes support this privacy property. FAC has the smallest credential size (0.544 KB) in this comparison and its overheads for $\textsf{Issue}$, $\textsf{Show}$, $\textsf{Verify}$ are the lowest or the second lowest among those supporting multi-show unlinkability.

\subsection{Evaluation of ACME and PriSrv}

Table \ref{Tab:PerACME} presents the computation cost (comp.) and communication cost (comm.) of ACME for different algorithms on desktop following the example in $\S$\ref{Sec:PriSrvOverview}, where the parameters are $n=10$, $k=2$, $\hat{m}=9$ and $|\mathcal{S}|=9$. The system setup time, performed on various curves, ranges from 20.526 ms to 33.344 ms. The sizes of master public key ($|\textsf{mpk}|$) and master secret key ($|\textsf{msk}|$) for BN256 are 4.128 KB and 1.6 KB, respectively. The credential key generation ($\textsf{CredKeyGen}$) and user key generation ($\textsf{UserKeyGen}$) cost no more than 118.622 ms and 9.102 ms, respectively. The credential issue ($\textsf{Issue}$) algorithm is efficient (39.383 ms) and the size of generated anonymous credential ($|\textsf{cred}|$) is merely 0.544 KB on BN256 curve, which is consistant with the theoretical analysis of FAC in $\S$\ref{SubSec:CompareAC}. The size of  attribute decryption key ($\textsf{DK}_{\vec{x}}$) and the size of policy decryption key ($\textsf{DK}_{f}$) are no more than 2.72 KB and 44.064 KB, respectively.
The computation costs for encryption and decryption are less than 188 ms and 232 ms, respectively, on BN256 curve. While the computation costs on MNT159 and MNT201 are significantly lower than those on BN256.

\renewcommand{\arraystretch}{1.2}
\begin{table}[thbp]\centering 
	\small
	\setlength{\tabcolsep}{0.6mm}
	\begin{tabular}{|c|c|c|c|c|}
		\hline \multirow{3}{*}{\textbf{Comp. (ms)}} & \multicolumn{3}{c|}{\textbf{Curve and Security Level}}\\
		\cline{2-4}
		& MNT159 & MNT201 & BN256 \\
		& (80-bit Security) & (90-bit Security) & (100-bit Security) \\
		\hline $\boldsymbol{\textsf{Setup}}$ & 20.526 & 26.882 & 33.344 \\ $\boldsymbol{\textsf{CredKeyGen}}$ & 98.261 & 105.883 & 118.622 \\ $\boldsymbol{\textsf{UserKeyGen}}$ & 6.153	& 7.582	& 9.102\\ $\boldsymbol{\textsf{Issue}}$ & 29.298 &	33.783  &  39.383 \\ $\boldsymbol{\textsf{DKGen}}$ & 21.63 & 18.64	& 15.75	\\ $\boldsymbol{\textsf{PolGen}}$ & 359.807 & 327.796	& 237.675 \\
		$\boldsymbol{\textsf{Enc}}$ & 146.931 &	167.337 & 187.822 \\
		$\boldsymbol{\textsf{Dec}}$ & 123.772& 188.346 & 231.214\\
		\hline\hline \textbf{Comm. (KB)} 
		& MNT159 & MNT201 & BN256 \\
		\hline $\boldsymbol{|\textsf{mpk}|/|\textsf{msk}|}$ &	1.044 / 1.2 & 1.332 / 1.36  & 4.128 / 1.6 \\ $\boldsymbol{|\textsf{pk}|/|\textsf{sk}|}$ & 0.91 / 0.18	& 1.158 / 0.204	& 3.408 / 0.24 \\
		$\boldsymbol{|\textsf{upk}|/|\textsf{usk}|}$ & 0.116 / 0.03	& 0.148 / 0.034	& 0.4 / 0.04\\
		$\boldsymbol{|\textsf{DK}_{\vec{x}}|/|\textsf{DK}_f|}$ & 0.86 / 13.932	& 1.1 / 17.82	& 2.72 / 44.064\\
		$\boldsymbol{|\textsf{cred}|/|\textsf{CT}|}$ &  0.172 / 164.34	&  0.220 / 212.964	& 0.544 / 537.984 \\
		\hline		
	\end{tabular}
	\caption{Performance of ACME}	
	\label{Tab:PerACME}	
\end{table}

\renewcommand{\arraystretch}{1.2}
\begin{table}[thbp]\centering 
	\small
	\setlength{\tabcolsep}{1.15mm}
	\begin{tabular}{|c|c|c|c|c|c|c|c|c|c|c|c|c|c|c|}
		\hline  \multirow{4}{*}{Device}& \multicolumn{6}{c|}{\textbf{Private Service Broadcast}} \\
		\cline{2-13}  & \multicolumn{2}{c|}{MNT159} & \multicolumn{2}{c|}{MNT201} & \multicolumn{2}{c|}{BN256} \\
		& \multicolumn{2}{c|}{(80-bit Security)} & \multicolumn{2}{c|}{(90-bit Security)} & \multicolumn{2}{c|}{(100-bit Security)} \\
		\cline{2-13}  & Comp. & Comm. & Comp. & Comm. & Comp. & Comm. \\
		\hline
		1 & 158.931 & 164.34 &	180.337 & 212.96 &	202.822 & 537.98\\
		2&	216.493 & 164.34 & 261.059 & 212.96 &	287.287 & 537.98\\
		3&	385.553 & 164.34 & 443.686 & 212.96 & 482.725 & 537.98\\
		4&	638.259 & 164.34 & 880.868 & 212.96 &	1188.392 & 537.98\\		
		\hline
		\hline  \multirow{4}{*}{Device}&
		\multicolumn{6}{c|}{\textbf{Anonymous Mutual Authentication}}\\
		\cline{2-13}  &  \multicolumn{2}{c|}{MNT159} & \multicolumn{2}{c|}{MNT201} & \multicolumn{2}{c|}{BN256} \\
		& \multicolumn{2}{c|}{(80-bit Security)} & \multicolumn{2}{c|}{(90-bit Security)} & \multicolumn{2}{c|}{(100-bit Security)} \\
		\cline{2-13}  & Comp. & Comm. & Comp. & Comm. & Comp. & Comm.  \\
		\hline
		1 &	429.282	& 164.45 &517.512& 213.09 & 673.039	& 538.83\\	
		2&	576.161& 164.45 &686.054& 213.09 &	854.177 & 538.83\\	
		3&	727.572& 164.45 &892.712& 213.09 &	972.163 & 538.83\\
		4&	1224.365& 164.45 &1832.187& 213.09 & 2711.013 & 538.83\\		
		\hline
	\end{tabular}
	\caption{Performance of PriSrv (ms/KB)}	
	\label{Tab:PerPriSvc}
\end{table}

%

Using the same example and parameter settings, Table \ref{Tab:PerPriSvc} provides a comprehensive evaluation of PriSrv on multiple hardware platforms with various elliptic curves and security levels. The communication overheads of the broadcast and mutual authentication phases are similar, as both of them are primarily determined by the size of the ACME ciphertext. The communication costs remain the same for different platforms, and the computation costs gradually increase from desktop to Raspberry Pi. The desktop, laptop and smartphone take less than 0.483 s for private service broadcast, and less than 0.973 s for anonymous mutual authentication. Raspberry Pi is relatively resource-limited, which takes 1.189 s and 2.712 s for private broadcast and authentication, respectively. The experimental results show that the broadcast and anonymous mutual authentication delays on the first three devices stay well below 1 s, which humans perceive the delays as an “immediate response” \cite{heinrich2021privatedrop,Stuart1991Information}, while the delays on Raspberry Pi are longer but not too significant.

\begin{figure}
	\centering
	\includegraphics[width=3.5in]{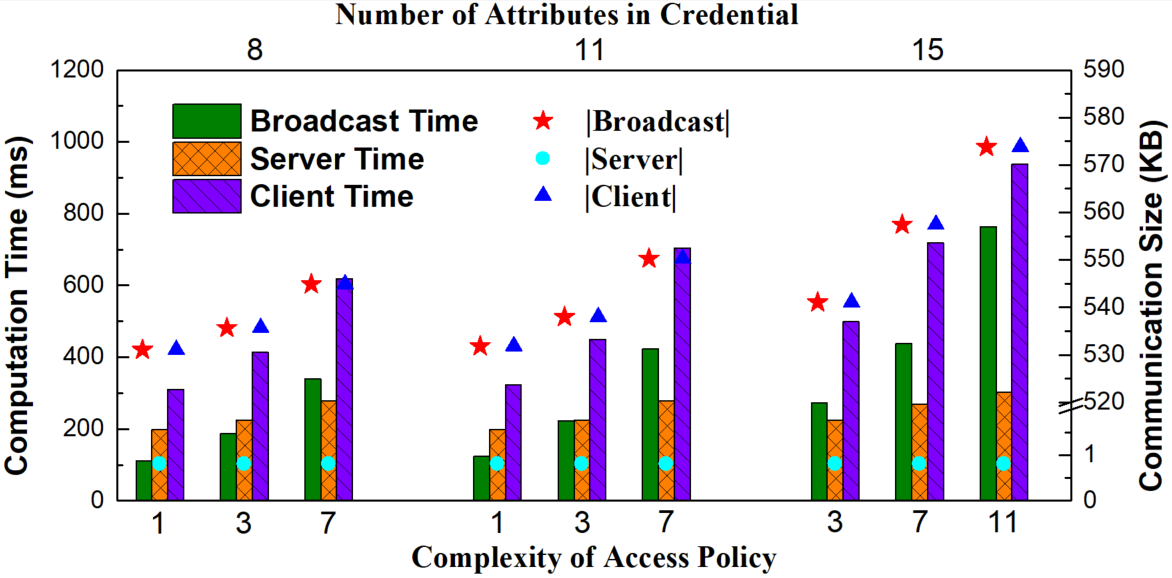}
	\caption{Computation/Communication cost of PriSrv}
	\label{Fig:PriSvcPerf}
\end{figure}

\begin{figure}
	\centering
	\includegraphics[width=3.5in]{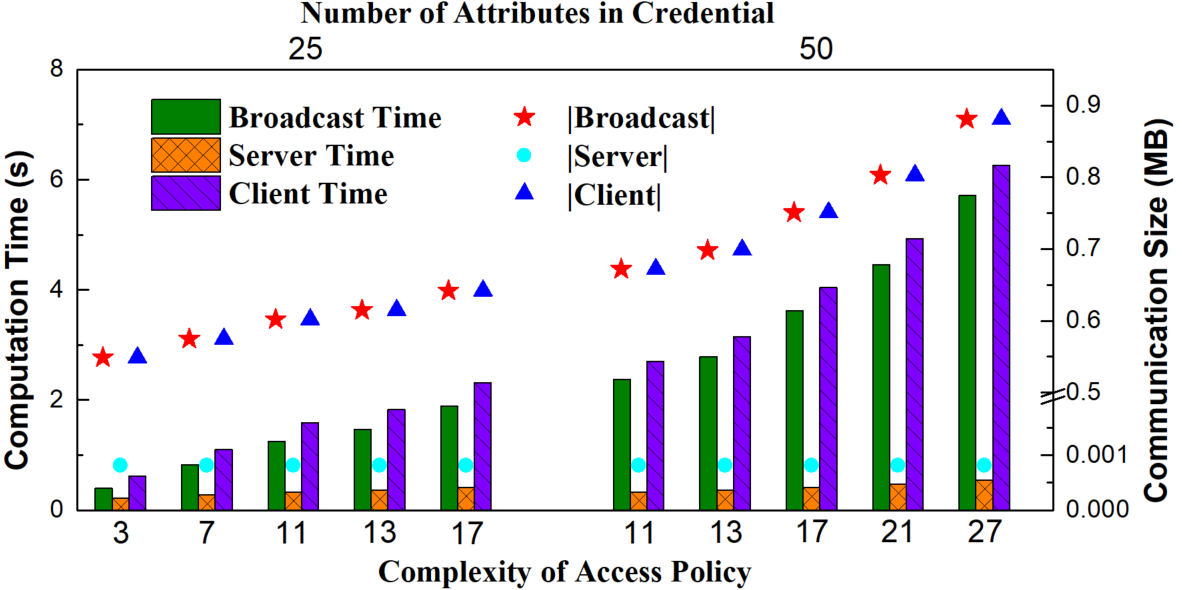}
	\caption{Performance of PriSrv with Complex Policies}
	\label{Fig:PriSvcScale}
\end{figure}

We further implement PriSrv in wireless environment by adapting an open-source project of Wi-Fi Alliance \cite{WiFi}, which implements IEEE 802.1X and enables the deployment of clients (running \textit{wpa$\_$supplicant} program of the project) and service providers (running \textit{hostapd} program). Experiments of PriSrv in wireless communication use two laptops running Ubuntu 20.04. We deploy one laptop as the service provider and the other as the client.
Fig. \ref{Fig:PriSvcPerf}-\ref{Fig:PriSvcScale} present the broadcast time ($T_B$), server's computation time ($T_S$) and client's computation time ($T_C$) during the anonymous mutual authentication phase, where the total mutual authentication time is $T_{MA}=T_S+T_C$. The left y-axis shows the computation time, and right y-axis indicates the communication overhead in the broadcast phase ($|\textsf{Broadcast}|$) and the communication overhead of service provider/client in the authentication phase ($|\textsf{Server}|$/$|\textsf{Client}|$). The performance of PriSrv varies with the attribute number $n$ (top x-axis) and the wire number $\hat{m}$ of $\textsf{NC}^1$ (i.e. number of shares for policy, bottom x-axis), where the matrix size is fixed to be $k=2$. 

In Fig. \ref{Fig:PriSvcPerf}, we set $n=8,11, 15$ and vary the complexity of access policy $\hat{m}$ among $\{1, 3, 7, 11\}$ for practicality test. For $n=15$, $\hat{m}=11$, we have $T_B=763.892$ ms, $T_S=302.973$ ms and $T_C=938.395$ ms, $|\textsf{Client}|$=573.852 KB and $|\textsf{Server}|=0.82$ KB. 
Fig. \ref{Fig:PriSvcScale} sets $n=25,50$ and varies $\hat{m}$ among $\{3,7, 11,13,17,21,27\}$ for testing complex policies involving large number of attributes. The computation time increases with the number of attributes and complexity of access policies. For $n=50$ and $\hat{m}=27$, the computation costs are $T_B=5.711$s, $T_S=0.549$s, $T_C=6.262$s. The communication cost in the broadcast phase grows from 0.549 MB to 0.881 MB. The transmission overhead of the server in the authentication phase remains relatively low (no more than 0.82 KB), while that of the client is mainly influenced by the ACME ciphertext, ranging from 0.549 MB to 0.881 MB.
The comprehensive evaluations demonstrate the efficiency of PriSrv in wireless communications.

\section{Conclusion}
\label{Sec:Conclusion}

This paper presented PriSrv, a privacy-enhanced service discovery protocol with high usability, for wireless communications. PriSrv enforces bilateral flexible policy control for anonymous mutual authentication, making it an ideal solution for enhancing privacy protection in popular wireless communication protocols such as EAP, mDNS, BLE, and AirDrop. 
PriSrv is built upon a novel primitive called anonymous credential-based matchmaking encryption (ACME), which extends the concept of ME proposed in CRYPTO'19 by offering selective attribute disclosure and eliminating the need for heavy cryptographic tools. ACME relies on a newly designed Fast Anonymous Credential (FAC) scheme to generate and verify authentication tokens that are unlinkable across multiple protocol sessions. Comprehensive experimental evaluations and comparisons demonstrated that ACME outperforms existing ME instantiations in terms of functionality and efficiency, which makes it a contribution of independent interests. Formal security models are provided to prove that PriSrv, ACME and FAC have desired security and privacy properties. 
Benchmarks on multiple hardware platforms demonstrated that PriSrv is suitable for interoperating with a wide range
of service discovery protocols with enhanced privacy protection and high usability.


\section*{Acknowledgment}

The authors would like to thank the shepherd and anonymous reviewers for their valuable comments and insightful suggestions. Yang Yang is supported by Lee Kong Chian Professor Fund, National Natural Science Foundation of China under Grant No. 62372110, and Fujian Provincial Natural Science of Foundation under Grant 2023J02008. Robert Deng is supported by AXA Research Fund. Guomin Yang is supported by Lee Kong Chian Fellowship awarded by Singapore Management University. HweeHwa Pang is supported by Lee Kong Chian Chair Professor Fund. Jian Weng is supported by National Key Research and Development Plan of China under Grant No. 2020YFB1005600, National Natural Science Foundation of China under Grant Nos. 61825203, 62332007 and U22B2028, Science and Technology Major Project of Tibetan Autonomous Region of China under Grant No. XZ202201ZD0006G, National Joint Engineering Research Center of Network Security Detection and Protection Technology, Guangdong Key Laboratory of Data Security and Privacy Preserving, Guangdong Hong Kong Joint Laboratory for Data Security and Privacy Protection, and Engineering Research Center of Trustworthy AI, Ministry of Education.



%


\clearpage
\appendix

\subsection{FAC: Security Model and Proof}
\label{Appendix:FAC}

\noindent\textsf{(1) Security Model}

Following the definitions in \cite{chase2014algebraic,sanders2020efficient}, we define  \textit{correctness}, \textit{unforgeability}, \textit{anonymity} and \textit{unlinkability} for anonymous credential scheme.
The lists in the security models are given in Table \ref{Tab:SecExplists}.

\textsf{Definition A.1}
\label{FAC:correctness}
Let $\mathcal{D}$ be the universe of user identity, and $\Omega$ be the universe of  attribute sets. Then an anonymous credential scheme $\mathcal{AC}$ is \textit{correct} for $\mathcal{D},\Omega$ if all $uid\in\mathcal{D}$, all $\vec{x}\subseteq\Omega$, for all security parameter $\lambda$, 
$$\textsf{Pr}\left[
\begin{array}{l}
\textsf{pp}\xleftarrow{\$}\textsf{Setup}(1^{\lambda}, 1^{n});(\textsf{pk},\textsf{sk})\xleftarrow{\$}\textsf{CredKeyGen}(\textsf{pp});\\
	(\textsf{upk},\textsf{usk})\xleftarrow{\$}\textsf{UserKeyGen}(\textsf{pp});\\
	\textsf{cred}\xleftarrow{\$}\left<\textsf{Issue.I}(\textsf{sk},\textsf{upk})\rightleftarrows\textsf{Issue.U}(uid,\vec{x},\textsf{usk})\right>;\\
	\textsf{tok}\xleftarrow{\$}\textsf{Show}(uid,\{x_i\}_{i\in\mathcal{I}},\textsf{cred},\textsf{usk},m):\\
	\textsf{Verify}(\textsf{tok},m)=0	
\end{array}
\right]$$
$\leq\nu(\lambda)$, where $\nu$ is a negligible function.

\textsf{Definition A.2}
\label{FAC:unforgeability}
An AC scheme satisfies \textit{unforgeability} if for any PPT adversary $\mathcal{A}$, there exists a negligible function $\nu$ such that $\textsf{Adv}_{\mathcal{AC}}^{\textsf{unforge}}(\lambda)\overset{\textsf{def}}{=}$
$$\textsf{Pr}\left[b=1~
\begin{array}{|l}
	\textsf{pp}\leftarrow\textsf{Setup}(1^{\lambda},1^n),\\
	(\textsf{pk,sk})\leftarrow\textsf{CredKeyGen}(\textsf{pp})\\
	(uid^*,\vec{x}^*,m^*,\textsf{cred}^*,\textsf{tok}^*)\leftarrow\mathcal{A}^{\mathcal{O}(\textsf{sk},\cdot)}(\textsf{pp},\textsf{pk})\\		
	b\leftarrow\textsf{Verify}(\textsf{tok}^*,m^*)\\
	\text{return } b~\text{if}\quad (m^*,\vec{x}^*,\textsf{cred}^*,\textsf{tok}^*)\notin\mathcal{L}_{\textsf{show}}\\
	\quad\quad\wedge~\textsf{cred}^*\notin\mathcal{L}_{\textsf{issue}}\wedge~(uid^*,\vec{x}^*)\notin\mathcal{L}_{\textsf{corrupt}}\\
	\text{else abort}			
\end{array}
\right]$$		
$\leq\nu(\lambda)$, where the oracle set $\mathcal{O}=\{\textsf{UserKeyGen},\textsf{Issue},$ $\textsf{Show},$ $\textsf{Corrupt}\}$ is implemented by $\textsf{UserKeyGen}(\textsf{pp},\cdot),$ $\textsf{Issue}(\textsf{sk},\textsf{st},\cdot)$, $\textsf{Show}(\textsf{pk},\cdot)$ and $\textsf{Corrupt}(\cdot)$.

\textsf{Definition A.3}
\label{FAC:anonymity}
An AC scheme $\mathcal{AC}$ satisfies \textit{anonymity} if for any PPT adversary $\mathcal{A}$, there exists a negligible function $\nu$ such that $\textsf{Adv}_{\mathcal{AC}}^{\textsf{anon}}(\lambda)\overset{\textsf{def}}{=}$	
$$\textsf{Pr}\left[b'=b~
\begin{array}{|l}
	\textsf{pp}\leftarrow\textsf{Setup}(1^{\lambda},1^n),\\
	(\textsf{pk,sk})\leftarrow\textsf{CredKeyGen}(\textsf{pp})\\
	(uid_0^*,uid_1^*,\vec{x}^*,m^*)\leftarrow\mathcal{A}^{\mathcal{O}(\textsf{sk},\cdot)}(\textsf{pp},\textsf{pk})\\
	\text{abort if } \exists d\xleftarrow{\$}\{0,1\}:\\
	(uid_d^*,\vec{x}^*,\textsf{cred}_d)\notin\mathcal{L}_{\textsf{honest}};~
	b\xleftarrow{\$}\{0,1\}\\
	\textsf{tok}^*\leftarrow\textsf{Show}(uid_b^*,\textsf{cred}_b,\textsf{usk}_b,\vec{x}^*,m^*)\\
	b'\leftarrow\mathcal{A}^{\mathcal{O}(\textsf{sk},\cdot)}(\textsf{pp},\textsf{tok}^*)\\
	\text{return } b'~\text{if}~~~ \textsf{cred}^*\notin\mathcal{L}_{\textsf{issue}}\wedge\\ (m^*,\vec{x}^*,\textsf{tok}^*)\notin\mathcal{L}_{\textsf{show}}~\wedge~(uid_{0/1}^*,\vec{x}^*)\notin\mathcal{L}_{\textsf{corrupt}}\\
	\text{else abort}			
\end{array}
\right]$$
$\leq\nu(\lambda)$, where the oracle set $\mathcal{O}=\{\textsf{UserKeyGen},\textsf{Issue},$ $\textsf{Show},$ $\textsf{Corrupt}\}$ is implemented by $\textsf{UserKeyGen}(\textsf{pp},\cdot),$ $\textsf{Issue}(\textsf{sk},\textsf{st},\cdot)$, $\textsf{Show}(\textsf{pk},\cdot)$ and $\textsf{Corrupt}(\cdot)$.

\textsf{Definition A.4}
\label{FAC:unlinkability}
An AC scheme $\mathcal{AC}$ satisfies \textit{unlinkability} if for any PPT adversary $\mathcal{A}$, there exists a negligible function $\nu$ such that $\textsf{Adv}_{\mathcal{AC}}^{\textsf{unlink}}(\lambda)\overset{\textsf{def}}{=}$	
$$\textsf{Pr}\left[b'=b~
\begin{array}{|l}
	\textsf{pp}\leftarrow\textsf{Setup}(1^{\lambda},1^n),\\
	(\textsf{pk,sk})\leftarrow\textsf{CredKeyGen}(\textsf{pp})\\
	(uid_0^*,uid_1^*,\mathcal{I}^*,\vec{x}_0^*,\vec{x}_1^*,m^*)\leftarrow\mathcal{A}^{\mathcal{O}(\textsf{sk},\cdot)}(\textsf{pp},\textsf{pk})\\
	\text{where } \mathcal{I}^*\subseteq[1,n],\vec{x}_d^*=\{x_i^{(d)}\}_{i\in[1,n]},d\in\{0,1\}\\	
	\text{abort if }\exists j\in\mathcal{I}^*: x_j^{(0)}\neq  x_j^{(1)}\\
	\text{or } \exists d\xleftarrow{\$}\{0,1\}:
	(uid_d^*,\vec{x}_d^*,\textsf{cred}_d)\notin\mathcal{L}_{\textsf{honest}},\\
	b\xleftarrow{\$}\{0,1\}\\
	\textsf{tok}_b^*\leftarrow\textsf{Show}(uid_b^*,\textsf{cred}_b^*,\textsf{usk}_b^*,\{x_i^{(b)}\}_{i\in\mathcal{I}^*},m^*)\\
	b'\leftarrow\mathcal{A}^{\mathcal{O}(\textsf{sk},\cdot)}(\textsf{pp},\textsf{tok}_b^*)\\
	\text{return } b'~\text{if }~\forall d'\in\{0,1\},~~~\textsf{cred}_{d'}^*\notin\mathcal{L}_{\textsf{issue}}\wedge\\ (m^*,\vec{x}_{d'}^*,\textsf{tok}_{d'}^*)\notin\mathcal{L}_{\textsf{show}}~\wedge~(uid_{d'}^*,\vec{x}_{d'}^*)\notin\mathcal{L}_{\textsf{corrupt}}\\
	\text{else abort}			
\end{array}
\right]$$
$\leq\nu(\lambda)$, where the oracle set $\mathcal{O}=\{\textsf{UserKeyGen},\textsf{Issue},$ $\textsf{Show},$ $\textsf{Corrupt}\}$ is implemented by $\textsf{UserKeyGen}(\textsf{pp},\cdot),$ $\textsf{Issue}(\textsf{sk},\textsf{st},\cdot)$, $\textsf{Show}(\textsf{pk},\cdot)$ and $\textsf{Corrupt}(\cdot)$.

It is easy to see that the security definition of \textit{unlinkability} implicitly implies that of \textit{anonymity}.

\begin{table}[thbp]\centering
	\footnotesize
	\setlength{\tabcolsep}{3.2mm}
	\begin{tabular}{ll}
		\hline
		\textbf{List} & \textbf{Description}\\
		\hline 
		$\mathcal{L}_{\textsf{registerd}}$ & registered users\\
		$\mathcal{L}_{\textsf{issue}}$ & credentials that have been issued\\
		$\mathcal{L}_{\textsf{show}}$ & credentials that have been shown\\
		$\mathcal{L}_{\textsf{honest}}$ & registered users that are honest\\
		$\mathcal{L}_{\textsf{corrupt}}$ & registered users that are corrupted\\		
		$\mathcal{L}_{\textsf{auth}}$ & authorized users in service discovery session\\
		$\mathcal{L}_{\textsf{exec}}$ & contains all messages that user or adversary  \\
		& exchanged during executions of protocol\\
		\hline
	\end{tabular} 
	\caption{Lists in Security Experiments}
	\label{Tab:SecExplists}	
\end{table}

\noindent\textsf{(2) Correctness Proof}

\textsf{Theorem A.1}
\label{Lemma:FAC-Correct}
The FAC scheme satisfies \textit{correctness}.

\begin{proof}
	For the correctness proof, we need to demonstrate the following properties: 1) the instantiation of SPK $\pi_1$ is correct in the $\mathsf{UserKeyGen}$ algorithm; 2) the credential created by the issuer is verified true by the user in the $\mathsf{Issue}$ algorithm; 3) the instantiation of SPK $\pi_2$ is correct in $\mathsf{Show}$ and $\mathsf{Verify}$ algorithms.
	
	For the \textit{first property}, The correctness of $\textsf{SPK}$ $\pi_1$ is verified as
	$\gamma=h^{\widetilde{\textsf{usk}}}=h^{\overline{\textsf{usk}}}(h^{\textsf{usk}})^{c}=h^{\overline{\textsf{usk}}}\textsf{upk}^{c}$, where $\overline{\textsf{usk}}=\widetilde{\textsf{usk}}-c\cdot \textsf{usk} \mod p$ and	
	$\textsf{upk}=h^{\textsf{usk}}$.

	For the \textit{second property}, a received credential is $\textsf{cred}= (\sigma_1,\sigma_2)$, $\sigma_1=h^r$, $\sigma_2=\textsf{upk}^{r\cdot y_0}\cdot h^{r(\tau+\sum_{i=1}^ny_ix_i+y_{n+1}\cdot uid)}$, and the user verifies it as $e(W\cdot Y_0^{\textsf{usk}}\cdot Y_{n+1}^{uid}\prod\nolimits_{i=1}^nY_i^{x_i},\sigma_1)\overset{?}{=}e(g,\sigma_2)$. This equation holds since
	\begin{eqnarray*}
		&&e(W\cdot Y_0^{\textsf{usk}}\cdot Y_{n+1}^{uid}\prod\nolimits_{i=1}^nY_i^{x_i},\sigma_1)\\
		&=&e(g^{\tau}\cdot (g^{y_0})^{\textsf{usk}}\cdot (g^{y_{n+1}})^{uid}\prod\nolimits_{i=1}^n(g^{y_i})^{x_i},h^r)\\
		&=&e(g,(h^{\textsf{usk}})^{r\cdot y_0}h^{r(\tau+\sum_{i=1}^ny_ix_i+y_{n+1}\cdot uid)})\\
		&=&e(g,\textsf{upk}^{r\cdot y_0}\cdot h^{r(\tau+\sum_{i=1}^ny_ix_i+y_{n+1}\cdot uid)})=e(g,\sigma_2).
	\end{eqnarray*}
	
	For the \textit{third property}, assume that $\textsf{cred}$ is a valid anonymous credential  on $uid$, $\vec{x}$ and $\textsf{usk}$, and we have
	$$e(W\cdot Y_0^{\textsf{usk}}\cdot Y_{n+1}^{uid}\prod\nolimits_{i=1}^nY_i^{x_i},\sigma_1)=e(g,\sigma_2),$$
	$$\Rightarrow e(Y_0^{\textsf{usk}}\cdot Y_{n+1}^{uid},\sigma_1)=e(g,\sigma_2)e(W\cdot\prod\nolimits_{i=1}^nY_i^{x_i},\sigma_1)^{-1}.$$
	
	The correctness of $\textsf{SPK}$ $\pi_2$ can be derived from
	\begin{eqnarray*}
		&&e(Y_0^{\overline{\textsf{usk}}}Y_{n+1}^{\overline{uid}},\bar{\sigma}_1)^{-1}\cdot\Lambda\\
		&=&e(Y_0^{\overline{\textsf{usk}}}Y_{n+1}^{\overline{uid}},\bar{\sigma}_1)^{-1}\cdot e(Y_0^{\widetilde{\textsf{usk}}}Y_{n+1}^{\widetilde{uid}},\bar{\sigma}_1)\\
		&=&e(Y_0^{\textsf{usk}}Y_{n+1}^{uid},\sigma_1)^{c\cdot t_2}\\
		&=&[e(g,\sigma_2)e(W\cdot\prod\nolimits_{i\in[1,n]}Y_i^{x_i},\sigma_1)^{-1}]^{c\cdot t_2}\\
		&=&[e(g,\bar{\sigma}_2)e(g,\bar{\sigma}_1)^{-t_1}e(W\cdot\prod\nolimits_{i\in[1,n]}Y_i^{x_i},\bar{\sigma}_1)^{-1}]^{c}\\
		&=&[e(g,\bar{\sigma}_2)e(W\cdot g^{t_1}\cdot\prod\nolimits_{i\in[1,n]\backslash\mathcal{I}}Y_i^{x_i}\prod\nolimits_{i\in\mathcal{I}}Y_i^{x_i},\bar{\sigma}_1)^{-1}]^{c}\\
		&=&[e(g,\bar{\sigma}_2)e(W\cdot T_1\cdot\prod\nolimits_{i\in\mathcal{I}}Y_i^{x_i},\bar{\sigma}_1)^{-1}]^{c}=[e(g,\bar{\sigma}_2)\cdot\Gamma]^{c}.
	\end{eqnarray*}
	On the other hand,
	\begin{eqnarray*}
		&&e(T_1,\prod\nolimits_{i\in\mathcal{I}'}X_i)\\
		&=&e(g^{t_1},\prod\nolimits_{i\in\mathcal{I}'}X_i)e(\prod_{j\in[1,n]\backslash\mathcal{I}}Y_j^{x_j},\prod\nolimits_{i\in\mathcal{I}'}X_i)\\
		&=&e(\prod_{i\in\mathcal{I}'}Y_i,h)^{t_1}e(\prod_{i\in\mathcal{I}',j\in[1,n]\backslash\mathcal{I}}Z_{i,j}^{x_j},h)\\
		&=&e((\prod_{i\in\mathcal{I}'}Y_i)^{t_1}\prod_{i\in\mathcal{I}',j\in[1,n]\backslash\mathcal{I}}Z_{i,j}^{x_j},h)=e(T_2,h).
	\end{eqnarray*}
	
	Therefore, a valid credential $\textsf{cred}$ and its SPK $\pi_2$ will be verified true.	
\end{proof}

\noindent\textsf{(3) Security Proof}

The security of FAC in
Theorem 
5.1 is proved in aspects of \textit{unforgeability}, \textit{anonymity} and \textit{unlinkability}.\vspace{2mm}

\noindent\textsf{3.1) Unforgeability Proof of FAC}

\textsf{Lemma A.1}
\label{Lemma:FAC-Unforge}
The FAC scheme is \textit{unforgeable} if the underlying unforgeable redactable signature (URS) \cite{sanders2020efficient} is unforgeable and the discrete logarithm (DL) assumption holds.

\begin{proof} 
	The proof reduces the unforgeability of FAC to the existential unforgeability of the unlinkable redactable signature (URS) scheme (in $\S$4 of \cite{sanders2020efficient}) and discrete logarithm (DL) assumption. Let $\mathcal{A}$ be a PPT adversary that wins the unforgeability game with probability $\epsilon$.	
	
	During the challenge phase, $\mathcal{A}$ returns a challenge user identifier $uid^*$, attribute set $\vec{x}^*$ and proves possession of a valid authentication token $\textsf{tok}^*$ for credential $\textsf{cred}^*$ on $(uid^*,\vec{x}^*)$. Obviously, it should be constrained that $(m^*,\vec{x}^*,\textsf{cred}^*,\textsf{tok}^*)\notin\mathcal{L}_{\textsf{show}}$,
	$\textsf{cred}^*\notin\mathcal{L}_{\textsf{issue}}$, $(uid^*,\vec{x}^*)\notin\mathcal{L}_{\textsf{corrupt}}$.
	Let $\textsf{usk}^*$ be the secret key whose knowledge should be proved by $\mathcal{A}$ when it generates a challenge credential $\textsf{cred}^*$ on $(uid^*,\vec{x}^*)$. Denote $\mathcal{L}_{\textsf{honest}}$ as a set of registered users that are honest and $\mathcal{L}_{\textsf{corrupt}}$ as a set of registered users that are corrupted. Denote $q$ as the number of honest user.  We define two types of adversaries $(\mathcal{A}_1,\mathcal{A}_2)$ that possess different resources for the attack: type-1 adversary $\mathcal{A}_1$: $\exists uid_{\beta^*}\in\mathcal{L}_{\textsf{honest}}$, s.t., $\textsf{usk}_{\beta^*}=\textsf{usk}^*$;
	type-2 adversary $\mathcal{A}_2$: $\forall uid_i\in\mathcal{L}_{\textsf{honest}}$, s.t., $\textsf{usk}_i\neq\textsf{usk}^*$.
	In the following, we prove the unforgeability of FAC with two propositions for the two types of adversaries.
	
	\textsf{Proposition A.1}
	Suppose type-1 adversary $\mathcal{A}_1$ is able to break the unforgeability of FAC with advantage $\epsilon_1$. Then, we can utilize $\mathcal{A}_1$ to solve the DL problem with advantage $\epsilon_1/q$, where $q$ is the number of honest users.
	
	\begin{proof}  	 
		The proof reduces the unforgeability of FAC to the security of DL assumption. 	
		Let $\mathcal{A}_1$ be a PPT adversary that wins the security game with probability $\epsilon_1$. Consider a simulator $\mathcal{S}$ which runs $\mathcal{A}_1$ as a subroutine and interacts with a challenger $\mathcal{C}$ for the DL-assumption.   	 
		According to the definition of type-1 adversary $\mathcal{A}_1$, there exists an index $\beta^*\in[1,q]$ for $\mathcal{A}_1$ to impersonate the $\beta^*$-th honest user in $\mathcal{L}_{\textsf{honest}}$. Then, the challenger $\mathcal{C}$ is requested to make a guess on $\beta^*$ from the $q$ honest users. If $\mathcal{A}_1$ is able to break the unforgeability of FAC, $\mathcal{C}$ could makes use of the advantage of $\mathcal{A}_1$ to solve the DL problem. 	 
		Let $g,h$ be the generators of  groups $G_1$ and $G_2$, respectively. Let $(h,h^a)$ be challenge tuple of DL assumption on group $G_2$.
		
		\textbf{Setup}. 
		$\mathcal{S}$ creates the public key  $\textsf{pk}=(W,\{X_i,Y_{i}\}_{i\in[0,n+1]},$ $\{Z_{i,j}\}_{0\leq i\neq j\leq n+1})$ following the $\textsf{Setup}$ and $\textsf{CredKeyGen}$ algorithms in FAC and forwards it to $\mathcal{A}_1$.	 	 
		
		\textbf{Query}. The adversary $\mathcal{A}_1$ adaptively makes the following queries.
		
		$-$ According to the definition of type-1 adversary $\mathcal{A}_1$, there exists an index $\beta^*\in[1,q]$ for $\mathcal{A}_1$ to impersonate the $\beta^*$-th honest user in $\mathcal{L}_{\textsf{honest}}$. For the $\textsf{UserKeyGen}$ queries on $(uid_{\beta^*},\vec{x}_{\beta^*})$, $\mathcal{S}$ implicitly sets $usk_{\beta^*}=a$, and sends $\textsf{upk}_{\beta^*}=h^a$ to $\mathcal{A}_1$. For the $\textsf{UserKeyGen}$ queries on $(uid_j,\vec{x}_j)$ with $j\neq i$, $\mathcal{S}$ generates user's secret key $\textsf{usk}_j\xleftarrow{\$}\mathbb{Z}_p^*$ and public key $\textsf{upk}_j=h^{\textsf{usk}_j}$, which are returned to $\mathcal{A}_1$. $\mathcal{S}$ adds $(uid,\vec{x},\textsf{upk})$ to a list $\mathcal{L}_{\textsf{honest}}$. If $(uid,\vec{x})$ already exists in $\mathcal{L}_{\textsf{honest}}$, $\mathcal{S}$ just replies with the same answer.

		$-$ For the $\textsf{Issue}$ queries on $(uid_j,\vec{x}_j)$ with restriction that $j\neq \beta^*$, $\mathcal{S}$ runs $\textsf{UserKeyGen}$ to generate $(\textsf{upk}_j,\textsf{usk}_j)$ if $(uid_j,\vec{x}_j)$ has not been queried beforehand. Otherwise, $\mathcal{S}$ extracts user's keys $(\textsf{upk}_j,\textsf{usk}_j)$ from $\mathcal{L}_{\textsf{honest}}$. Then, $\mathcal{S}$ queries $\mathcal{C}$ on $(uid_j,\vec{x}_j)$ and obtains a URS signature $\sigma_j=(\widetilde{\sigma}_1,\widetilde{\sigma}_2)$, where $\widetilde{\sigma}_1\xleftarrow{\$}G_2$ and $\widetilde{\sigma}_2\leftarrow\widetilde{\sigma}_1^{\tau+\sum_{i=1}^n y_i\cdot x_i}$. $\mathcal{S}$ implicitly sets $\sigma_1=\widetilde{\sigma}_1=h^r$, computes
		\begin{eqnarray*}
			\sigma_2&=&(\widetilde{\sigma}_1)^{\textsf{usk}_j\cdot y_0}(\widetilde{\sigma}_1)^{y_{n+1}\cdot uid}\widetilde{\sigma}_2\\
			&=&(h^\textsf{usk}_j)^{r\cdot y_0}h^{r(\tau+\sum_{i=1}^ny_ix_i+y_{n+1}\cdot uid)}\\
			&=&\textsf{upk}^{r\cdot y_0}\cdot h^{r(\tau+\sum_{i=1}^ny_ix_i+y_{n+1}\cdot uid)},
		\end{eqnarray*}		
		and returns $\textsf{cred}_j\leftarrow(\sigma_1,\sigma_2)$ to $\mathcal{A}_1$.		
		$\mathcal{S}$ inserts $(uid_j,\vec{x}_j,\textsf{upk}_j,$ $\textsf{cred}_j)$ into a list $\mathcal{L}_{\textsf{issue}}$.
		
		$-$ To answer the $\textsf{Show}$ query on $(uid_j,\{x_i\}_{i\in\mathcal{I}},\textsf{cred}_j,m_j)$ with restriction that $j\neq \beta^*$, $\mathcal{S}$ runs $\textsf{UserKeyGen}$ to generate $(\textsf{upk}_j,\textsf{usk}_j)$ if $uid_j$ has not been queried beforehand. Otherwise, $\mathcal{S}$ extracts user's keys $(\textsf{upk}_j,\textsf{usk}_j)$ from $\mathcal{L}_{\textsf{honest}}$. Then, 	
		$\mathcal{S}$ answers the query by executing the $\textsf{Show}$ algorithm and returns the token $\textsf{tok}$ to $\mathcal{A}_1$.		
		$\mathcal{S}$ inserts $(uid_j,\{x_i\}_{i\in\mathcal{I}},\textsf{cred}_j,m_j,$ $\textsf{tok}_j)$ into a list $\mathcal{L}_{\textsf{show}}$.
		
		$-$ To answer the $\textsf{Corrupt}$ request on $(uid_j,\vec{x}_j)\in\mathcal{L}_{\textsf{honest}}$ with restriction that $j\neq \beta^*$, simulator $\mathcal{S}$ returns the corresponding user secret key $\textsf{usk}_j$, credential $\textsf{cred}_j$ and the token $\textsf{tok}_j$ to $\mathcal{A}_1$, which are recorded in $\mathcal{L}_{\textsf{honest}}$, $\mathcal{L}_{\textsf{issue}}$ and  $\mathcal{L}_{\textsf{show}}$, respectively. $\mathcal{S}$ inserts $(uid_j,\vec{x}_j)$ into $\mathcal{L}_{\textsf{corrupt}}$. If $(uid_j,\vec{x}_j)$ does exist in these lists, $\mathcal{S}$ returns $\bot$.
		
		\textbf{Challenge}. Adversary $\mathcal{A}_1$ outputs a challenge tuple $(\vec{x}^*,$ $m^*,\textsf{cred}^*,\textsf{tok}^*)$ with attributes $\vec{x}^*$, a message $m^*$ and an authentication token $\textsf{tok}^*$ for the $\beta^*$-th honest user with $uid_{\beta^*}$. The restriction is that $(m^*,\vec{x}^*,\textsf{cred}^*,\textsf{tok}^*)\notin\mathcal{L}_{\textsf{show}}$,
		$\textsf{cred}^*\notin\mathcal{L}_{\textsf{issue}}$, $(uid^*,\vec{x}^*)\notin\mathcal{L}_{\textsf{corrupt}}$. We say that $\mathcal{A}_1$ wins the game if $\mathcal{FAC}.\textsf{Verify}$ $(\textsf{tok}^*,m^*)=1$.
		
		One can note that this game is perfectly simulated if the guess on $\beta^*\in[1,q]$ is correct, which occurs with probability $1/q$. In this case, adversary $\mathcal{A}_1$ is succeed with advantage $\epsilon_1$ to prove knowledge of $\textsf{usk}_{\beta^*}=a$ when it shows a valid credential. $\mathcal{S}$ sends the challenge tuple to $\mathcal{C}$. Then, $\mathcal{C}$ runs the extractor of the proof of knowledge to recover $a$, which is a solution to the DL problem. Therefore, the probability for $\mathcal{C}$ to break the DL assumption is $\epsilon_1/q$.
	\end{proof}

	\textsf{Proposition A.2}
	Suppose type-2 adversary $\mathcal{A}_2$ is able to break the unforgeability of FAC with advantage $\epsilon_2$. Then, we can utilize $\mathcal{A}_2$ to break the existential unforgeability of URS in \cite{sanders2020efficient}  with advantage $\epsilon_2$.
	
	\begin{proof} 
		The proof reduces the unforgeability of FAC to the existential unforgeability of the unlinkable redactable signature (URS) scheme in Section 4 of \cite{sanders2020efficient}. As the unforgeability of URS relies on the DL assumption, this proposition follows.	
		Let type-2 adversary $\mathcal{A}_2$ be a PPT adversary that wins the unforgeability game with probability $\epsilon_2$. Consider a simulator $\mathcal{S}$ which runs $\mathcal{A}_2$ as a subroutine and interacts with a unforgeability game challenger $\mathcal{C}$ for the URS scheme in \cite{sanders2020efficient}.	
		
		\textbf{Setup}. 
		$\mathcal{S}$ generates the public parameter $\mathsf{pp}=(g,h,n)$ and sends it to $\mathcal{C}$, where $g,h$ are generators of $G_1$, $G_2$, respectively, and $n$ is the attribute number.
		$\mathcal{C}$ generates the public key $\widetilde{\mathsf{pk}}=(W,\{X_i,Y_{i}\}_{i\in[1,n]},\{Z_{i,j}\}_{1\leq i\neq j\leq n})$ of URS, and transmits it to simulator $\mathcal{S}$, where $W=g^{\tau}$, $X_i=h^{y_i}$, $Y_i=g^{y_i}$ for $i\in[1,n]$, and $Z_{i,j}=g^{y_i\cdot y_j}$ for $1\leq i\neq j\leq n$. Note that the secret key $\widetilde{\mathsf{sk}}=(\tau,\{y_i\}_{i\in[1,n]})$ of URS is unknown to $\mathcal{S}$.	
		$\mathcal{S}$ selects random elements $y_0,y_{n+1}\xleftarrow{\$}\mathbb{Z}_p^*$, and implicitly sets the secret key of FAC as $\textsf{sk}=(\tau,\{y_i\}_{i\in[0,n+1]})$. $\mathcal{S}$ calculates $X_i=h^{y_i}$, $Y_i=g^{y_i}$ for $i=\{0,n+1\}$, computes $Z_{i,n+1}=Y_i^{y_{n+1}}=g^{y_i\cdot y_{n+1}}$ for $1\leq i\leq n$, $Z_{0,j}=Y_j^{y_0}=g^{y_0\cdot y_j}$ for $1\leq j\leq n$, and $Z_{0,n+1}=g^{y_0\cdot y_{n+1}}$.
		$\mathcal{S}$ sets the public key of FAC as $\textsf{pk}=(W,\{X_i,Y_{i}\}_{i\in[0,n+1]},\{Z_{i,j}\}_{0\leq i\neq j\leq n+1})$ and forwards it to $\mathcal{A}_2$. 		 	 
		
		\textbf{Query}. The adversary $\mathcal{A}_2$ adaptively makes the following queries.
		
		$-$ For the $\textsf{UserKeyGen}$ queries on $(uid,\vec{x})$, $\mathcal{S}$ generates user's secret key $\textsf{usk}\xleftarrow{\$}\mathbb{Z}_p^*$ and public key $\textsf{upk}=h^\textsf{usk}$, which are returned to $\mathcal{A}_2$. $\mathcal{S}$ adds $(uid,\vec{x},\textsf{upk},\textsf{usk})$ to a list $\mathcal{L}_{\textsf{honest}}$. If $(uid,\vec{x})$ already exists in $\mathcal{L}_{\textsf{honest}}$, $\mathcal{S}$ just replies with the same answer.

		$-$ For the $\textsf{Issue}$ queries on $(uid,\vec{x})$, $\mathcal{S}$ runs $\textsf{UserKeyGen}$ to generate $(\textsf{upk},\textsf{usk})$ if $(uid,\vec{x})$ has not been queried beforehand. Otherwise, $\mathcal{S}$ extracts user's keys $(\textsf{upk},\textsf{usk})$ from $\mathcal{L}_{\textsf{honest}}$. Then, $\mathcal{S}$ queries $\mathcal{C}$ on $(uid,\vec{x})$ and obtains a URS signature $\sigma=(\widetilde{\sigma}_1,\widetilde{\sigma}_2)$, where $\widetilde{\sigma}_1\xleftarrow{\$}G_2$ and $\widetilde{\sigma}_2\leftarrow\widetilde{\sigma}_1^{\tau+\sum_{i=1}^n y_i\cdot x_i}$. $\mathcal{S}$ implicitly sets $\sigma_1=\widetilde{\sigma}_1=h^r$, computes
		\begin{eqnarray*}
			\sigma_2&=&(\widetilde{\sigma}_1)^{\textsf{usk}\cdot y_0}(\widetilde{\sigma}_1)^{y_{n+1}\cdot uid}\widetilde{\sigma}_2\\
			&=&(h^\textsf{usk})^{r\cdot y_0}h^{r(\tau+\sum_{i=1}^ny_ix_i+y_{n+1}\cdot uid)}\\
			&=&\textsf{upk}^{r\cdot y_0}\cdot h^{r(\tau+\sum_{i=1}^ny_ix_i+y_{n+1}\cdot uid)},	
		\end{eqnarray*}		
		and returns $\textsf{cred}\leftarrow(\sigma_1,\sigma_2)$ to $\mathcal{A}_2$. $\mathcal{S}$ inserts $(uid,\vec{x},\textsf{upk},\textsf{cred})$ into a list $\mathcal{L}_{\textsf{issue}}$.
		
		$-$ The inputs of $\textsf{Show}$ query are  $(uid,\{x_i\}_{i\in\mathcal{I}},\textsf{cred},m)$. A $\textsf{Show}$ query can only be made for a credential that has been created in the $\textsf{Issue}$ query since the latter uses the $\mathcal{O}_{Sign^*}$ oracle of the unforgeability game of URS scheme \cite{sanders2020efficient} as subroutine. Then, 	
		$\mathcal{S}$ answers the query by executing the $\textsf{Show}$ algorithm and returns the token $\textsf{tok}$ to $\mathcal{A}_2$. $\mathcal{S}$ inserts $(uid,\{x_i\}_{i\in\mathcal{I}},\textsf{cred},m,\textsf{tok})$ into a list $\mathcal{L}_{\textsf{show}}$.
		
		$-$ To answer $\textsf{Corrupt}$ on $(uid,\vec{x})$, simulator $\mathcal{S}$ returns the corresponding user secret key $\textsf{usk}$, credential $\textsf{cred}$ and the token $\textsf{tok}$ to $\mathcal{A}_2$, which are recorded in $\mathcal{L}_{\textsf{honest}}$, $\mathcal{L}_{\textsf{issue}}$ and  $\mathcal{L}_{\textsf{show}}$, respectively. $\mathcal{S}$ inserts $(uid,\vec{x})$ into $\mathcal{L}_{\textsf{corrupt}}$. If $(uid,\vec{x})$ does exist in these lists, $\mathcal{S}$ returns $\bot$.
		
		\textbf{Challenge}. Adversary $\mathcal{A}_2$ outputs a challenge tuple $(uid^*,$ $\vec{x}^*,m^*,\textsf{cred}^*,\textsf{tok}^*)$, which associates with challenge secret key $\textsf{usk}^*$. Since we are simulating a type-2 adversary $\mathcal{A}_2$, it is requested that $\textsf{usk}^*$ should be different from $\textsf{usk}_i$ for any honest user $i$.	
		The constraints also include that $(m^*,\vec{x}^*,\textsf{cred}^*,\textsf{tok}^*)\notin\mathcal{L}_{\textsf{show}}$,
		$\textsf{cred}^*\notin\mathcal{L}_{\textsf{issue}}$, $(uid^*,\vec{x}^*)\notin\mathcal{L}_{\textsf{corrupt}}$. We say that $\mathcal{A}_2$ wins the game if $\mathcal{FAC}.\textsf{Verify}$ $(\textsf{tok}^*,m^*)=1$.
		
		If $\mathcal{A}_2$ is succeed with advantage $\epsilon_2$ to prove knowledge of $\textsf{usk}^*$ when it shows a valid credential. $\mathcal{S}$ sends the challenge tuple to $\mathcal{C}$. Then, $\mathcal{C}$ runs the extractor of proof of knowledge to recover $\textsf{usk}^*$. $\mathcal{C}$ parses $\textsf{cred}^*\leftarrow(\sigma_1^*,\sigma_2^*)$ and calculates $\widetilde{\sigma}_1^*=\sigma_1^*=h^r$, computes
		$\widetilde{\sigma}_2^*=\sigma_2^*(\sigma_1^*)^{-\textsf{usk}^*\cdot y_0}(\sigma_1^*)^{y_{n+1}\cdot uid^*}=(\widetilde{\sigma}_1^*)^{\tau+\sum_{i=1}^n y_i\cdot x_i^*}$. Therefore, $\mathcal{C}$ obtains a valid forgery $\sigma^*=(\widetilde{\sigma}_1^*,\widetilde{\sigma}_2^*)$ for the URS scheme in \cite{sanders2020efficient} with advantage $\epsilon_2$.
	\end{proof}

	The proofs for two propositions against $\mathcal{A}=(\mathcal{A}_1,\mathcal{A}_2)$ conclude the proof for unforgeability of FAC.
\end{proof}\vspace{5mm}

\noindent\textsf{3.2) Anonymity and Unlinkability of FAC}

\textsf{Lemma A.2}
\label{Lemma:FAC-Ano}
The FAC scheme satisfies \textit{anonymity} and \textit{unlinkability} under the decisional Diffie–Hellman (DDH) assumption.

\begin{proof}		
	Suppose a PPT adversary $\mathcal{A}$ is able to break the anonymity of FAC with advantage $\epsilon$. Then, we can utilize $\mathcal{A}$ to solve the DDH problem with advantage $\epsilon$.	
	Consider a simulator $\mathcal{S}$ which runs $\mathcal{A}$ as a subroutine and interacts with a challenger $\mathcal{C}$ for the DDH-assumption.   
	Let $g,h$ be the generators of  groups $G_1$ and $G_2$, respectively. Let $(h,h^a,h^b,h^c)$ be a challenge tuple of DDH assumption on group $G_2$. It is required to decide whether $c=a\cdot b$ or $c\xleftarrow{\$}\mathbb{Z}_p^*$.
	
	\textbf{Setup}. 
	$\mathcal{S}$ creates the public key  $\textsf{pk}=(W,\{X_i,Y_{i}\}_{i\in[0,n+1]},$ $\{Z_{i,j}\}_{0\leq i\neq j\leq n+1})$ following the $\textsf{Setup}$ and $\textsf{CredKeyGen}$ algorithms in FAC and forwards it to $\mathcal{A}$.	 	 
	
	\textbf{Query}. $\mathcal{A}$ adaptively makes the following queries.
	
	$-$ For the $\textsf{UserKeyGen}$ queries on $(uid,\vec{x})$, $\mathcal{S}$ generates user's secret key $\textsf{usk}\xleftarrow{\$}\mathbb{Z}_p^*$ and public key $\textsf{upk}=h^\textsf{usk}$, which are returned to $\mathcal{A}$. $\mathcal{S}$ adds $(uid,\vec{x},\textsf{upk},\textsf{usk})$ to a list $\mathcal{L}_{\textsf{honest}}$. If $(uid,\vec{x})$ already exists in $\mathcal{L}_{\textsf{honest}}$, $\mathcal{S}$ just replies with the same answer.
	
	$-$ For the $\textsf{Issue}$ queries on $(uid,\vec{x})$, $\mathcal{S}$ runs $\textsf{UserKeyGen}$ to generate $(\textsf{upk},\textsf{usk})$ if $(uid,\vec{x})$ has not been queried beforehand. Otherwise, $\mathcal{S}$ extracts user's keys $(\textsf{upk},\textsf{usk})$ from $\mathcal{L}_{\textsf{honest}}$. Since $\mathcal{S}$ creates the issuer's secret key $\textsf{sk}$ by itself in Setup phase and knows user's secret key $\textsf{usk}$, $\mathcal{S}$ executes the $\mathsf{Issue}$ protocol to obtain $\textsf{cred}\leftarrow(\sigma_1,\sigma_2)$, which is returned to $\mathcal{A}$. $\mathcal{S}$ inserts $(uid,\vec{x},\textsf{upk},\textsf{cred})$ into a list $\mathcal{L}_{\textsf{issue}}$.
	
	$-$ The inputs of $\textsf{Show}$ query are  $(uid,\{x_i\}_{i\in\mathcal{I}},\textsf{cred},m)$. A $\textsf{Show}$ query can only be made for a credential that has been created in the $\textsf{Issue}$ query. $\mathcal{S}$ extracts $(uid,\vec{x},\textsf{upk},\textsf{cred})$ from $\mathcal{L}_{\textsf{issue}}$. Then, 	
	$\mathcal{S}$ answers the query by executing the $\textsf{Show}$ algorithm and returns the token $\textsf{tok}$ to $\mathcal{A}$. $\mathcal{S}$ inserts $(uid,\{x_i\}_{i\in\mathcal{I}},\textsf{cred},m,\textsf{tok})$ into a list $\mathcal{L}_{\textsf{show}}$.
	
	$-$ To answer $\textsf{Corrupt}$ on $(uid,\vec{x})$, simulator $\mathcal{S}$ returns the corresponding user secret key $\textsf{usk}$, credential $\textsf{cred}$ and the token $\textsf{tok}$ to $\mathcal{A}$, which are recorded in $\mathcal{L}_{\textsf{honest}}$, $\mathcal{L}_{\textsf{issue}}$ and  $\mathcal{L}_{\textsf{show}}$, respectively. $\mathcal{S}$ inserts $(uid,\vec{x})$ into $\mathcal{L}_{\textsf{corrupt}}$. If $(uid,\vec{x})$ does exist in these lists, $\mathcal{S}$ returns $\bot$.
	
	\textbf{Challenge}. In this phase, $\mathcal{A}$ outputs two challenge users with attributes $(uid_0^*,\vec{x}_0^*)$, $( uid_1^*,\vec{x}_1^*)$, and message $m^*$. The simulator $\mathcal{S}$ flips a random coin $\bar{b}\in\{0,1\}$ and generates challenge  $\textsf{tok}^*$ for user $uid_{\bar{b}}^*$, where  $\textsf{tok}^*\leftarrow(\{x_i^*\}_{i\in\mathcal{I}^*},T_1^*,T_2^*,\bar{\sigma}_1^*,\bar{\sigma}_2^*,\pi_2^*)$, $T_1^*=g^{\alpha}$, $T_2^*=(T_1^*)^{\sum_{i\in{\mathcal{I}^*}'} y_i}$,
	$\bar{\sigma}_1^*=h^b$, $\bar{\sigma}_2^*=(h^c)^{y_0}\cdot(\bar{\sigma}_1^*)^{\alpha+\tau+\sum_{i\in\mathcal{I}^*}y_ix_i^*+y_{n+1}\cdot uid_{\bar{b}}^*})$, $\pi_2^*$ is a simulated knowledge of $a$, the disclosed attribute set $\mathcal{I}^*\subseteq[1,n]$, $\alpha\in_R \mathbb{Z}_p^*$.
	The restriction is that
	the disclosed attributes in $\vec{x}_0^*$ and $\vec{x}_1^*$ are the same,
	$\textsf{cred}_{0/1}^*\notin\mathcal{L}_{\textsf{issue}}$, $(m^*,\vec{x}_{0/1}^*,\textsf{tok}_{0/1}^*)\notin\mathcal{L}_{\textsf{show}}$, $(uid_{0/1}^*,\vec{x}_{0/1}^*)\notin\mathcal{L}_{\textsf{corrupt}}$.

	\textbf{Guess}. The adversary $\mathcal{A}$ makes a guess $\bar{b}'\in\{0,1\}$ on the identity of the user from $(uid_0^*, uid_1^*)$.		
	$\mathcal{A}$ wins the game if $\bar{b}'=\bar{b}$.	
	
	$\mathcal{S}$ sends the guess result of $\mathcal{A}$ to $\mathcal{C}$. 	
	It is noted that if $c=ab$, by setting $t_1=\alpha-\sum_{i\in[1,n]\backslash\mathcal{I}^*}y_ix_i^*$, 
	One can see that  $\textsf{tok}^*$ is  distributed as in the FAC scheme.
	Else, $c$ is a random number in $\mathbb{Z}_p^*$ and $\bar{\sigma}_2^*$ is a random element in $\mathbb{G}_2$. Since $(T_1^*,T_2^*,\bar{\sigma}_1^*)$ are independent of $a$ and $\{x_i^*\}_{i\in[n]\backslash\mathcal{I}^*}$, $\mathcal{A}$ cannot succeed in this game with non-negligile advantage.
	If $\mathcal{A}$ is able to win the security game with advantage $\epsilon$, $\mathcal{C}$ can makes use of  $\mathcal{A}$ to solve the DDH problem with advantage $\epsilon$.	
\end{proof}

%

\subsection{ACME: Security Model and Proof}
\label{Appendix:ACME}

\noindent\textsf{(1) Security Model}

This section defines  \textit{correctness}, \textit{privacy}, \textit{authenticity}, \textit{anonymity} and \textit{unlinkability} for ACME scheme.

\textsf{Definition B.1.}
\label{ACME:correctness}
Let $\mathcal{D}$ be the universe of user identity, and $\Omega$ be the universe of attributes. An anonymous credential-based matchmaking encryption encryption scheme $\mathcal{ACME}$ is \textit{correct} for $\mathcal{D},\Omega$ if all $uid\in\mathcal{D}$, all $\vec{x}\subseteq\Omega$ for all security parameter $\lambda$,
\small 
$$\textsf{Pr}\left[
	\begin{array}{l}	
	(\textsf{mpk},\textsf{msk})\xleftarrow{\$}\textsf{Setup}(1^{\lambda},1^n);\\
	(\textsf{pk},\textsf{sk})\xleftarrow{\$}\textsf{CredKeyGen}(\textsf{mpk});\\
	(\textsf{upk},\textsf{usk})\xleftarrow{\$}\textsf{UserKeyGen}(\textsf{mpk});\\
	\textsf{cred}_{\textsf{snd}}\xleftarrow{\$}\left<\textsf{Issue.I}(\textsf{sk},\textsf{upk})\rightleftarrows\textsf{Issue.U}(uid,\vec{x}_{\textsf{snd}},\textsf{usk})\right>;\\
	\textsf{DK}_{\vec{x}_{\textsf{rcv}}}\xleftarrow{\$}\textsf{DKGen}(\mathsf{msk},\vec{x}_{\textsf{rcv}});\\
	\textsf{DK}_{f_{\textsf{rcv}}}\leftarrow\textsf{PolGen}(\mathsf{msk},f_{\textsf{rcv}});\\
	\textsf{CT}_{\vec{x}_{\textsf{snd}},f_{\textsf{snd}}}\xleftarrow{\$}\mathsf{Enc}(\textsf{cred}_{\textsf{snd}},\vec{x}_{\textsf{snd}},f_{\textsf{snd}},M):\\
	 f_{\textsf{rcv}}(\vec{x}_{\textsf{snd}}^{(out)})=1 \wedge f_{\textsf{snd}}(\vec{x}_{\textsf{rcv}}^{(out)})=1 \wedge\\
	\mathsf{Dec}(\textsf{DK}_{\vec{x}_{\textsf{rcv}}},\textsf{DK}_{f_{\textsf{rcv}}},\textsf{CT}_{\vec{x}_{\textsf{snd}},f_{\textsf{snd}}})=\bot
	\end{array}
	\right]$$
\normalsize
$\leq\nu(\lambda)$, where $\nu$ is a negligible function.

\textsf{Definition B.2.}
\label{ME:privacy}
An ACME scheme $\mathcal{ACME}$ satisfies \textit{privacy} if for any PPT adversary $\mathcal{A}=(\mathcal{A}_1,\mathcal{A}_2)$, there exists a negligible function $\nu$ such that $\textsf{Adv}_{\mathcal{ACME}}^{\textsf{priv}}(\lambda)\overset{\textsf{def}}{=}$
\small
$$\textsf{Pr}\left[b'=b~
\begin{array}{|l}
	(\textsf{mpk},\textsf{msk})\leftarrow\textsf{Setup}(1^{\lambda},1^n),b\xleftarrow{\$}\{0,1\}\\
	(M_0^*,M_1^*, \textsf{cred}_{\textsf{snd}_0}^*,\textsf{cred}_{\textsf{snd}_1}^*, \vec{x}_{\textsf{snd}_0}^*, \vec{x}_{\textsf{snd}_1}^*,f_{\textsf{snd}}^*)\\
	\quad\quad\quad\quad\quad\quad\quad\quad\quad\quad\leftarrow\mathcal{A}_1^{\mathcal{O}_1,\mathcal{O}_2,\mathcal{O}_3}(\textsf{mpk})\\	
	\textsf{CT}^*\leftarrow\textsf{Enc}(\textsf{cred}_{\textsf{snd}_b}^*,\vec{x}_{\textsf{snd}_b}^*,f_{\textsf{snd}}^*,M_b)\\
	b'\leftarrow\mathcal{A}_2^{\mathcal{O}_1,\mathcal{O}_2,\mathcal{O}_3}(\textsf{mpk},\textsf{CT}^*)\\
\end{array}
\right]$$
\normalsize
$\leq\nu(\lambda),$ where oracles $\mathcal{O}_1$, $\mathcal{O}_2$, $\mathcal{O}_3$ are implemented by $\textsf{Issue}$ $(\textsf{msk},\cdot)$, $\textsf{DKGen}(\textsf{msk},\cdot)$, $\textsf{PolGen}(\textsf{msk},\cdot)$, respectively. 
It is required that $\mathcal{O}_2$ and $\mathcal{O}_3$  are not queried for attributes and policies that can satisfy $(\vec{x}_{\textsf{snd}_0}^*, f_{\textsf{snd}}^*)$ or $(\vec{x}_{\textsf{snd}_1}^*, f_{\textsf{snd}}^*)$. It is also required that for public attributes $\vec{x}_{\textsf{snd}_0}^{(out)*}$/ $\vec{x}_{\textsf{snd}_1}^{(out)*}$ in $\vec{x}_{\textsf{snd}_0}^*$ / $\vec{x}_{\textsf{snd}_1}^*$, we have $\vec{x}_{\textsf{snd}_0}^{(out)*} = \vec{x}_{\textsf{snd}_1}^{(out)*}$.

This model only captures security under chosen plaintext attacks (CPA). We can extend the above definition by introducing another decryption oracle $\mathcal{O}_4$ which can decrypt ciphertexts except the challenge ciphertext $\textsf{CT}^*$ to capture security under chosen-ciphertext attacks (CCA). 

\textsf{Definition  B.3.}
\label{ME:authenticity}
An ACME scheme $\mathcal{ACME}$ satisfies \textit{authenticity} if for any PPT adversary $\mathcal{A}$, there exists a negligible function $\nu$ such that $\textsf{Adv}_{\mathcal{ACME}}^{\textsf{auth}}(\lambda)\overset{\textsf{def}}{=}$
\small
$$\textsf{Pr}\left[
\begin{array}{l}
	(\textsf{mpk},\textsf{msk})\leftarrow\textsf{Setup}(1^{\lambda},1^n)\\
	(\textsf{CT}_{\vec{x}_{\textsf{snd}},f_{\textsf{snd}}},\vec{x}_{\textsf{rcv}}, f_{\textsf{rcv}})\leftarrow\mathcal{A}^{\mathcal{O}_1,\mathcal{O}_2,\mathcal{O}_3}(\textsf{mpk})\\
	\textsf{DK}_{\vec{x}_{\textsf{rcv}}}\leftarrow\textsf{DKGen}(\textsf{msk},\vec{x}_{\textsf{rcv}})\\
	\textsf{DK}_{ f_{\textsf{rcv}}}\leftarrow\textsf{PolGen}(\textsf{msk}, f_{\textsf{rcv}})\\
	M=\textsf{Dec}(\textsf{DK}_{\vec{x}_{\textsf{rcv}}},\textsf{DK}_{ f_{\textsf{rcv}}},\textsf{CT}_{\vec{x}_{\textsf{snd}},f_{\textsf{snd}}})\\
	\forall\vec{x}\in \mathcal{Q}_{\mathcal{O}_1,\mathcal{O}_2}:( f_{\textsf{rcv}}(\vec{x}^{(out)})=0)\wedge (M\neq\bot)\\
\end{array}
\right]$$
\normalsize
$\leq\nu(\lambda),$ where oracles $\mathcal{O}_1$, $\mathcal{O}_2$, $\mathcal{O}_3$ are implemented by $\textsf{Issue}$ $(\textsf{msk},\cdot)$, $\textsf{DKGen}(\textsf{msk},\cdot)$, $\textsf{PolGen}(\textsf{msk},\cdot)$.

ACME also satisfy the security properties of anonymous credential, namely \textit{anonymity} (Def. A.3) and \textit{unlinkability} (Def. A.4), against an entity who can decrypt the ciphertext.

\noindent\textsf{(2) Security Proof}

\textsf{Theorem 6.2.}	
The proposed ACME scheme achieves privacy, authenticity, anonymity and unlinkability if the $MDDH_k$ assumption holds and the underlying FAC is secure.

The security proofs of authenticity, anonymity and unlinkability of ACME follows those of unforgeability, anonymity
and unlinkability of FAC.
Next, we prove that the proposed ACME scheme achieves privacy under the $MDDH_k$ assumption.

\noindent\textbf{Proof Intuition.} In our proposed ACME scheme, the message $M$ and the FAC token $\textsf{tok}_{\textsf{snd}}$ (corresponding to the private attributes) are encrypted using a symmetric key $K$ which is encapsulated in
$\textsf{ct}_0=e([\widetilde{\textbf{s}}^{\top}\textbf{A}+\textbf{s}^{\top}\textbf{A}]_1,[\textbf{v}]_2)\cdot K.$ 
Hence, the privacy of both $M$ and $\textsf{tok}_{\textsf{snd}}$ is based on the confidentiality of the symmetric key $K$. 

As shown in the correctness, when $f_{\textsf{rcv}}(\vec{x}_{\textsf{snd}}^{(out)})=1$ (corresponding to KP-ABE), we have
$$\frac{e\big(\textsf{ct}_2',\prod\nolimits_{j\in\mathcal{S}_r}\textsf{dk}_j^{\omega_j}\big)}{e\big(\textsf{ct}_1',\prod\nolimits_{j\in\mathcal{S}_r}\big(\prod_{i:x_{s,i}^{(out)}=1}\textsf{dk}_{i,j}\big)^{\omega_j}\big)} = ([\widetilde{\textbf{s}}^{\top}\textbf{A}\textbf{v}]_T)^{-1}$$
and when $f_{\textsf{snd}}(\vec{x}_{\textsf{rcv}}^{(out)})=1$ (corresponding to CP-ABE), we have
$$\frac{e(\prod\nolimits_{j\in\mathcal{S}_s}(\prod\nolimits_{i:x_{r,i}^{(out)}=1}\textsf{ct}_{i,j})^{\mu_j},\textsf{dk}_2)}{e(\textsf{ct}_1,\textsf{dk}_1)\cdot e(\prod\nolimits_{j\in\mathcal{S}_s}\widetilde{\textsf{ct}}_{j}^{\mu_j},\textsf{dk}_3)} = ([\textbf{s}^{\top}\textbf{A}\textbf{v}]_T)^{-1}.$$
Hence, the proof for the confidentiality of $K$ essentially follows the same proof techniques used in the underlying (dual) CP-ABE and KP-ABE schemes \cite{kowalczyk2019compact,katsumata2020compact}. Specifically, if $f_{\textsf{rcv}}(\vec{x}^{(out)*}_{\textsf{snd}})\ne1$, then the confidentiality of $K$ is ensured by the security of the KP-ABE scheme; otherwise, since according to the security game, the adversary is not allowed to obtain keys for $f_{\textsf{rcv}}$ and $\vec{x}_{\textsf{rcv}}$ such that $f_{\textsf{rcv}}(\vec{x}^{(out)*}_{\textsf{snd}}) = 1 \wedge f^*_{\textsf{snd}}(\vec{x}^{(out)}_{\textsf{rcv}})=1$, the security is ensured by that of the CP-ABE scheme.

The dual ABE schemes in  \cite{kowalczyk2019compact,katsumata2020compact} both applied a sequence of games and a hybrid argument in the security proofs. 
The initial game is the same as the original security game whereas in the last game, the encrypted symmetric key $K$ is replaced by a random key. Here we follow the same game sequences defined in \cite{kowalczyk2019compact,katsumata2020compact} by considering two cases: if the adversary would not query a key for $f_{\textsf{rcv}}$ such that $f_{\textsf{rcv}}(\vec{x}^{(out)*}_{\textsf{snd}}) = 1$, then we follow the transitions of the keys and ciphertexts in the proof of KP-ABE and use normal keys and ciphertexts for the CP-ABE componments; otherwise, we perfom the transitions in the opposite way. Below we outline the crucial steps of the proof.

\begin{proof} 		
	Let $\equiv$ denote that two distributions are identically distributed, and $\approx_c$ represent that two distributions are computationally indistinguishable.
	
	The security of ACME scheme is proved by a series of hybrid games, depending on whether the adversary would query a key for $f_{\textsf{rcv}}$ such that $f_{\textsf{rcv}}(\vec{x}^{(out)*}_{\textsf{snd}}) = 1$. 
	
	
	\noindent\textit{Case 1: the adversary queries a key for $f_{\textsf{rcv}}$ such that $f_{\textsf{rcv}}(\vec{x}^{(out)*}_{\textsf{snd}}) = 1$}. Note that in this case the adversary cannot query a key for $\vec{x}_{\textsf{rcv}}$ such that $f^*_{\textsf{snd}}(\vec{x}_{\textsf{rcv}}^{(out)})=1$.
	
	A ciphertext (under access policy $f$ and attributes $\vec{x}$) can be in one of the following forms:
	
	- \textsf{Normal}: A normal ciphertext is generated by $\textsf{Enc}$.
	
	- \textsf{SF}: An \textsf{SF} ciphertext is the same as \textsf{Normal} ciphertext, except that
	$\textbf{s}^{\top}\textbf{A},\textbf{s}_j^{\top}\textbf{A}$
	are replaced with $\textbf{c}^{\top}$, $\textbf{c}_j^{\top}$,
	where $\textbf{c},\textbf{c}_j\leftarrow\mathbb{Z}_p^{2k}$. Let $\widetilde{\textbf{c}} := \widetilde{\textbf{s}}^{\top}\textbf{A}$ as in the normal ciphertext, then $\textsf{CT}_{\vec{x},f}:=$
	$$(
	\textsf{ct}_0=e([{\widetilde{\textbf{c}}^{\top}}+\boxed{\textbf{c}^{\top}}]_1,[\textbf{v}]_2)\cdot K,$$
	$$\textsf{ct}_1'=[{\widetilde{\textbf{c}}^{\top}}]_1,\textsf{ct}_2'=\big[{\widetilde{\textbf{c}}^{\top}}\sum\nolimits_{i:x_i^{(out)}=1}\textbf{W}_i\big]_1,$$
	$$\textsf{ct}_1=[\boxed{\textbf{c}^{\top}}]_1,\{\widetilde{\textsf{ct}}_{j}=[\boxed{\textbf{c}_j^{\top}}]_1,\textsf{ct}_{\rho(j),j}=[\textbf{u}_j^{\top}+\boxed{\textbf{c}_j^{\top}}\textbf{W}_{\rho(j)}]_1,$$
	$$\textsf{ct}_{i,j}=\big[\boxed{\textbf{c}_j^{\top}}\textbf{W}_i\big]_1\}).$$
	
	A secret key  $\textsf{DK}_{f}$ (for policy $f$) follows its normal form in this case. A secret key (for attributes $\vec{x}$) can be in one of the following forms:
	
	- \textsf{Normal}: A normal secret key is generated by $\textsf{DKGen}$.
	
	- \textsf{SF}: An \textsf{SF} key is sampled as a \textsf{Normal} key, except $\textbf{v}$ is replaced with $\textbf{v}+\textbf{A}^{\bot}\boldsymbol{\delta}^{(q)}$, where a fresh $\boldsymbol{\delta}^{(q)}\leftarrow\mathbb{Z}_p^k$ is chosen per \textsf{SF} key and $\textbf{A}^{\bot}$ is any fixed $\textbf{A}^{\bot}\in\mathbb{Z}_p^{2k\times k}\backslash\{\textbf{0}\}$ such that $\textbf{A}\textbf{A}^{\bot}=\textbf{0}$. That is $\textsf{DK}_{\vec{x}}:=$
	\begin{eqnarray*}
		(\textsf{dk}_1&=&[\boxed{\textbf{v}+\textbf{A}^{\bot}\boldsymbol{\delta}^{(q)}}+\textbf{U}_0\textbf{B}\textbf{r}]_2,\textsf{dk}_2=[\textbf{B}\textbf{r}]_2,\\
		\textsf{dk}_3&=&[\sum\nolimits_{i:x_i^{(out)}=1}\textbf{W}_i\textbf{B}\textbf{r}]_2).
	\end{eqnarray*}
	
	- \textsf{P-Normal}: A \textsf{P-Normal} key as the same as a \textsf{Normal} key, except $\textbf{B}\textbf{r}$ is replaced with $\textbf{d}\leftarrow\mathbb{Z}_p^{k}$. That is $\textsf{DK}_{\vec{x}}:=$
	$$\big(\textsf{dk}_1=[\textbf{v}+\textbf{U}_0\boxed{\textbf{d}}]_2,\textsf{dk}_2=[\boxed{\textbf{d}}]_2,\textsf{dk}_3=[\sum\nolimits_{i:x_i^{(out)}=1}\textbf{W}_i\boxed{\textbf{d}}]_2\big).$$
	
	- \textsf{P-SF}: A \textsf{P-SF} key is the same as an \textsf{SF} key, except $\textbf{B}\textbf{r}$ is replaced with $\textbf{d}\leftarrow\mathbb{Z}_p^{k}$. That is $\textsf{DK}_{\vec{x}}:=$
	$$(\textsf{dk}_1=[\boxed{\textbf{v}+\textbf{A}^{\bot}\boldsymbol{\delta}^{(q)}}+\textbf{U}_0\boxed{\textbf{d}}]_2,\textsf{dk}_2=[\boxed{\textbf{d}}]_2,$$
	$$\textsf{dk}_3=[\sum\nolimits_{i:x_i^{(out)}=1}\textbf{W}_i\boxed{\textbf{d}}]_2).$$

	Next, we define the hybrid sequence for the proof. Assume the adversary $\mathcal{A}$ makes at most $Q_x$ attribute decryption key queries. 
	
	- $\textsf{H}_0$: This is the real game where all secret keys and ciphertexts are \textsf{Normal}.
	
	- $\textsf{H}_1$: This game is the same as $\textsf{H}_0$ except that the challenge ciphertext is \textsf{SF}.
	
	- $\textsf{H}_{2,\ell,1}$: This game is the same as $\textsf{H}_1$ except that the $\ell$-th attribute decryption key is \textsf{P-Normal}, the first $\ell-1$ attribute decryption keys are \textsf{SF} and the last $Q_x-\ell$ attribute decryption keys are \textsf{Normal}, where $\ell=0,\cdots,Q_x$.
	
	- $\textsf{H}_{2,\ell,2}$: This game is the same as $\textsf{H}_{2,\ell,1}$ except the $\ell$'th attribute decryption key is \textsf{P-SF}, where $\ell=0,\cdots,Q_x$.
	
	- $\textsf{H}_{2,\ell,3}$: This game is the same as $\textsf{H}_{2,\ell,2}$ except the $\ell$'th attribute decryption key is \textsf{SF}, where $\ell=0,\cdots,Q_x$.

	
	- $\textsf{H}_3$: This game is the same as $\textsf{H}_{2,Q_x,3}$ except that the message encryption symmetric key $K$ to be encrypted is replaced by a random $\widetilde{K}$.

	Let $\mathcal{A}$ be a PPT adversary, and $\textsf{Adv}_{\textsf{xxx}}$ be the advantage of $\mathcal{A}$ in game $\textsf{H}_{\textsf{xxx}}$. Also, define $\textsf{H}_1\equiv\textsf{H}_{2,0,1}$.
	To complete the proof for Case 1, we prove lemmas E.2-E.6 in the following.

	\textsf{Lemma B.1.}
	Under the $MDDH^{2m+1}_{k}$ assumption on $G_1$, we have 
	$$|Pr[\left<\mathcal{A},\textsf{H}_0\right>=1]-Pr[\left<\mathcal{A},\textsf{H}_1\right>=1]|= \textsf{negl}(\lambda).$$
	\begin{proof}
		Assume that $\mathcal{A}$ distinguishes $\textsf{H}_0$ and $\textsf{H}_1$ with non-negligible advantage. Then, we can construct another adversary $\mathcal{B}$ to break the $MDDH^{2m+1}_k$ assumption. On input a $MDDH^{2m+1}_{k}$  challenge 
		$([\textbf{A}]_1,[\textbf{Z}]_1)$, where either $
		\textbf{Z}^{\top}=\textbf{S}^{\top}\textbf{A}$ or $
		\textbf{Z}=\textbf{C}$, for $
		\textbf{S},\textbf{C}\leftarrow\mathbb{Z}^{(2m+1)\times k}_p$. $\mathcal{B}$ proceeds as below.
		
		\textbf{Setup.} $\mathcal{B}$ chooses generators $g\leftarrow G_1$, $h\leftarrow G_2$, user's attribute number $n$, and sets $\textsf{pp}=(g,h,n)$. $\mathcal{B}$ runs $\mathcal{FAC}.\textsf{CredKeyGen}$ to create $(\textsf{pk},\textsf{sk})$. $\mathcal{B}$ selects 
		$\textbf{B}\leftarrow\mathbb{Z}_p^{k\times k}$,
		$\textbf{U}_0,\textbf{W}_i\leftarrow\mathbb{Z}_p^{2k\times k}$, $\textbf{v}\leftarrow\mathbb{Z}_p^{2k}$ and sets $\textsf{mpk}$, $\textsf{msk}$ as in the scheme, where the elements $[\textbf{A}\textbf{W}_i]_1$ in $\textsf{mpk}$ can be derived from $[\textbf{A}]_1$ and $\textbf{W}_i$. 
		
		\textbf{Issue Query.} $\mathcal{B}$ firstly executes $\mathcal{FAC}.\textsf{UserKeyGen}$ to create user's public/secret keys $\textsf{upk}/\textsf{usk}$.  $\mathcal{B}$ can response to any credential issue query since credential issuer's secret key $\textsf{sk}$ is generated by $\mathcal{B}$.
		
		\textbf{Attribute Decryption Key Query.}  $\mathcal{B}$ can response to any attribute decryption key query since $\textsf{msk}$ is generated by $\mathcal{B}$. 
		
		\textbf{Policy Decryption Key Query.}  $\mathcal{B}$ can response to any policy decryption key query since $\textsf{msk}$ is generated by $\mathcal{B}$.
		
		\textbf{Challenge.}  After the secret key queries, $\mathcal{A}$ requests for the challenge ciphertext corresponding to 
		symmetric keys $(K_0,K_1)$,  attributes $\vec{x}$ and formula $f$. $\mathcal{B}$ flips a random coin $\textsf{coin}\leftarrow\{0,1\}$ and constructs the challenge ciphertext for $K_{\textsf{coin}}$. 
		$\mathcal{B}$ computes $(\{\textbf{u}_j^{\top}\},\rho)\leftarrow\textsf{share}(f,{\textbf{z}}^{\top}_{2m+1}\textbf{U}_0)$ and sets the challenge ciphertext as $\textsf{CT}_{\vec{x},f}:=$
		$$(
		\textsf{ct}_0=e([\widetilde{\textbf{z}}^{\top}+\textbf{z}^{\top}_{2m+1}]_1,[\textbf{v}]_2)\cdot K_{\textsf{coin}},$$
		$$\textsf{ct}_1'=[\widetilde{\textbf{z}}^{\top}]_1,\textsf{ct}_2'=[\widetilde{\textbf{z}}^{\top}\sum\nolimits_{i:x_i^{(out)}=1}\textbf{W}_i]_1,$$
		$$\textsf{ct}_1=[\textbf{z}^{\top}_{2m+1}]_1,\{\widetilde{\textsf{ct}}_{j}=[\textbf{z}_j^{\top}]_1,\textsf{ct}_{\rho(j),j}=[\textbf{u}_j^{\top}+\textbf{z}_j^{\top}\textbf{W}_{\rho(j)}]_1,$$
		$$\textsf{ct}_{i,j}=[\textbf{z}_j^{\top}\textbf{W}_i]_1\}),$$
		where $(\widetilde{\textbf{z}}, \textsf{ct}_1', \textsf{ct}_2')$ are computed normally and note that $|\{\textbf{u}_j\}|\leq 2m$.
		
		\textbf{Guess.}  $\mathcal{A}$ halts the game with a guess $\textsf{coin}'\leftarrow\{0,1\}$. $\mathcal{B}$ outputs 1 if $\textsf{coin}'=\textsf{coin}$, and 0 otherwise.	
		It is straight forward to see that if $\textbf{Z}^{\top}=\textbf{S}^{\top}\textbf{A}
		$, the challenge ciphertext is \textsf{Normal} and $\mathcal{B}$ simulates $\textsf{H}_0$;	
		if $\textbf{Z}^{\top}=\textbf{C}^{\top}
		$, the challenge ciphertext is \textsf{SF} and $\mathcal{B}$ simulates $\textsf{H}_1$.
	\end{proof}
	
	\textsf{Lemma B.2.}
	\label{Lemma:2}
	Under the $MDDH_{k}$ assumption on $G_2$, we have 
	$$|Pr[\left<\mathcal{A},\textsf{H}_{2,\ell-1,3}\right>=1]-Pr[\left<\mathcal{A},\textsf{H}_{2,\ell,1}\right>=1]|= \textsf{negl}(\lambda).$$
	\begin{proof}
		Assume that $\mathcal{A}$ distinguishes $\textsf{H}_{2,\ell-1,3}$ and $\textsf{H}_{2,\ell,1}$ with non-negligible advantage. Then, we can construct another adversary $\mathcal{B}$ to break the $MDDH_k$ assumption. 	
		On input $MDDH_{k}$ challenge $([\textbf{B}]_2,[\textbf{z}]_2)$, where either $\textbf{z}=\textbf{B}\textbf{r}$ for $\textbf{r}\leftarrow\mathbb{Z}_p^{k}$, or $\textbf{z}=\textbf{d}$ for $\textbf{d}\leftarrow\mathbb{Z}_p^{k+1}$. $\mathcal{B}$ proceeds as below.
		
		\textbf{Setup.} $\mathcal{B}$ chooses generators $g\leftarrow G_1$, $h\leftarrow G_2$, user's attribute number $n$, and sets $\textsf{pp}=(g,h,n)$. Next, $\mathcal{B}$ selects $\textbf{A}\leftarrow\mathbb{Z}_p^{k\times 2k}$, $\textbf{U}_0,\textbf{W}_i\leftarrow\mathbb{Z}_p^{2k\times (k+1)}$, $\textbf{v}\leftarrow\mathbb{Z}_p^{2k}$, and forms $\textsf{mpk}$ with these parameters as in the scheme. $\mathcal{B}$ computes $\textbf{A}^{\bot}\in\mathbb{Z}_p^{2k\times k}$ such that $\textbf{A}\textbf{A}^{\bot}=\textbf{0}$, which is used for responding secret key queries.  $\mathcal{B}$ runs $\mathcal{FAC}.\textsf{CredKeyGen}$ to create $(\textsf{pk},\textsf{sk})$.
		
		\textbf{Issue Query.} $\mathcal{B}$ firstly executes $\mathcal{FAC}.\textsf{UserKeyGen}$ to create user's public/secret keys $\textsf{upk}/\textsf{usk}$.  $\mathcal{B}$ can response to any credential issue query since credential issuer's secret key $\textsf{sk}$ is generated by $\mathcal{B}$.

		\textbf{Attribute Decryption Key.} $\mathcal{B}$ simulates attribute decryption keys as below.
		
		- For the first $\ell-1$ attribute decryption key queries, say the $q$-th request is for $\vec{x}$, $\mathcal{B}$ samples $\boldsymbol{\delta}^{(q)},\textbf{r}^{(q)}\in\mathbb{Z}_p^{k}$, and creates (\textsf{SF}) attribute decryption key $\textsf{DK}_{\vec{x}}:=$
		$$\big([\textbf{v}+\textbf{A}^{\bot}\boldsymbol{\delta}^{(q)}+\textbf{U}_0\textbf{B}\textbf{r}^{(q)}]_2,[\textbf{B}\textbf{r}^{(q)}]_2,[\sum\nolimits_{i:x_i=1}\textbf{W}_i\textbf{B}\textbf{r}^{(q)}]_2\big).$$
		
		- For the last $Q_x-\ell$ attribute decryption key queries, $\mathcal{B}$ proceeds as before for the first $\ell-1$ keys except substituting $\textbf{v}+\textbf{A}^{\bot}\boldsymbol{\delta}^{(q)}$ with $\textbf{v}$. It is obvious that it becomes a $\textsf{Normal}$ key.
		
		- For the $\ell$th attribute decryption key request, $\mathcal{A}'$ creates the secret key 
		$\textsf{DK}_{\vec{x}}:=\big([\textbf{v}+\textbf{U}_0\textbf{z}]_2,[\textbf{z}]_2,[\sum\nolimits_{i:x_i=1}\textbf{W}_i\textbf{z}]_2\big).$
		
		\textbf{Policy Decryption Key.}  $\mathcal{B}$ simulates any policy decryption key normally since the elements for generating policy decryption key are generated by $\mathcal{B}$.
		
		\textbf{Challenge.} After the secret key queries, $\mathcal{A}$ requests for the challenge ciphertext corresponding to 
		symmetric keys $(K_0,K_1)$,  attributes $\vec{x}$ and formula $f$. $\mathcal{B}$ flips a random coin $\textsf{coin}\leftarrow\{0,1\}$ and constructs the challenge ciphertext for 
		$K_{\textsf{coin}}$. Sample $\textbf{c},\textbf{c}_j\leftarrow\mathbb{Z}_p^{2k}$ for each $j$, compute $(\{\textbf{u}_j^{\top}\},\rho)\leftarrow\textsf{share}(f,\textbf{c}^{\top}\textbf{U}_0)$  and return (\textsf{SF}) challenge ciphertext $\textsf{CT}_{\vec{x},f}:=$
		$$\big(
		\textsf{ct}_0=e([\widetilde{\textbf{z}}^{\top}+\textbf{c}^{\top}]_1,[\textbf{v}]_2)\cdot K_{\textsf{coin}},$$
		$$\textsf{ct}_1'=[\widetilde{\textbf{z}}^{\top}]_1,\textsf{ct}_2'=[\widetilde{\textbf{z}}^{\top}\sum\nolimits_{i:x_i^{(out)}=1}\textbf{W}_i]_1,$$
		$$\textsf{ct}_1=[\textbf{c}^{\top}]_1,\widetilde{\textsf{ct}}_{j}=[\textbf{c}_j^{\top}]_1,\textsf{ct}_{\rho(j),j}=[\textbf{u}_j^{\top}+\textbf{c}_j^{\top}\textbf{W}_{\rho(j)}]_1,$$
		$$\textsf{ct}_{i,j}=[\textbf{c}_j^{\top}\textbf{W}_i]_1\big).$$
		
		\textbf{Guess.}  $\mathcal{A}$ halts the game with a guess $\textsf{coin}'\leftarrow\{0,1\}$. $\mathcal{B}$ outputs 1 if $\textsf{coin}'=\textsf{coin}$, and 0 otherwise.	
		It is straight forward to see that if $\textbf{z}=\textbf{B}\textbf{r}$, then the $\ell$th attribute decryption key is \textsf{Normal} and $\mathcal{A}'$ has simulated $\textsf{H}_{2,\ell-1,3}$;	
		if $\textbf{z}=\textbf{d}$, then the $\ell$th attribute decryption key is \textsf{P-Normal} and $\mathcal{A}'$ has simulated $\textsf{H}_{2,\ell,1}$.
	\end{proof}

	\textsf{Lemma B.3.}
	\label{Lemma:3}
	Under the $MDDH_{k}$ assumption, we have 
	$$|Pr[\left<\mathcal{A},\textsf{H}_{2,\ell,1}\right>=1]-Pr[\left<\mathcal{A},\textsf{H}_{2,\ell,2}\right>=1]|= \textsf{negl}(\lambda).$$
	\begin{proof}
		Assume that $\mathcal{A}$ distinguishes $\textsf{H}_{2,\ell,1}$ and $\textsf{H}_{2,\ell,2}$ with non-negligible advantage. Then, we can construct another adversary $\mathcal{B}$ that distinguishes the oracles in $\textsf{G}_{\beta}^{1-\textsf{ABE}}$ of \cite{kowalczyk2019compact}, which implies an attacker against the $MDDH_{k}$ assumption. Given $\mu^{(0)}$ as an input and equipped with oracles $\mathcal{O}_{\textsf{F},\beta}$, $\mathcal{O}_X$ and $\mathcal{O}_E$ (defined in $\textsf{G}_{\beta}^{1-\textsf{ABE}}$ of \cite{kowalczyk2019compact}), $\mathcal{B}$ proceeds as below.	
		
		\textbf{Setup.} $\mathcal{B}$ chooses generators $g\leftarrow G_1$, $h\leftarrow G_2$, user's attribute number $n$, and sets $\textsf{pp}=(g,h,n)$. Next, $\mathcal{B}$ chooses $\textbf{A}\leftarrow\mathbb{Z}_p^{k\times 2k}$,
		$\textbf{B}\leftarrow\mathbb{Z}_p^{k\times k}$, $\widetilde{\textbf{U}}_0,\widetilde{\textbf{W}}_i\leftarrow\mathbb{Z}_p^{2k\times k}$ for $i\in[1,n]$, and $\widetilde{\textbf{v}}\leftarrow\mathbb{Z}_p^{2k}$, computes $\textbf{A}^{\bot}\in\mathbb{Z}_p^{2k\times k}\backslash\{\textbf{0}\}$, $\textbf{b}^{\bot}\leftarrow\mathbb{Z}_p^{k}$ such that $\textbf{A}\textbf{A}^{\bot}=\textbf{0}$ and $(\textbf{b}^{\bot})^{\top}\textbf{B}=\textbf{0}$ and implicitly defines
		$$\textbf{v}:=\widetilde{\textbf{v}}-\frac{\mu^{(0)}((\textbf{b}^{\bot})^{\top}\textbf{d})}{\textbf{c}^{\top}\textbf{A}^{\bot}\textbf{u}}\textbf{A}^{\bot}\textbf{u},$$
		$$\textbf{U}_0:=\widetilde{\textbf{U}}_0+\frac{\mu^{(\beta)}}{\textbf{c}^{\top}\textbf{A}^{\bot}\textbf{u}}\textbf{A}^{\bot}\textbf{u}(\textbf{b}^{\bot})^{\top},$$
		$$\textbf{W}_i:=\widetilde{\textbf{W}}_i+\textbf{A}^{\bot}\textbf{w}_i(\textbf{b}^{\bot})^{\top},$$
		where $\textbf{w}_i\in\mathbb{Z}_p^{k}$, $\mu^{(\beta)}\in\mathbb{Z}_p$ are chosen by the $\textsf{G}_{\beta}^{\textsf{1-ABE}}$ game in \cite{kowalczyk2019compact}, 
		$\textbf{c}\leftarrow\mathbb{Z}_p^{2k}$ is selected for generating the challenge ciphertext, $\textbf{d}\leftarrow\mathbb{Z}_p^{k+1}$ is selected for generating the $\ell$th secret key, and $\textbf{u}\leftarrow\mathbb{Z}_p^{k}$. Note that $\mathcal{B}$ can compute $\textbf{v}$ since it has $\mu^{(0)}$ from the game and knows all other vectors.	
		Then, $\mathcal{B}$ creates
		$$\mathsf{mpk}:= (\textbf{pp},[\textbf{A}]_1,[\textbf{A}\widetilde{\textbf{U}}_0]_1,[\textbf{A}\widetilde{\textbf{W}}_1]_1,\cdots,[\textbf{A}\widetilde{\textbf{W}}_n]_1,e([\textbf{A}]_1,[\widetilde{\textbf{v}}]_2)).$$ $\mathcal{B}$ runs $\mathcal{FAC}.\textsf{CredKeyGen}$ to create $(\textsf{pk},\textsf{sk}).$
		
		\textbf{Issue Query.} $\mathcal{B}$ firstly executes $\mathcal{FAC}.\textsf{UserKeyGen}$ to create user's public/secret keys $\textsf{upk}/\textsf{usk}$.  $\mathcal{B}$ can response to any credential issue query since credential issuer's secret key $\textsf{sk}$ is generated by $\mathcal{B}$.

		\textbf{Attribute Decryption Key.} $\mathcal{B}$ simulates attribute decryption key as below.
		
		- For the first $\ell-1$ attribute decryption key queries, say the $q$th request is for $\vec{x}$, $\mathcal{B}$ samples $\boldsymbol{\delta}^{(q)},\textbf{r}^{(q)}\leftarrow\mathbb{Z}_p^{k}$, and creates the (\textsf{SF}) attribute decryption key:
		$\textsf{DK}_{\vec{x}}:=$
		$$(\textsf{dk}_1=[\textbf{v}+\textbf{A}^{\bot}\boldsymbol{\delta}^{(q)}+\underbrace{\widetilde{\textbf{U}}_0\textbf{B}\textbf{r}^{(q)}}_{=\textbf{U}_0\textbf{B}\textbf{r}^{(q)}}]_2,$$
		$$\textsf{dk}_2=[\textbf{B}\textbf{r}^{(q)}]_2,~~~\textsf{dk}_3=[\sum\nolimits_{i:x_i^{(out)}=1}\underbrace{\widetilde{\textbf{W}}_i\textbf{B}\textbf{r}^{(q)}}_{=\textbf{W}_i\textbf{B}\textbf{r}^{(q)}}]_2).$$
		
		- For the last $Q_x-\ell$ attribute decryption key queries, $\mathcal{B}$ proceeds as before for the first $\ell-1$ keys except using just $\textbf{v}$ instead of $\textbf{v}+\textbf{A}^{\bot}\boldsymbol{\delta}^{(q)}$. It is easy to see that it forms a \textsf{Normal} attribute decryption key.
		
		- For the $\ell$th attribute decryption key query for $\vec{x}$, queries $\mathcal{O}_{\textsf{X}}(\vec{x})\rightarrow(\{\textbf{w}_i\}_{x_i=1})$ and creates the attribute decryption key:
		$\textsf{DK}_{\vec{x}}:=$ $$(\textsf{dk}_1=[\underbrace{\widetilde{\textbf{v}}+\widetilde{\textbf{U}}_0\textbf{d}}_{=\textbf{v}+\frac{(\mu^{(0)}-\mu^{(\beta)})((\textbf{b}^{\bot})^{\top}\textbf{d})}{(\textbf{c}^{\top}\textbf{A}^{\bot}\textbf{u})}\textbf{A}^{\bot}\textbf{u}+\textbf{U}_0\textbf{d}}]_2,~~~\textsf{dk}_2=[\textbf{d}]_2,$$
		$$\textsf{dk}_3=[\sum\nolimits_{i:x_i^{(out)}=1}\underbrace{(\widetilde{\textbf{W}}_i+\textbf{A}^{\bot}\textbf{w}_i(\textbf{b}^{\bot})^{\top})\textbf{d}}_{=\textbf{W}_i\textbf{d}}]_2).$$
		
		We claim that if $\beta=0$, then the $\ell$th key is a \textsf{P-Normal} attribute decryption key since $\textbf{v}+\frac{(\mu^{(0)}-\mu^{(0)})((\textbf{b}^{\bot})^{\top}\textbf{d})}{(\textbf{c}^{\top}\textbf{A}^{\bot}\textbf{u})}\textbf{A}^{\bot}\textbf{u}=\textbf{v}$; and
		if $\beta=1$, then the $\ell$th key is a \textsf{P-SF} key since $\boldsymbol{\delta}^{(\ell)}=\frac{(\mu^{(0)}-\mu^{(1)})((\textbf{b}^{\bot})^{\top}\textbf{d})}{(\textbf{c}^{\top}\textbf{A}^{\bot}\textbf{u})}\textbf{u}$.
		
		\textbf{Policy Decryption Key.}  $\mathcal{B}$ can simulate any policy key normally since the elements for generating policy decryption key are generated by $\mathcal{B}$.
		
		\textbf{Challenge.} When $\mathcal{A}$ requests a challenge ciphertext for 
		symmetric keys $(K_0,K_1)$,  attributes $\vec{x}$ and formula $f$, $\mathcal{B}$ flips a random coin $\textsf{coin}\leftarrow\{0,1\}$ and constructs the challenge ciphertext for 
		$K_{\textsf{coin}}$.  $\mathcal{B}$ queries $\mathcal{O}_{\textsf{F}}(f)\rightarrow\big(\{[\mu_j+\textbf{r}_j^{\top}\textbf{w}_{\rho(j)}]_1,[\textbf{r}_j]_1\}\big)$ on input $f$. $\mathcal{B}$ samples $\widetilde{\textbf{c}}_j\leftarrow\mathbb{Z}_p^{k}$ for each $j$, defines $\textbf{A}_C^{\bot}:=\left[\begin{matrix}
			(\textbf{A}^{\bot})^{\top}\\\textbf{M}
		\end{matrix}\right]\in\mathbb{Z}_p^{2k\times 2k}$ for a choice of $\textbf{M}$ that makes $\textbf{A}_C^{\bot}$ invertible, computes $(\{\widetilde{\textbf{u}}_j^{\top}\},\rho)\leftarrow\textsf{share}(f,\textbf{c}^{\top}\widetilde{\textbf{U}}_0)$, 
		$[\textbf{c}_j]_1:=\left[(\textbf{A}_C^{\bot})^{-1}\left(\begin{matrix}
			\textbf{r}_j\\ \widetilde{\textbf{c}}_j
		\end{matrix}\right)\right]_1$, and constructs the  (\textsf{SF}) challenge ciphertext $\textsf{CT}_{\vec{x},f}:=$
		$$\big(
		\textsf{ct}_0=e([\widetilde{\textbf{z}}^{\top}+\textbf{c}^{\top}]_1,[\textbf{v}]_2)\cdot K_{\textsf{coin}},$$ $$\textsf{ct}_1'=[\widetilde{\textbf{z}}^{\top}]_1,
		\textsf{ct}_2'=[\widetilde{\textbf{z}}^{\top}\sum\limits_{i:x_i^{(out)}=1}\underbrace{\widetilde{\textbf{W}}_i+\textbf{A}^{\bot}\textbf{w}_i(\textbf{b}^{\bot})^{\top}}_{=\textbf{W}_i}]_1,$$
		$$\textsf{ct}_1=[\textbf{c}^{\top}]_1,~~~\widetilde{\textsf{ct}}_{j}=[\textbf{c}_j^{\top}]_1,$$
		$$\textsf{ct}_{\rho(j),j}=[\widetilde{\textbf{u}}_j^{\top}+(\mu_j+\textbf{r}_j^{\top}\textbf{w}_{\rho(j)})(\textbf{b}^{\bot})^{\top}+\textbf{c}_j^{\top}\widetilde{\textbf{W}}_{\rho(j)}]_1),$$		
		$$\textsf{ct}_{i,j}=\textbf{r}_j^{\top}\textbf{w}_i(\textbf{b}^{\bot})^{\top}+\textbf{c}_j^{\top}\widetilde{\textbf{W}}_i]_1).$$
		We can deduce that 
		$$\textsf{ct}_{\rho(j),j}=[\underbrace{\widetilde{\textbf{u}}_j^{\top}+\mu_j(\textbf{b}^{\bot})^{\top}}_{\equiv\textsf{share}(f,\textbf{c}^{\top}\textbf{U}_0)}+\underbrace{\textbf{c}_j^{\top}\widetilde{\textbf{W}}_{\rho(j)}+\textbf{r}_j^{\top}\textbf{w}_{\rho(j)}(\textbf{b}^{\bot})^{\top}}_{=\textbf{c}_j^{\top}\textbf{W}_{\rho(j)}}]_1.$$
		It should be noted that $\{\mu_j(\textbf{b}^{\bot})^{\top}\}$ is distributed like the output of $\textsf{share}(f,\mu^{b}(\textbf{b}^{\bot})^{\top})$, and therefore due to linearity and the fact that $\textbf{c}^{\top}\textbf{U}_0=\textbf{c}^{\top}\widetilde{\textbf{U}}_0+\mu^{(b)}(\textbf{b}^{\bot})^{\top}$, then $\{\widetilde{\textbf{u}}_j^{\top}+\mu_j(\textbf{b}^{\bot})^{\top}\}$ is distributed like $\textsf{share}(f,\textbf{c}^{\top}\widetilde{\textbf{U}}_0+\mu^{(b)}(\textbf{b}^{\bot})^{\top})\equiv\textsf{share}(f,\textbf{c}^{\top}\textbf{U}_0)$. Also, note that $\textbf{c}_j^{\top}\textbf{W}_i=\textbf{c}_j^{\top}\widetilde{\textbf{W}}_i+\textbf{r}_j\textbf{w}_{\rho(j)}\textbf{b}^{\bot}$ since $\textbf{c}_j^{\top}\textbf{A}^{\bot}=\textbf{r}_j$.
		
		\textbf{Guess.}  $\mathcal{A}$ halts the game with a guess $\textsf{coin}'\leftarrow\{0,1\}$. $\mathcal{B}$ outputs 1 if $\textsf{coin}'=\textsf{coin}$, and 0 otherwise.	
		
		Putting everything together, we can see that  $\mathcal{B}$ simulates $\textsf{H}_{2,
			\ell,1}$ when $\beta=0$; and $\textsf{H}_{2,
			\ell,2}$ when $\beta=1$.
	\end{proof}

	\textsf{Lemma B.4.}
	\label{Lemma:4}
	Under the $MDDH_{k}$ assumption, we have 
	$$|Pr[\left<\mathcal{A},\textsf{H}_{2,\ell,2}\right>=1]-Pr[\left<\mathcal{A},\textsf{H}_{2,\ell,3}\right>=1]|= \textsf{negl}(\lambda).$$	
	\begin{proof}
		Omitted, since this proof is similar to that of Lemma E.3. We need to substitute $\textbf{v}$ with $\textbf{v}+\textbf{A}^{\bot}\boldsymbol{\delta}^{(\ell)}$ for the $\ell$th attribute decryption key query, where $\boldsymbol{\delta}^{(\ell)}$ is a random element.
	\end{proof}

	\textsf{Lemma B.5.}
	\label{Lemma:5}
	We have
	$$|Pr[\left<\mathcal{A},\textsf{H}_{2,Q_x, 3}\right>=1]-Pr[\left<\mathcal{A},\textsf{H}_{3}\right>=1]|\leq 1/p$$ unconditionally.
	\begin{proof}
		These two hybrids are identically distributed conditioned on $\textbf{c}^{\top}\textbf{A}^{\bot}\neq\textbf{0}$. To see this, consider two ways of choosing $\textbf{v}:\textbf{v}=\widetilde{\textbf{v}}\leftarrow\mathbb{Z}_p^{2k}$ and $\textbf{v}:\textbf{v}=\widetilde{\textbf{v}}+\textbf{A}^{\bot}\widetilde{\textbf{m}}$ for an independently random $\widetilde{\textbf{m}}\leftarrow\mathbb{Z}_p^{k}$. Note that both result in $\textbf{v}$ having a uniform distribution.
		
		Using $\widetilde{\textbf{v}}$ to simulate hybrid $\textsf{H}_{2,Q_x,3}$ obviously results in $\textsf{H}_{2,Q_x,3}$  (where $\textbf{v}=\widetilde{\textbf{v}}$). However, using the identically distributed $\textbf{v}=\widetilde{\textbf{v}}+\textbf{A}^{\bot}\widetilde{\textbf{m}}$ to simulate $\textsf{H}_{2,Q_x,3}$  results in $\textsf{H}_3$ with $\widetilde{K}=K_{\textsf{coin}}\cdot[ \textbf{c}^{\top}\textbf{A}^{\bot}\widetilde{\textbf{m}}]_{T}$. Note that the information of $\widetilde{\textbf{m}}$ is not leaked to $\mathcal{A}$ from the secret key queries since $\widetilde{\textbf{m}}$ is blinded by random value $\boldsymbol{\delta}^{(i)}$ for each key. Therefore, $\widetilde{K}$ is distributed uniformly at random over $G_T$ as long as $\textbf{c}^{\top}\textbf{A}^{\bot}\neq \textbf{0}$.
		
		Since $\textbf{c}$ is chosen at random and independent from $\widetilde{\textbf{s}}^{\top}\textbf{A}$ and  $\textbf{A}^{\bot}\neq\textbf{0}$, so $\textbf{c}^{\top}\textbf{A}^{\bot}=\textbf{0}$ with probability $1/p$, and since we know that $\textsf{H}_{2,Q_x,3}\equiv\textsf{H}_3$ conditioned on $\textbf{c}^{\top}\textbf{A}^{\bot}\neq\textbf{0}$, then the lemma follows.
	\end{proof}
	
	\noindent\textit{Case 2: the adversary does not query a key for $f_{\textsf{rcv}}$ such that $f_{\textsf{rcv}}(\vec{x}^{(out)*}_{\textsf{snd}}) = 1$}.
	
	A ciphertext (under access policy $f$ and attributes $\vec{x}$) can be in one of the following forms:
	
	- \textsf{Normal}: A normal ciphertext is generated by $\textsf{Enc}$.
	
	- \textsf{SF}: An \textsf{SF} ciphertext is the same as \textsf{Normal} ciphertext, except that
	$\widetilde{\textbf{s}}^{\top}\textbf{A}$
	is replaced with 
	$\widetilde{\textbf{c}}^{\top}\leftarrow\mathbb{Z}_p^{2k}$. 
	That is $\textsf{CT}_{\vec{x},f}:=$
	$$(
	\textsf{ct}_0=e([\boxed{\widetilde{\textbf{c}}^{\top}}+{\textbf{c}^{\top}}]_1,[\textbf{v}]_2)\cdot K,$$
	$$\textsf{ct}_1'=[\boxed{\widetilde{\textbf{c}}^{\top}}]_1,\textsf{ct}_2'=\big[\boxed{\widetilde{\textbf{c}}^{\top}}\sum\nolimits_{i:x_i^{(out)}=1}\textbf{W}_i\big]_1,$$
	$$\textsf{ct}_1=[{\textbf{c}^{\top}}]_1,\{\widetilde{\textsf{ct}}_{j}=[{\textbf{c}_j^{\top}}]_1,\textsf{ct}_{\rho(j),j}=[\textbf{u}_j^{\top}+{\textbf{c}_j^{\top}}\textbf{W}_{\rho(j)}]_1,$$
	$$\textsf{ct}_{i,j}=[{\textbf{c}_j^{\top}}\textbf{W}_i]_1\}),$$
	where $\textbf{c}^{\top} = \textbf{s}^{\top}\textbf{A}$ and $\textbf{c}_j^{\top} = \textbf{s}_j^{\top}\textbf{A}$ are same as in the original ciphertext.
	
	A secret key (for attributes $\vec{x}$) follows its normal form.
	A secret key (for a Boolean formula $f$) can be one of the following forms:
	
	- \textsf{Normal}: A normal secret key is generated by $\textsf{PolGen}$.
	
	- \textsf{SF}: An \textsf{SF} key is sampled as a \textsf{Normal} key, except $\textbf{v}$ is replaced with $\textbf{v}+\delta\textbf{a}^{\bot}$, where a fresh $\delta$ is chosen per \textsf{SF} key and $\textbf{a}^{\bot}$ is any fixed $\textbf{a}^{\bot}\in\mathbb{Z}_p^{2k}\backslash\{\textbf{0}\}$. That is $\textsf{DK}_f:=$
	$$\big(\textsf{dk}_j=[\textbf{r}_j]_2,\textsf{dk}_{\rho(j),j}=[\textbf{v}_j+\textbf{W}_{\rho(j)}\textbf{r}_j]_2,\textsf{dk}_{i,j}=[\textbf{W}_i\textbf{r}_j]_2\big).$$
	where $(\{\textbf{v}_j\}_{j\in[\hat{m}]},\rho)\xleftarrow{\$}\textsf{share}(f,\boxed{\textbf{v}+\delta\textbf{a}^{\bot}})$, $\textbf{r}_j\xleftarrow{\$}\mathbb{Z}_p^k$.	
	
	Next, we define hybrid sequences for the proof. Assume the adversary $\mathcal{A}$ makes at most $Q_x$ attribute decryption key queries and $Q_f$ policy decryption key queries. 
	
	- $\textsf{H}_0$: This is the real game where all secret keys and ciphertexts are \textsf{Normal}.
	
	- $\textsf{H}_1$: This game is the same as $\textsf{H}_0$ except that the challenge ciphertext is \textsf{SF}.
	
	- $\textsf{H}_{2,\ell}$: This game is the same as $\textsf{H}_{1}$ except that the first $\ell$ policy decryption keys are \textsf{SF} and the remaining $Q_f-\ell$ keys are \textsf{Normal}, where $\ell=0,\cdots,Q_f$.
	
	- $\textsf{H}_3$: This game is the same as $\textsf{H}_{2,Q_f}$ except that the message encryption symmetric key $K$ to be encrypted is replaced by a random $\widetilde{K}$.

	\textsf{Lemma B.6.}
	\label{Lemma:6}
	Under the $MDDH_{k}$ assumption on $G_1$, we have 
	$$|Pr[\left<\mathcal{A},\textsf{H}_0\right>=1]-Pr[\left<\mathcal{A},\textsf{H}_1\right>=1]|= \textsf{negl}(\lambda).$$
	
	\begin{proof}
		If $\mathcal{A}$ can distinguish $\textsf{H}_0$ from $\textsf{H}_1$ with non-negligible advantage, then we can construct an algorithm $\mathcal{B}$ that can solve the $MDDH_{k}$ assumption. On input a $MDDH_{k}$  challenge $([\textbf{A}]_1,[\widetilde{\textbf{z}}]_1)$,
		where either $\widetilde{\textbf{z}}^{\top}=\widetilde{\textbf{s}}^{\top}\textbf{A}$
		or $\widetilde{\textbf{z}}=\widetilde{\textbf{c}}$
		for $\widetilde{\textbf{s}} \leftarrow\mathbb{Z}_p^{k}$
		and $\widetilde{\textbf{c}}\leftarrow\mathbb{Z}_p^{2k}$. $\mathcal{B}$ proceeds as in the proof of Lemma E.2 except that an SF challenge ciphertext is returned to $\mathcal{A}$. $\mathcal{B}$ flips a random coin $\textsf{coin}\leftarrow\{0,1\}$ and constructs the challenge ciphertext for 
		$K_{\textsf{coin}}$ by computing $(\{\textbf{u}_j^{\top}\},\rho)\leftarrow\textsf{share}(f,{\textbf{z}}^{\top}\textbf{U}_0)$, where $\textbf{z}^{\top}=\textbf{s}^{\top}\textbf{A}$, and $\textbf{z}_j^{\top}=\textbf{s}_j^{\top}\textbf{A}$ as in the normal ACME construction, and sets the challenge ciphertext as $\textsf{CT}_{\vec{x},f}:=$
		$$\big(
		\textsf{ct}_0=e([\widetilde{\textbf{z}}^{\top}+\textbf{z}^{\top}]_1,[\textbf{v}]_2)\cdot K_{\textsf{coin}},$$
		$$\textsf{ct}_1'=[\widetilde{\textbf{z}}^{\top}]_1,\textsf{ct}_2'=[\widetilde{\textbf{z}}^{\top}\sum\nolimits_{i:x_i^{(out)}=1}\textbf{W}_i]_1,$$
		$$\textsf{ct}_1=[\textbf{z}^{\top}]_1,\{\widetilde{\textsf{ct}}_{j}=[\textbf{z}_j^{\top}]_1,\textsf{ct}_{\rho(j),j}=[\textbf{u}_j^{\top}+\textbf{z}_j^{\top}\textbf{W}_{\rho(j)}]_1.$$
		$$\textsf{ct}_{i,j}=[\textbf{z}_j^{\top}\textbf{W}_i]_1\}\big).$$
		
		\textbf{Guess.}  $\mathcal{A}$ halts the game with a guess $\textsf{coin}'\leftarrow\{0,1\}$. $\mathcal{B}$ outputs 1 if $\textsf{coin}'=\textsf{coin}$, and 0 otherwise.	
		It is straight forward to see that if $\widetilde{\textbf{z}}^{\top}=\widetilde{\textbf{s}}^{\top}\textbf{A}
		$, the challenge ciphertext is \textsf{Normal} and $\mathcal{B}$ simulates $\textsf{H}_0$;	
		if $\widetilde{\textbf{z}}^{\top}=\widetilde{\textbf{c}}^{\top}
		$, the challenge ciphertext is \textsf{SF} and $\mathcal{B}$ simulates $\textsf{H}_1$.
	\end{proof}
	
	\textsf{Lemma B.7.}
	\label{Lemma:7}
	Under the $MDDH_{k}$ assumption on $G_2$, we have 
	$$|Pr[\left<\mathcal{A},\textsf{H}_{2,\ell-1}\right>=1]-Pr[\left<\mathcal{A},\textsf{H}_{2,\ell}\right>=1]|= \textsf{negl}(\lambda).$$
	\begin{proof}
		Assume that $\mathcal{A}$ distinguishes $\textsf{H}_{2,\ell-1}$ and $\textsf{H}_{2,\ell}$ with non-negligible advantage. Then, we can construct another adversary $\mathcal{B}$ that distinguishes the oracles in $\textsf{G}_{\beta}^{1-\textsf{ABE}}$ of \cite{katsumata2020compact}, which implies an attacker against the $MDDH_{k}$ assumption. Given $\mu^{(0)}$ as an input and equipped with oracles $\mathcal{O}_{\textsf{F},\beta}$, $\mathcal{O}_X$ and $\mathcal{O}_E$ (defined in $\textsf{G}_{\beta}^{1-\textsf{ABE}}$ of \cite{katsumata2020compact}), $\mathcal{B}$ proceeds as below.	
		
		\textbf{Setup.} $\mathcal{B}$ chooses generators $g\leftarrow G_1$, $h\leftarrow G_2$, user's attribute number $n$, and sets $\textsf{pp}=(g,h,n)$. Next, $\mathcal{B}$ chooses $\textbf{A}\xleftarrow{\$}\mathbb{Z}_p^{k\times 2k}$,
		$\widetilde{\textbf{U}}_0,\widetilde{\textbf{W}}_i\xleftarrow{\$}\mathbb{Z}_p^{2k\times k}$ for $i\in[1,n]$, and $\widetilde{\textbf{v}}\xleftarrow{\$}\mathbb{Z}_p^{2k}$, computes $\textbf{a}^{\bot}\in\mathbb{Z}_p^{2k}\backslash\{\textbf{0}\}$ such that $\textbf{A}\textbf{a}^{\bot}=\textbf{0}$. It then sets $\widetilde{\textbf{W}}_0:=\textbf{0}$ and implicitly defines
		$\textbf{v}:=\widetilde{\textbf{v}}+\mu^{(0)}\textbf{a}^{\bot},~\textbf{W}_i:=\widetilde{\textbf{W}}_i+\textbf{a}^{\bot}\textbf{w}_i^{\top},$
		where $\textbf{w}_i\in\mathbb{Z}_p^{k}$, $\mu^{(0)}\in\mathbb{Z}_p$ are chosen by the $\textsf{G}_{\beta}^{\textsf{1-ABE}}$ game in \cite{katsumata2020compact}.	
		Then, $\mathcal{B}$ creates
		$$\mathsf{mpk}:= (\textbf{pp},[\textbf{A}]_1,[\textbf{A}\widetilde{\textbf{U}}_0]_1,[\textbf{A}\widetilde{\textbf{W}}_1]_1,\cdots,[\textbf{A}\widetilde{\textbf{W}}_n]_1,e([\textbf{A}]_1,[\widetilde{\textbf{v}}]_2)).$$
		
		$\mathcal{B}$ runs $\mathcal{FAC}.\textsf{CredKeyGen}$ to create $(\textsf{pk},\textsf{sk})$.
		
		\textbf{Issue Query.} $\mathcal{B}$ firstly executes $\mathcal{FAC}.\textsf{UserKeyGen}$ to create user's public/secret keys $\textsf{upk}/\textsf{usk}$.  $\mathcal{B}$ can response to any credential issue query since credential issuer's secret key $\textsf{sk}$ is generated by $\mathcal{B}$.
		
		\textbf{Attribute Decryption Key Query.}  $\mathcal{B}$ simulates any attribute decryption key normally.
		%
		%
		
		\textbf{Policy Decryption Key Query.} $\mathcal{B}$ responds to $\mathcal{A}$'s policy decryption key queries as below.
		
		- For the first $\ell-1$ policy decryption key queries, say for formula $f$ of size $m$, $\mathcal{B}$ computes:
		$$(\{\textbf{v}_j\}_{j\in[\hat{m}]},\rho)\xleftarrow{\$}\textsf{share}(f,\underbrace{\widetilde{\textbf{v}}+\widetilde{\delta}\textbf{a}^{\bot}}_{=\textbf{v}+\delta\textbf{a}^{\bot}}),$$
		where $\widetilde{\delta}\xleftarrow{\$}\mathbb{Z}_p$ is drawn independently for each key (here, the per-key $\delta=\widetilde{\delta}-\mu^{(0)}$ implicitly). Next, for each $j\in[\hat{m}]$, it queries $\mathcal{O}_{\textsf{E}}\rightarrow([\textbf{r}_j]_2,\{[\textbf{w}_i^{\top}\textbf{r}_j]_2\}_{i\in[n]})$ and forms the \textsf{SF} policy decryption key as
		$\textsf{DK}_f:=$
		$$(\textsf{dk}_j=[\textbf{r}_j]_2,\textsf{dk}_{\rho(j),j}=[\underbrace{\textbf{v}_j+\widetilde{\textbf{W}}_{\rho(j)}\textbf{r}_j+\textbf{a}^{\bot}\textbf{w}_{\rho(j)}^{\top}\textbf{r}_j}_{\textbf{v}_j+\textbf{W}_{\rho(j)}\textbf{r}_j}]_2,$$
		$$\textsf{dk}_{i,j}=[\underbrace{\widetilde{\textbf{W}}_i\textbf{r}_j+\textbf{a}^{\bot}\textbf{w}_i^{\top}\textbf{r}_j}_{=\textbf{W}_i\textbf{r}_j}]_2).$$
		Then, it returns $\textsf{DK}_f$ to $\mathcal{A}$.
		
		- For the last $Q_f-\ell$ policy decryption key queries, say for formula $f$ of size $m$, $\mathcal{B}$ proceeds as before for the first $\ell-1$ policy decryption keys except		
		$$(\{\textbf{v}_j\}_{j\in[\hat{m}]},\rho)\xleftarrow{\$}\textsf{share}(f,\underbrace{\widetilde{\textbf{v}}+\mu^{(0)}\textbf{a}^{\bot}}_{=\textbf{v}}),$$ 		
		It is easy to see that it forms a \textsf{Normal} policy decryption key.
		
		- For the $\ell$-th policy decryption key query, say for formula $f$ of size $m$, $\mathcal{B}$ computes $(\{\textbf{v}_j\}_{j\in[\hat{m}]},\rho)\xleftarrow{\$}\textsf{share}(f,\widetilde{\textbf{v}})$, queries $\mathcal{O}_{\textsf{F},\beta}(f)\rightarrow(\{[\textbf{r}_j]_2,[\mu_j+\textbf{w}_{\rho(j)}\textbf{r}_j]_2,\{[\textbf{w}_i^{\top}\textbf{r}_j]_2\}_{i\in[n]\backslash\{\rho(j)\}}\}_{j\in[\hat{m}]})$ and uses these components to return:
		$\textsf{DK}_f:=$ $$(\textsf{dk}_j=[\textbf{r}_j]_2,\textsf{dk}_{\rho(j),j}=[\underbrace{\textbf{v}_j+\widetilde{\textbf{W}}_{\rho(j)}\textbf{r}_j+\textbf{a}^{\bot}(\mu_j+\textbf{w}_{\rho(j)}^{\top}\textbf{r}_j)}_{=(\textbf{v}_j+\mu_j\textbf{a}^{\bot})+\textbf{W}_{\rho(j)}\textbf{r}_j}]_2,$$
		$$\textsf{dk}_{i,j}=[\underbrace{\widetilde{\textbf{W}}_i\textbf{r}_j+\textbf{a}^{\bot}\textbf{w}_i^{\top}\textbf{r}_j}_{=\textbf{W}_i\textbf{r}_j}]_2).$$
		
		We claim that if $\beta=0$, then $\textsf{DK}_f$ is a \textsf{Normal} policy decryption key, and
		if $\beta=1$, then $\textsf{DK}_f$ is a \textsf{SF} policy key. This follows from the fact that thanks to linearity, the shares
		$(\{\textbf{v}_j+\mu_j\textbf{a}^{\bot}\}_{j\in[\hat{m}]},\rho)$, where $(\{\textbf{v}_j\}_{j\in[\hat{m}]},\rho)\xleftarrow{\$}\textsf{share}(f,\widetilde{\textbf{v}})$, $(\{\mu_j\}_{j\in[\hat{m}]},\rho)\xleftarrow{\$}\textsf{share}(f,\mu^{(\beta)})$, are identically distributed to $\textsf{share}(f,\widetilde{\textbf{v}}+\mu^{(\beta)}\textbf{a}^{\bot})$. The claim follows the fact that $\textbf{v}=\widetilde{\textbf{v}}+\mu^{(0)}\textbf{a}^{\bot}$, where we set $\delta:=\mu^{(1)}-\mu^{(0)}$ is a fresh random value for the key.
		
		\textbf{Challenge.} When $\mathcal{A}$ requests a challenge ciphertext for 
		symmetric keys $(K_0,K_1)$,  attributes $\vec{x}$ and formula $f$, $\mathcal{B}$ flips a random coin $\textsf{coin}\leftarrow\{0,1\}$ and constructs the challenge ciphertext for 
		$K_{\textsf{coin}}$.  $\mathcal{B}$ queries $\mathcal{O}_{\textsf{X}}$ on input $\vec{x}$ to obtain $\{\textbf{w}_i\}_{i:x_i=1}$. $\mathcal{B}$ computes 
		$\textbf{c}^{\top}=\textbf{s}^{\top}\textbf{A}$, $\textbf{c}_j^{\top}=\textbf{s}_j^{\top}\textbf{A}$, $(\{\textbf{u}_j^{\top}\},\rho)\leftarrow\textsf{share}(f,{\textbf{c}}^{\top}\textbf{U}_0)$ normally,
		samples $\widetilde{\textbf{c}}\xleftarrow{\$}\mathbb{Z}_p^{2k}$,
		and constructs the  challenge ciphertext $\textsf{CT}_{\vec{x},f}:=$
		$$\big(
		\textsf{ct}_0=e([\widetilde{\textbf{c}}^{\top}+\textbf{c}^{\top}]_1,[\widetilde{\textbf{v}}+\mu^{(0)}\textbf{a}^{\bot}]_2)\cdot K_{\textsf{coin}},$$
		$$\textsf{ct}_1'=[\widetilde{\textbf{c}}^{\top}]_1,
		\textsf{ct}_2'=[\widetilde{\textbf{c}}^{\top}\sum\limits_{i:x_i^{(out)}=1}\underbrace{\widetilde{\textbf{W}}_i+\textbf{a}^{\bot}\textbf{w}_i^{\top}}_{=\textbf{W}_i}]_1,$$
		$$\textsf{ct}_1=[\textbf{c}^{\top}]_1,~~~\widetilde{\textsf{ct}}_{j}=[\textbf{c}_j^{\top}]_1,$$
		$$\textsf{ct}_{\rho(j),j}=[{\textbf{u}}_j^{\top}+\textbf{c}_j^{\top}\underbrace{(\widetilde{\textbf{W}}_{\rho(j)}+\textbf{a}^{\bot}\textbf{w}_{\rho(j)}^{\top})}_{=\textbf{W}_{\rho(j)}}]_1),$$
		$$\textsf{ct}_{i,j}=[\textbf{c}_j^{\top}\underbrace{(\widetilde{\textbf{W}}_i+\textbf{a}^{\bot}\textbf{w}_i^{\top})}_{=\textbf{W}_i}]_1).$$

		\textbf{Guess.}  $\mathcal{A}$ halts the game with a guess $\textsf{coin}'\leftarrow\{0,1\}$. $\mathcal{B}$ outputs 1 if $\textsf{coin}'=\textsf{coin}$, and 0 otherwise.	
		
		Putting everything together, we can see that  $\mathcal{B}$ simulates $\textsf{H}_{2,
			\ell-1}$ when $\beta=0$; and $\textsf{H}_{2,
			\ell}$ when $\beta=1$.
	\end{proof}
	
	\textsf{Lemma B.8.}
	\label{Lemma:8}
	We have
	$$|Pr[\left<\mathcal{A},\textsf{H}_{2,Q_f}\right>=1]-Pr[\left<\mathcal{A},\textsf{H}_{3}\right>=1]|\leq 1/p$$ unconditionally.
	
	\begin{proof}
		The two hybrids are identically distributed conditioned on $\widetilde{\textbf{c}}^{\top}\textbf{a}^{\bot} \ne 0$. To see this, consider two ways to sample $\textbf{v}$: as $\widetilde{\textbf{v}}\xleftarrow{\$}\mathbb{Z}_p^{2k}$ and  as $\widetilde{\textbf{v}}+\widetilde{m}\textbf{a}^{\bot}$ for an independent $\widetilde{m} \xleftarrow{\$} \mathbb{Z}_p$. Both result in $\textbf{v}$ having a	uniform distribution.
		
		Using $\widetilde{\textbf{v}}$ to simulate hybrid $\textsf{H}_{2,Q_f}$ obviously results in $\textsf{H}_{2,Q_f}$  (where $\textbf{v}=\widetilde{\textbf{v}}$). However, using the identically distributed $\textbf{v}=\widetilde{\textbf{v}}+\widetilde{m}\textbf{a}^{\bot}$ to simulate $\textsf{H}_{2,Q_f}$  results in $\textsf{H}_3$ with $\widetilde{K}=K_{\textsf{coin}}\cdot[\widetilde{\textbf{c}}^{\top}\widetilde{m}\textbf{a}^{\bot}]_{T}$ and re-defined randomness $\widetilde{\delta}_j = \delta_j + \widetilde{m}$ for all the keys. Note that the information of $\widetilde{m}$ is not leaked to $\mathcal{A}$ from the secret key queries since $\widetilde{m}$ is blinded by random value ${\delta}_j$ for each key. Therefore, $\widetilde{K}$ is distributed uniformly at random over $G_T$ as long as $[\widetilde{\textbf{c}}^{\top}\textbf{a}^{\bot}]_{T}\neq {0}$.
		
		Since $\widetilde{\textbf{c}}$ is chosen at random and independent from $\textbf{a}^{\bot}\neq\textbf{0}$, so $[\widetilde{\textbf{c}}^{\top}\textbf{a}^{\bot}]_{T} = 0$ with probability $1/p$, and since we know that $\textsf{H}_{2,Q_f}\equiv\textsf{H}_3$ conditioned on $[\widetilde{\textbf{c}}^{\top}\textbf{a}^{\bot}]_{T} \neq 0$, then the lemma follows.
	\end{proof}
	
	This completes the proof of Theorem 6.2. 
\end{proof}\vspace{3mm}

\subsection{PriSrv: Security Model and Proof}\vspace{2mm}
\label{Appendix:PriSrv}

\noindent\textsf{(1) Security Model of PriSrv}

We formalize the security model for PriSrv, which includes the service discovery with bilateral control, key secrecy and bilateral anonymity, by following the
Canetti-Krawczyk model for authenticated key-exchange (AKE) in \cite{canetti2001analysis,canetti2002security,zhang2016practical} and the service discovery model in \cite{wu2016privacy}.\\

\noindent\textsf{1.1) Service Discovery Security}

The service discovery \textit{security} captures \textit{service discovery with bilateral control} and \textit{AKE security}.
The framework of PriSrv contains two sub-protocols: a private broadcast protocol that announces the service type, server's identifier, as well as other relevant information \textit{in a privacy-preserving manner}; and an anonymous mutual authentication protocol with bilateral policy control. Compared with traditional mutual authentication settings, a remarkable difference in PriSrv is that multiple clients can respond to a service provider's broadcast message if their credentials satisfy the service authorization policy. 

\noindent\textbf{Protocol participants}. The participants of PriSrv includes a set of clients $\textsf{C}=\{C_1,\cdots,C_{n_1}\}$ and a set of service providers $\textsf{S}=\{S_1,\cdots,S_{n_2}\}$.

\noindent\textbf{Long-term Keys}. Each $C_i\in\textsf{C}$ and $S_j\in\textsf{S}$ hold long-term secret keys for bilateral authentication and message decryption.

\noindent\textbf{Session and Pairing}. Denote the $\rho$-th instance of participant $U\in \textsf{C} \cup \textsf{S}$ as $U^{\rho}$, which is modeled as a PPT Turing machine. 
A participant $U^{\rho}$ can be activated to initiate a \textit{session} with a broadcast identifier $bid_{U}^{\rho}$, a session identifier $sid_{U}^{\rho}$,  attributes $\vec{x}_{U}^{\rho}$, and a policy $f_{U}^{\rho}$.
A client instance $C_i^{\rho}$ and a service provider instance $S_j^{\delta}$ are said to be \textit{paired} if their session instances $(C_i^{\rho},bid_{C_i}^{\rho},sid_{C_i}^{\rho},\vec{x}_{C_i}^{\rho},f_{C_i}^{\rho})$ and $(S_j^{\delta},bid_{S_j}^{\delta},sid_{S_j}^{\delta},\vec{x}_{S_j}^{\delta},f_{S_j}^{\delta})$ satisfy $bid_{C_i}^{\rho}=bid_{S_j}^{\delta}$, $sid_{C_i}^{\rho}$ $=sid_{S_j}^{\delta}$,
$f_{S_j}^{\delta}(\vec{x}_{C_i}^{\rho(out)})=1$, $f_{C_i}^{\rho}(\vec{x}_{S_j}^{\delta(out)})=1$. A completed \textit{session} contains a tuple $(C_i^{\rho},bid_{C_i,S_j}^{\rho,\delta},sid_{C_i,S_j}^{\rho,\delta},S_j^{\delta},SSK_{C_i,S_j}^{\rho,\delta})$, where $bid_{C_i,S_j}^{\rho,\delta}=bid_{C_i}^{\rho}=bid_{S_j}^{\delta}$, $sid_{C_i,S_j}^{\rho,\delta}=sid_{C_i}^{\rho}=sid_{S_j}^{\delta}$. 

\noindent\textbf{Adversary Capability}. We capture all of the adversary's attack capabilities in real world to have full control over the public network communication, including revealing some secrets in the protocol, intercepting or tampering with the channel messages, replaying, delaying, injecting or dropping data packets, interleaving messages from different sessions, etc.

\noindent\textbf{Protocol Execution}. An adversary $\mathcal{A}$ is modeled as a PPT machine with a distinguished query tape to issue a set of session exposure queries for gaining the ephemeral and long-term secrets possessed by participants.

$\bullet$ $\textsf{Send}(U^{\rho},M)$: transmits a message $M$ to $U^{\rho}$, who executes the protocol and returns the operation result to adversary $\mathcal{A}$. If the message in the query causes the protocol to execute or abort, it will be made known to $\mathcal{A}$.

$\bullet$ $\textsf{Execute}(C_i^{\rho},S_j^{\delta})$: executes a complete protocol between $C_i^{\rho}$ and $S_j^{\delta}$. The adversary captures all messages transmitted over the public network. Hence, the query to $\textsf{Execute}$ oracle models passive eavesdropping capability of the adversary.

$\bullet$ $\textsf{RevealBroadcast}(S_j^{\delta},bid)$: returns the semi-static state in a service provider $S_j^{\delta}$ that is maintained
for the lifetime of its current broadcast with identifier $bid$, including the attributes $\vec{x}_{S_j^{\delta}}$ and authorization policy $ f_{S_j^{\delta}}$. The revealed state does not involve the long-term secrets. $\mathcal{A}$ is allowed to make query for any service provider $S_j\in\textsf{S}$.

$\bullet$ $\textsf{RevealState}(U^{\rho},bid,sid)$: returns the local state associated with the targeted session, which does not contain the long-term secrets.

$\bullet$ $\textsf{RevealKey}(U^{\rho},bid,sid)$: outputs the secret session key associated with a targeted session.

$\bullet$ $\textsf{Corrupt}(U)$: returns all information (including ephemeral and long-term secrets) held by $U$.

$\bullet$ $\textsf{TestSession}(U^{\rho},bid,sid)$: This oracle is used to model key secrecy. A random bit $b\in\{0,1\}$ is selected to respond this query. If $b=1$, the target session key is returned to $\mathcal{A}$. Otherwise, a randomly value picked from the secret session key space is returned.\vspace{2mm}

\noindent
\textbf{Session Exposure}.
A session $(U^{\rho},bid,sid)$ is said to be \textit{exposed} if the adversary makes the following queries.

- The adversary makes a $\textsf{RevealKey}$ query on the session.

- The adversary makes a $\textsf{Corrupt}$ query on $U$, or any partner with $(\vec{x},f)$ satisfying $f_{U}(\vec{x}^{(out)}) = 1 \wedge f(\vec{x}_U^{(out)}) = 1$, before the session has expired.

- $U$ is the client in the protocol and the adversary has made a $\textsf{RevealState}$ query on the session.

- $U$ is the server in the protocol and the adversary has made a $\textsf{RevealState}$ query on the session, and also made a $\textsf{RevealBroadcast}$ query on the session or a $\textsf{Corrupt}$ query on $U$ before the session has expired.\vspace{2mm}

\noindent
\textbf{Session Freshness}.
A session $(U^{\rho},bid,sid)$ is said to be \textit{fresh} if itself is not exposed and all its matching sessions are not exposed.

	\textsf{Definition C.1}
	Let $\textsf{Succ}_{\textsf{PriSrv}}^{\textsf{Sec}}(\mathcal{A})$ denote the event that $\mathcal{A}$ makes a single $\textsf{TestSession}$ query with the restriction that the queried session $(U^{\rho},bid,sid)$ is fresh, and finally outputs a bit $b'=b$, where $b$ is the random value selected in the $\textsf{TestSession}$ query. A private service discovery  protocol $\textsf{PriSrv}$ is secure if for any PPT adversary $\mathcal{A}$, there exists a negligible function $\nu$ such that $\textsf{Adv}_{\textsf{PriSrv}}^{\textsf{Sec}}(\mathcal{A})\overset{\textsf{def}}{=}2\textsf{Pr}[\textsf{Succ}_{\textsf{PriSrv}}^{\textsf{Sec}}(\mathcal{A})]-1\leq\nu(\lambda).$\\

\noindent\textsf{1.2) Bilateral Anonymity}

The bilateral anonymity property implies that no PPT service provider (or client) can learn anything about another participant's identifier and private attributes unless it satisfies the latter's authorization policy. The adversary is permitted to compromise multiple participants. This property should hold provided that the compromised participants do not satisfy the target's policy. We include the registration query oracle in the security model for the bilateral anonymity proof. In the following, we firstly define the security game to prove the client anonymity, which captures the property that no adversary can distinguish the interactions with $C_{i_0}^*$ or $C_{i_1}^*$ (challenge clients). Let $n$ be the number of parties participating in the protocol execution experiment, which are denoted as $(P_1,\cdots,P_n)$. A special test party $P_T$ is introduced at the beginning of the experiment whose identity is kept confidential from the adversary.  We introduce two experiments $\textsf{Exp}_0$ and $\textsf{Exp}_1$, and select random $b\in\{0,1\}$ at the beginning of the game. The experiment $\textsf{Exp}_b$ proceeds as below.

\noindent\textbf{Setup Phase}. 	At the beginning of the experiment, adversary $\mathcal{A}$ submits a set of identities (with attributes and policies) $\{(uid_{j},\vec{x}_{j},f_{j})\}_{j=1}^n$ for parties $(P_1,\cdots,P_n)$. For each $j\in[n]$, the challenger sets up anonymous credential and long-term secret attribute/policy keys for party $P_j$.  $\mathcal{A}$ also submits two challenge clients $C_{i_0}^*,$ $C_{i_1}^*\in\textsf{C}$, where $C_{i_0}^*$ possesses $(uid_{i_0}^*,\vec{x}_{i_0}^*,f_{i_0}^*)$ and $C_{i_1}^*$ has $(uid_{i_1}^*,\vec{x}_{i_1}^*,f_{i_1}^*)$. It is required that $\vec{x}_{i_0}^* = \vec{x}_{i_1}^*$ and $f_{i_0}^* = f_{i_1}^*$. Note that the challenge tuples
$(uid_{i_0}^*,\vec{x}_{i_0}^*,f_{i_0}^*)$ and  $(uid_{i_1}^*,\vec{x}_{i_1}^*,f_{i_1}^*)$ are distinct from $\{(uid_{j},\vec{x}_{j},f_{j})\}_{j=1}^n$. Then, the challenger associates $(uid_{i_b}^*,\vec{x}_{i_b}^*,f_{i_b}^*)$ with $P_T$, and executes the setup algorithm for $P_T$ that is defined in the protocol.

\noindent\textbf{Protocol Execution}. Adversary $\mathcal{A}$ is allowed to issue the following queries.

$\bullet$ $\textsf{Reg}(U^{\rho},uid_U^{\rho},\vec{x}_U^{\rho}, f_U^{\rho})$: If $U^{\rho}\in(P_1,\cdots,P_n)$ with user identifier $uid_U^{\rho}$, attributes $\vec{x}_U^{\rho}$ and policy $f_U^{\rho}$ is unregistered, it executes as the protocol definition, and returns the result to $\mathcal{A}$.

$\bullet$ $\textsf{Send}$,
$\textsf{RevealBroadcast}$, $\textsf{RevealState}$,
$\textsf{RevealKey}$ and $\textsf{Corrupt}$ are the same as the definition in ``Service Discovery Security" model.

$\bullet$ $\textsf{Challenge}$. In $\textsf{Exp}_b$,
the instance $C_{i_b}^*$ with $(uid_{i_b}^*,\vec{x}_{i_b}^*,f_{i_b}^*)$ executes PriSrv by following the protocol steps.
The restriction is that $\mathcal{A}$ does not issue any of the following queries:

- $\textsf{Reg}$ query on $(uid_{i_0}^*,\vec{x}_{i_0}^*, f_{i_0}^*)$ or $(uid_{i_1}^*,\vec{x}_{i_1}^*, f_{i_1}^*)$ or any service provider whose attributes and policy satisfy $(\vec{x}_{i_0}^*,f_{i_0}^*)$ or $(\vec{x}_{i_1}^*,f_{i_1}^*)$; 

- $\textsf{RevealState}$ or $\textsf{RevealKey}$ query on any $(P_T,bid,sid)$ or its matching session; 

- $\textsf{Corrupt}$ query on $C_{i_0}^{*}$ or $C_{i_1}^{*}$ or any service provider whose attributes and policy satisfy $(\vec{x}_{i_0}^*,f_{i_0}^*)$ or $(\vec{x}_{i_1}^*,f_{i_1}^*)$. 

- If $\mathcal{A}$ associates a policy $(\vec{x}_{S_j}, f_{S_j})$ with a  service provider session $(S_j^\delta, bid, sid)$, it is required that either both $C_{i_0}^*$ and $C_{i_1}^*$ satisfy the policy, or neither of them satisfies the policy.

In any above query, if the query causes an instance to accept or termination, these outputs will be shown to $\mathcal{A}$.

\noindent\textbf{Output phase}. $\mathcal{A}$ outputs a guess $b'\in\{0,1\}$ for $b$.\vspace{2mm}

A service discovery protocol achieves client anonymity if no adversary can distinguish the experiments $\textsf{Exp}_0$ and $\textsf{Exp}_1$ with non-negligible advantage greater than 1/2, which captures the property that no active adversary can distinguish communications with a client $C_{i_0}^*$ from those with a client $C_{i_1}^*$. Here is the formal definition.

	\textsf{Definition C.2}
	\label{Def:anon-C}
	Let $\textsf{Succ}_{\textsf{PriSrv}}^{\textsf{anon-C}}(\mathcal{A})$ denote the event that $\mathcal{A}$ outputs a bit $b'=b$. $\textsf{PriSrv}$ satisfies \textit{\textbf{client anonymity}} if for any PPT adversary $\mathcal{A}$, there exists a negligible function $\nu$ such that $\textsf{Adv}_{\textsf{PriSrv}}^{\textsf{anon-C}}(\mathcal{A})\overset{\textsf{def}}{=}2\textsf{Pr}[\textsf{Succ}_{\textsf{PriSrv}}^{\textsf{anon-C}}(\mathcal{A})]-1\leq\nu(\lambda).$

The security game for \textit{service provider anonymity} is similar to that for the client anonymity, except for exchanging their roles and restriction that the adversary is not allowed to query $\textsf{Revealbroadcast}$ for the challenge sessions. The concrete security model is omitted for brevity.

	\textsf{Definition C.3}
	\label{Def:anon-S}
	Let $\textsf{Succ}_{\textsf{PriSrv}}^{\textsf{anon-S}}(\mathcal{A})$ denote the event that $\mathcal{A}$ outputs a bit $b'=b$. $\textsf{PriSrv}$ satisfies \textit{\textbf{service provider anonymity}} if for any PPT adversary $\mathcal{A}$, there exists a negligible function $\nu$ such that $\textsf{Adv}_{\textsf{PriSrv}}^{\textsf{anon-S}}(\mathcal{A})\overset{\textsf{def}}{=}2\textsf{Pr}[\textsf{Succ}_{\textsf{PriSrv}}^{\textsf{anon-S}}(\mathcal{A})]-1\leq\nu(\lambda).$

	\textsf{Definition C.4}
	\label{Def:anon-CS}
	$\textsf{PriSrv}$ satisfies \textit{\textbf{bilateral anonymity}} if it achieves \textit{client anonymity} and \textit{service provider anonymity}.\\

\noindent\textsf{(2) Security Proof of PriSrv}

\textsf{Theorem 7.1.}
Suppose that the DDH assumption holds, $\mathcal{ACME}$ is secure, $\mathcal{MAC}$ is unforgeable, and $H$ is a random oracle, then  $\textsf{PriSrv}$ is a secure service discovery protocol and satisfies bilateral anonymity.

We utilize three lemmas to prove the security of PriSrv in Theorem 7.1, which demonstrate PriSrv is secure and private in extended Canetti-Krawzyk key-exchange model (Lemma C.1), and it provides anonymity for both the client and the server (Lemma C.2 and Lemma C.3).

The security proof of PriSrv requires the underlying ACME scheme should be CCA secure.  There are standard (and efficient) generic approaches (e.g., the Fujisaki-Okamoto transformation \cite{fujisaki1999secure}) to transform our ACME construction to achieve CCA security.


\textsf{Lemma C.1.}
	\label{lemma:PriSrvPrivacy}
	(\textbf{Service discovery privacy with bilateral control}.) Suppose that DDH assumption holds on $G_1$ and $G_2$, $\mathcal{ACME}$ is secure, $\mathcal{MAC}$ is unforgeable, and $H$ is random oracle, then $\textsf{PriSrv}$ is a secure  service discovery protocol with bilateral control.

\begin{proof}
	Following similar proofs for key-exchange and service discovery protocols proposed in \cite{canetti2001analysis,canetti2002security,zhang2016practical,wu2016privacy}, we assume selective security in the adversary's choice of the test session, i.e., at the beginning of the security game, the adversary commits to the following:
	
	$\bullet$ The test session $(\alpha^*,bid,sid,\vec{x}_{\alpha^*}, f_{\alpha^*})$.
	
	$\bullet$ The peer's identity, attributes and policy $(\beta^*,\vec{x}_{\beta^*}, f_{\beta^*})$.
	
	$\bullet$ Whether $\alpha^*$ is the initializer or the responder of the test session.	
	
	Note that a selective security proof can be converted to an adaptive one at a security loss that increases polynomially in the number of parties and the number of sessions the adversary initiates. 
	
	We define a simulator $\mathcal{S}=\mathcal{S}(\mathcal{A})$. On input the number of parties $n$ and an adversary $\mathcal{A}$, the simulator $\mathcal{S}$ simulates a series of security games for the protocol.	
	In the selective security setting, the adversary begins by committing to a test session $(\overline{\alpha^*},\overline{bid},\overline{sid},\overline{\vec{x}_{\alpha^*}}, \overline{f_{\alpha^*}})$, the peer $(\overline{\beta^*},\overline{\vec{x}_{\beta^*}}, \overline{f_{\beta^*}})$ in the test session, and whether $\overline{\alpha^*}$ was the initiator or the responder in the test session. Then, $\mathcal{S}$ initializes the $n$ parties by generating anonymous credential, attribute decryption key and policy decryption key for each of them. When $\mathcal{A}$ activates a party,  $\mathcal{S}$ executes as in the protocol on behalf of the parties, and outputs the corresponding messages to $\mathcal{A}$ as well as the public outputs of each session.
	
	\textbf{Description of the simulator}. We introduce several variants of $\mathcal{S}$, which are generally denoted as $\overline{\mathcal{S}}$. The simulator $\overline{\mathcal{S}}$ behaves similarly to $\mathcal{S}$ except the following differences.
	
	(1) At the beginning of the simulation, the simulator chooses four exponents $\overline{z},\overline{x}_1,\overline{x}_2,\overline{y}\xleftarrow{\$} \mathbb{Z}_p^*$ and a random session key $\overline{SSK}$. The specification of the keys will determine the different variants of the simulator $\overline{\mathcal{S}}$.
	
	(2) In the selective security model, the adversary $\mathcal{A}$ commits to a test session $(\overline{\alpha^*},\overline{bid},\overline{sid},\overline{\vec{x}_{\alpha^*}}, \overline{f_{\alpha^*}})$, the peer $(\overline{\beta^*},\overline{\vec{x}_{\beta^*}}, \overline{f_{\beta^*}})$, and the role of $\overline{\alpha^*}$ in the session at the beginning of the experiment. Let $\overline{S}\in\{\overline{\alpha^*},\overline{\beta^*}\}$ denote the server $\mathcal{A}$ commits to for a test session, and  $\overline{C}\in\{\overline{\alpha^*},\overline{\beta^*}\}$  the client to which it commits.

	The simulator $\overline{\mathcal{S}}$ simulates the execution of the PriSrv security game as $\mathcal{S}$, except for the following differences.
	
	$\bullet$ If the adversary $\mathcal{A}$ activates $\overline{S}$ to initiate the broadcast $(\overline{S},\overline{bid},$ $\overline{\vec{x}_{S}}, \overline{f_{S}})$, the simulator uses $\overline{z}$ as the semi-static DH exponent in the broadcast.
	
	$\bullet$ If the adversary $\mathcal{A}$ activates $\overline{C}$ to initiate the session $(\overline{C},\overline{bid},\overline{sid},$ $\overline{\vec{x}_{C}}, \overline{f_{C}})$, the simulator uses $\overline{x}_1,\overline{x}_2$ as the ephemeral DH exponents of $\overline{C}$.
	
	$\bullet$ If the adversary $\mathcal{A}$ activates $\overline{S}$ as a responder to the session $(\overline{S},\overline{C},\overline{bid},\overline{sid},\overline{\vec{x}_{S}}, \overline{f_{S}},\overline{\vec{x}_{C}}, \overline{f_{C}})$, the simulator uses $\overline{y}$ as the ephemeral DH exponent of $\overline{S}$.	
	
	$\bullet$ 
	It uses $\overline{SSK}$ in place of $SSK$ whenever the shares $(h^{\overline{z}},g^{\overline{x}_1},h^{\overline{x}_2},g^{\overline{y}})$ are used to derive the session key (that is, when the simulator needs to compute  $H(\overline{X}_1^{\overline{y}},\overline{X}_2^{\overline{z}})$).
	
	(3) At the end of the protocol, $\mathcal{A}$ outputs a bit $b\in\{0,1\}$. The simulator $\overline{S}$ outputs the same bit.	
	
	The security proof contains two cases, depending on whether $\mathcal{A}$ compromises the server's semi-static broadcast secret or not. We say the adversary is admissible as long as it is not the situation that both the server's epheral DH secret and the server's semi-static broadcast DH secret are compromised, which is similar to the security analysis in \cite{krawczyk2016optls}. Specifically, the two cases in our proof are given below.
	
	$\bullet$ \textbf{Case 1}: $\mathcal{A}$ neither issues a $\textsf{RevealBroadcast}$ query on $(\overline{S},\overline{bid},$ $\overline{\vec{x}_{S}}, \overline{f_{S}})$ nor corrupt $\overline{S}$ before the broadcast session expires (i.e., the semi-static broadcast DH secret is not compromised).
	
	$\bullet$ \textbf{Case 2}: $\mathcal{A}$ does not issue a $\textsf{RevealState}$ query on $(\overline{S},\overline{bid},$ $\overline{\vec{x}_{S}}, \overline{f_{S}})$.

	For each case, we define a series of hybrid games to show that each consecutive pair of hybrid games are computationally indistinguishable.
	Before the formal case analysis, we prove the following proposition.
	
	\textsf{Proposition C.1.}
		\label{Prop: ACME}
		Suppose a session $(\alpha^*,bid,sid,\vec{x}_{\alpha^*}, f_{\alpha^*})$ completes  with a peer $(\beta^*,\vec{x}_{\beta^*}, f_{\beta^*})$. Assume that neither $\alpha^*$ nor $\beta^*$ has been corrupted before the completion of $(\alpha^*,bid,sid,$ $\vec{x}_{\alpha^*}, f_{\alpha^*})$. Then, assuming that $\mathcal{ACME}$ has authenticity, the following statements hold:
		
		(1) If $\alpha^*$ is the client and $\beta^*$ is the server, then $\mathcal{A}$ initiated a broadcast $(\beta^*,bid,\vec{x}_{\beta^*}, f_{\beta^*})$ and $\alpha^*$ must have been activated to initiate a session $(\alpha^*,bid,sid,\vec{x}_{\alpha^*}, f_{\alpha^*})$ with the broadcast message if and only if $f_{\alpha^*}(\vec{x}_{\beta^*}^{(out)}) = 1~\wedge~f_{\beta^*}(\vec{x}_{\alpha^*}^{(out)}) = 1$.
		
		(2) If $\alpha^*$ is the server and $\beta^*$ is the client, the session $(\beta^*,bid, sid, \vec{x}_{\beta^*}, f_{\beta^*})$ cannot complete with a peer session $(\alpha'^*, bid, sid, \vec{x}_{\alpha'^*}, f_{\alpha'^*})$ such that $f_{\alpha'^*}(\vec{x}_{\beta^*}^{(out)})\neq 1~\vee~f_{\beta^*}(\vec{x}_{\alpha'^*}^{(out)})\neq 1$.	

	\begin{proof}
		We prove the two cases separately.\vspace{2mm}
		
		(1) When $\alpha^*$ is activated to initialize a session $(\alpha^*,bid,sid,$ $\vec{x}_{\alpha^*}, f_{\alpha^*})$ with a broadcast message $(bid',\textsf{CT}_B)$, it decrypts the broadcast ciphertext utilizing its private attribute and policy keys. If the decryption fails, it indicates that $f_{\alpha'^*}(\vec{x}_{\beta^*}^{(out)})\neq 1~\vee~f_{\beta^*}(\vec{x}_{\alpha'^*}^{(out)})\neq 1$. Otherwise, $\alpha^*$ is an intended client to obtain the broadcast messages, who checks whether $bid'=bid$ and verifies the authenticity of $\textsf{CT}_B$ for further communication. Since $(\alpha^*,bid,sid,$ $\vec{x}_{\alpha^*}, f_{\alpha^*})$ completes with $\beta^*$, it must be the case that $f_{\alpha^*}(\vec{x}_{\beta^*}^{(out)})$ $=1~\wedge~f_{\beta^*}(\vec{x}_{\alpha^*}^{(out)})= 1$. Since $\beta^*$ has not been corrupted before the completion with $(\alpha^*,bid,sid,$ $\vec{x}_{\alpha^*}, f_{\alpha^*})$, it generates at most one broadcast ciphertext containing $bid$. Thus, if $(\alpha^*,bid,sid,$ $\vec{x}_{\alpha^*}, f_{\alpha^*})$ completes with $\beta^*$, it must have been initialized with broadcast message output by $(\beta^*,bid,\vec{x}_{\beta^*}, f_{\beta^*})$ since this is the only message that contains a valid ACME ciphertext from $\beta^*$ with the broadcast identifier $bid$. Otherwise, $\mathcal{A}$ can be used to break the authenticity of $\mathcal{ACME}$.
		
		(2) If $\beta^*$ is activated to initiate the session $(\beta^*,bid,\vec{x}_{\beta^*}, f_{\beta^*})$ and the session completes with a peer $\alpha'^*$ with $(\vec{x}_{\alpha'^*}, f_{\alpha'^*})$ such that $f_{\alpha'^*}(\vec{x}_{\beta^*}^{(out)})\neq 1~\vee~f_{\beta^*}(\vec{x}_{\alpha'^*}^{(out)})\neq 1$. In this case, an honest $\beta^*$ would never succeed to decrypt the broadcast ciphertext and then generate the ACME ciphertext  $CT_{\beta^*}$. Therefore, any adversary that can cause $(\alpha^*,bid,sid,\vec{x}_{\alpha^*}, f_{\alpha^*})$ to complete with peer $\beta^*$, and have $(\beta^*,bid,sid,\vec{x}_{\beta^*}, f_{\beta^*})$ complete with peer $\alpha'^*$ can break the authenticity of $\mathcal{ACME}$.
	\end{proof}
	
	Next we consider the two possible cases and prove that the adversary's advantage in both cases is negligible.\vspace{2mm}

	\textbf{Case 1: $\mathcal{A}$ does not compromise the broadcast session and $z$ is not disclosed.}
	
	In this case, the security proof relies on the server's broadcast secret for the privacy of the session. A series of hybrid games are defined.
	
	$\bullet$ \textbf{Hybrid} $\textsf{H}_0$: This game is the same as a real interaction with the $\textsf{PriSrv}$ protocol. A random bit $b\in\{0,1\}$ is selected. When $b=1$, the real session key is returned as a response to the $\textsf{TestSession}$ query. Otherwise, a random key from the key space is returned as the session key.
	
	$\bullet$ \textbf{Hybrid} $\textsf{H}_1$: This game is the same as $\textsf{H}_0$, except that $Z^{x_2}=X_2^z$ is replaced by a random value in group $G_2$.

	
	$\bullet$ \textbf{Hybrid} $\textsf{H}_2$: This game is the same as $\textsf{H}_1$, except that $\bar{\mathcal{S}}$ also aborts if the session $(\overline{\beta^*},\overline{bid},\overline{sid},\overline{\vec{x}_{\beta^*}}, \overline{f_{\beta^*}})$ does not match $(\overline{\alpha^*},\overline{bid},$ $\overline{sid},\overline{\vec{x}_{\alpha^*}}, \overline{f_{\alpha^*}})$. 
	
	$\bullet$ \textbf{Hybrid} $\textsf{H}_3$: This game is the same as $\textsf{H}_2$, except that $\overline{SSK}$ is replaced by a random number from the secret session key space.\vspace{2mm}


	In the following, we show that each consecutive pair of hybrid games described above are computationally indistinguishable.

	\textsf{Claim C.1.}
		Hybrids $\textsf{H}_0$ and $\textsf{H}_1$ are computationally indistinguishable if the DDH assumption holds in group $G_2$.

	\begin{proof}
		Let $(\overline{\alpha^*},\overline{bid},\overline{sid},\overline{\vec{x}_{\alpha^*}}, \overline{f_{\alpha^*}})$ be the session and $(\overline{\beta^*},$ $\overline{\vec{x}_{\beta^*}}, \overline{f_{\beta^*}})$ be the peer that the adversary commits to at the beginning of the experiment. By definition, this means that $(\overline{\alpha^*},\overline{\beta^*},\overline{bid},\overline{sid},\overline{\vec{x}_{\alpha^*}}, \overline{f_{\alpha^*}},$ $\overline{\vec{x}_{\beta^*}}, \overline{f_{\beta^*}})$ is the public output of the test session.
		
		Let $\mathcal{A}$ be a distinguisher between $\textsf{H}_0$ and $\textsf{H}_1$. We use $\mathcal{A}$ to build a DDH adversary $\mathcal{B}$ as below. $\mathcal{B}$ is given a DDH challenge tuple $(h^{b_1},h^{b_2},h^{\gamma_2})$ over group $G_2$, where $\gamma_2=b_1b_2$ or $\gamma_2$ is a random number from $\mathbb{Z}_p^*$.
		
		At the beginning of the simulation, $\mathcal{B}$ generates anonymous credential, private attribute and policy key (in the same manner as the simulator $\overline{\mathcal{S}}$) for each of the $n$ parties.			
		$\mathcal{B}$ begins the simulation of the security game for  $\mathcal{A}$. In the following, we use $\overline{C}\in\{\overline{\alpha^*},\overline{\beta^*}\}$ to denote the client and $\overline{S}\in\{\overline{\alpha^*},\overline{\beta^*}\}$ to denote the server in the test session. 
		
		$\bullet$ \textbf{Server broadcast queries}. If the adversary activates a server $\overline{S}$ to initiate the test broadcast session $(\overline{S},\overline{bid},\overline{\vec{x}_s},\overline{f_s})$, the simulator uses $h^{b_1}$ from the DDH challenge instance as the semi-static DH share in the broadcast message. For other broadcast queries, $\mathcal{B}$ selects a random DH exponent $z$ to constructs the broadcast ciphertext exactly as in the real protocol. 
		
		$\bullet$ \textbf{Client initialization queries}. When $\mathcal{A}$ activates a party $\alpha^*$ to initiate a session $(\alpha^*,bid,sid,\vec{x}_{\alpha^*},f_{\alpha^*})$, if $(\alpha^*,bid,sid,$ $\vec{x}_{\alpha^*},f_{\alpha^*})\neq(\overline{C},\overline{bid},\overline{sid},\overline{\vec{x}_c},\overline{f_c})$, $\mathcal{B}$ selects a random DH exponent and generates the message exactly as in the real scheme. Otherwise, if $(\alpha^*,bid,sid,$ $\vec{x}_{\alpha^*},f_{\alpha^*})=(\overline{C},\overline{bid},\overline{sid},\overline{\vec{x}_c},\overline{f_c})$, $\mathcal{B}$ sets $h^{b_2}$ from the DDH challenge instance to be the DH share $X_2=h^{x_2}$ in its message. The other computation steps follow the real experiment.
		
		$\bullet$ \textbf{Server response queries}. When $\mathcal{A}$ activates a server $S$ to respond to a session $(\alpha^*,bid,sid,\vec{x}_{\alpha^*},f_{\alpha^*})$, $\mathcal{B}$ selects a random DH exponent $y$ and generates the message exactly as in the real scheme.
		
		$\bullet$ \textbf{Client finish queries}. When a client receives a response message for session $(\alpha^*,bid,sid,\vec{x}_{\alpha^*},f_{\alpha^*})$, if $(\alpha^*,bid,sid,$ $\vec{x}_{\alpha^*},$ $f_{\alpha^*})\neq(\overline{C},\overline{bid},\overline{sid},\overline{\vec{x}_c},\overline{f_c})$, $\mathcal{B}$ constructs the outputs as in the real scheme (this is feasible since $\mathcal{B}$ selects the client's ephemeral DH share in this case). Otherwise, if $(\alpha^*,bid,sid,$ $\vec{x}_{\alpha^*},f_{\alpha^*})=(\overline{C},\overline{bid},\overline{sid},\overline{\vec{x}_c},\overline{f_c})$, $\mathcal{B}$ runs the other computation steps following the real experiment except that it sets $SSK=H(X_1^y,h^{\gamma_2})$, where $X_1^y=Y^{x_1}$ (generated by $\mathcal{B}$) and $h^{\gamma_2}$ is from the DDH challenge instance.
		
		$\bullet$ $\textsf{RevealState}$ \textbf{and} $\textsf{RevealKey}$ \textbf{queries}. These are handled exactly as in $\textsf{H}_{0}$. 
		
		$\bullet$ $\textsf{Corrupt}$ \textbf{queries}. If $\mathcal{A}$ corrupts a party $U$, $\mathcal{B}$ sends the anonymous credential and private attribute/policy keys of $U$ to $\mathcal{A}$, as well as ephemeral secrets in the local storage of $U$.

		$\mathcal{B}$ perfectly simulates $\textsf{H}_0$ if $\gamma_2=b_1b_2$, and $\mathcal{B}$ simulates $\textsf{H}_1$ if $\gamma_2$ is a random number. Then, if $\mathcal{A}$ can distinguish $\textsf{H}_0$ from $\textsf{H}_1$, $\mathcal{B}$ can succeed in the DDH game on
		$G_2$ with the same advantage. 
	\end{proof}

		\textsf{Claim C.2.}
		Hybrids $\textsf{H}_1$ and $\textsf{H}_2$ are computationally indistinguishable if the ACME algorithm is private and MAC is unforgeable.
	
	\begin{proof}
		If an adversary $\mathcal{A}$ outputs a session $(\alpha^*,bid,sid,$ $\vec{x}_{\alpha^*},$ $f_{\alpha^*})$ in the $\textsf{TestSession}$ query, there must be a partner instance $(\beta^*,bid,sid,\vec{x}_{\beta^*}, f_{\beta^*})$. Otherwise, we can make use of the adversary $\mathcal{A}$ to break the privacy of 
		$\mathcal{ACME}$ or the unforgeability of MAC.
		
		We define an adversary $\mathcal{B}_0$ such that in the $\textsf{TestSession}$ query $\mathcal{B}_0$ outputs a session $(\alpha^*,bid,sid,\vec{x}_{\alpha^*}, f_{\alpha^*})$, which has a matching session $(\beta^*,bid,sid,\vec{x}_{\beta^*}, f_{\beta^*})$. Given an adversary $\mathcal{A}$ against the $\textsf{PriSrv}$ protocol in the security game, we build an adversary $\mathcal{B}_0$ as follows.
		
		Adversary $\mathcal{B}_0$ answers all queries made by $\mathcal{A}$ using its own oracles. If $\mathcal{A}$ outputs an instance $(\alpha^*,bid,sid,\vec{x}_{\alpha^*}, f_{\alpha^*})$ that has no matching session,  $\mathcal{B}_0$ aborts without any output. Otherwise, adversary $\mathcal{B}_0$ outputs $(\alpha^*,bid,sid,\vec{x}_{\alpha^*}, f_{\alpha^*})$ in the $\textsf{TestSession}$ query, and returns to adversary $\mathcal{A}$ the response it receives. 
		
		Let $E$ denote the event that the instance $(\alpha^*,bid,sid,\vec{x}_{\alpha^*},$ $ f_{\alpha^*})$ in the $\textsf{TestSession}$ query output by adversary $\mathcal{A}$ does not have a matching session. If event $E$ does not happen,  $\mathcal{B}_0$ and adversary $\mathcal{A}$ are the same. Otherwise,  we can construct an encryption-aided forger 	$\mathcal{B}_1$ who aims to produce a forgery $\mathcal{MAC}.\textsf{MAC}(K^*,M)$ for a secret key $K^*$ that is encapsulated in an ACME ciphertext $\textsf{CT}^*$ \cite{bellare1998modular}.

		$\mathcal{B}_1$ is given $\textsf{pp}$, $\textsf{mpk}$ and access to an oracle $\mathcal{O}_{\textsf{Issue}}(\cdot)$ which creates anonymous credential for user $U$, an oracle $\mathcal{O}_{\textsf{DKGen}}(\cdot)$ which creates attribute decryption key for $\vec{x}$, an oracle $\mathcal{O}_{\textsf{PolGen}}(\cdot)$ which creates policy decryption key for $f$, an oracle $\mathcal{O}_{\textsf{Dec}}(\cdot)$ which decrypts ciphertexts.
		Assume that adversary $\mathcal{A}$ performs at most $q_{I}$ activations of parties with an incoming message. 
		
		Forger $\mathcal{B}_1$ randomly chooses $\ell\leftarrow[1,q_I]$ and simulates the security game for adversary $\mathcal{A}$ in the following cases.
		
		$\bullet$ If adversary $\mathcal{A}$ does not make a $\textsf{TestSession}$ query with an activation of $\alpha^*$, forger $\mathcal{B}_1$ aborts.
		
		$\bullet$  If $\beta^*$ is not the matching session of $\alpha^*$, forger $\mathcal{B}_1$ aborts.
		
		$\bullet$ If $(\alpha^*,bid,sid,\vec{x}_{\alpha^*}, f_{\alpha^*})$ is not the $\ell$-th activation, forger $\mathcal{B}_1$ aborts.
		
		$\bullet$ If adversary $\mathcal{A}$ makes a $\textsf{Corrupt}$ query on $\beta^*$ or any partner $\beta$ with $(\vec{x}_{\beta},f_{\beta})$ satisfying $f_{\alpha^*}(\vec{x}_{\beta}^{(out)}) = 1 \wedge f_{\beta}(\vec{x}_{\alpha^*}^{(out)}) = 1$, before the session has expired, forger $\mathcal{B}_1$ aborts.
		
		$\bullet$ In the $\ell$-th activation, $\mathcal{B}_1$ uses $(\textsf{DK}_{\vec{x}_{\alpha^*}},\textsf{DK}_{ f_{\alpha^*}})$ to derive the broadcast message $MSG_{B}^*\leftarrow\mathcal{ACME}.\textsf{Dec}(\textsf{DK}_{\vec{x}_{\alpha^*}},$ $\textsf{DK}_{ f_{\alpha^*}},$ $\textsf{CT}_{\beta^*})$, where $MSG_{B}=(bid, Z,\cdots,K_{\alpha^*})$. $\mathcal{B}_1$ generates the ephemeral DH shares $(X_1,X_2)$ for $(\alpha^*,bid,sid,\vec{x}_{\alpha^*}, f_{\alpha^*})$. $\mathcal{B}_1$ sets $M_{\alpha^*}=(``\alpha^*\rightarrow\beta^*",bid,sid,X_1,X_2,Z)$ and creates the tag $\sigma_{\alpha^*}$ using $K_{\alpha^*}$ on $M_{\alpha^*}$. $\mathcal{B}_1$  asks its challenger to return $$\textsf{CT}_{\alpha^*}=\mathcal{ACME}.\textsf{Enc}(\textsf{cred}_{\alpha^*},\vec{x}_{\alpha^*}, f_{\alpha^*},MSG_{\alpha^*})$$ where $MSG_{\alpha^*}=(K^*,M_{\alpha^*})$ and $K^*$ is chosen by $\mathcal{B}_1$'s challenger. 		
		Forger $\mathcal{B}_1$ sets $\textsf{CT}^* = \textsf{CT}_{\alpha^*}$.
		
		$\bullet$ If adversary $\mathcal{A}$ sends $(\textsf{CT},\cdots)$ to $\beta^*$ where $\textsf{CT}\neq \textsf{CT}^*$, forger $\mathcal{B}_1$ makes a query to its decryption oracle $\mathcal{O}_{\textsf{Dec}}(\cdot)$ on input $\textsf{CT}$, and proceeds as usual after getting the response from $\mathcal{O}_{\textsf{Dec}}(\cdot)$.
		
		$\bullet$ If adversary $\mathcal{A}$ sends $(\textsf{CT}^*,M)$ to $\beta^*$, forger $\mathcal{B}_1$ issues a query to its oracle $\mathcal{O}_{\textsf{MAC}}$ to generate the tag $\sigma^*$ with regards to $K^*$ and $M$. The restriction is that $M$ does not contain $(bid,sid)$ of challenge session.
		
		$\bullet$ When adversary $\mathcal{A}$ sends the tag $\sigma^*$ to the $\ell$-th activation, forger $\mathcal{B}_1$ outputs the tag $\sigma^*$ and the corresponding message as its forgery. 
		
		Therefore, we have
		$\epsilon=Pr[\mathcal{B}_1 ~\text{succeeds}]\geq\frac{1}{q_I}Pr[E].$
		
		Given a forger $\mathcal{B}_1$, we now construct another adversary $\mathcal{B}_2$ against the anonymous credential matchmaking encryption scheme $\mathcal{ACME}$ in the security game, which is given the public parameter $\textsf{pp}$ and has access to the attribute/policy decryption key generation and decryption oracle.	
		When forger $\mathcal{B}_1$ asks for a challenger with input participant $\alpha^*$, adversary $\mathcal{B}_2$ randomly chooses two keys $K_0$ and $K_1$, and asks its challenger with inputs $(K_0, M_{\alpha^*})$ and $(K_1, M_{\alpha^*})$. After obtaining the challenge $\textsf{CT}^*$ (with respect to $K_0$ or $K_1$), adversary $\mathcal{B}_2$ sets $\textsf{CT}^*$ as forger $\mathcal{B}_1$'s challenge. When forger $\mathcal{B}_1$ makes a query with a ciphertext $\textsf{CT} \ne \textsf{CT}^*$, adversary $\mathcal{B}_2$ makes a decryption query with input $\textsf{CT}$ to its challenger. When forger $\mathcal{B}_1$ makes an $\mathcal{O}_{\textsf{MAC}}$ query on a message $M$, adversary $\mathcal{B}_2$ returns $\mathcal{MAC}.\textsf{MAC}(K_0,M)$ to forger $\mathcal{B}_1$. Finally, if forger $\mathcal{B}_1$ successfully makes a forgery $\mathcal{MAC}.\textsf{MAC}(K_0,M_{\alpha^*})$, $\mathcal{B}_2$ outputs 0 meaning that $\textsf{CT}^*$ is an encryption of $(K_0,M_{\alpha^*})$. Otherwise, if forger $\mathcal{B}_1$ fails to make a forgery, adversary $\mathcal{B}_2$ outputs 1, meaning that $\textsf{CT}^*$ is an encryption of $(K_1,M_{\alpha^*})$. Hence, we have
		\begin{eqnarray*}
			&&\textbf{Adv}_{\mathcal{B}_2}^{\textsf{ACME}}(\lambda)\\
			&=&Pr[\mathcal{B}_2~\text{outputs}~0|b=0]\cdot Pr[b=0]+\\
			&&Pr[\mathcal{B}_2~\text{outputs}~0|b=1]\cdot Pr[b=1]-\frac{1}{2}\\
			&=&\frac{1}{2}Pr[\mathcal{B}_1 ~\text{succeeds}|b=0]+\\
			&&\frac{1}{2}(1-\frac{1}{2}Pr[\mathcal{B}_1 ~\text{succeeds}|b=1])-\frac{1}{2}\\
			&=&\frac{1}{2}(Pr[\mathcal{B}_1 ~\text{succeeds}|b=0]-Pr[\mathcal{B}_1 ~\text{succeeds}|b=1])\\
			&=&\frac{1}{2}(\epsilon-\textbf{Adv}_{\mathcal{B}_1}^{\textsf{MAC}}(\lambda)).
		\end{eqnarray*}	
		The last line of the above equation is concluded from when $b=0$, forger $\mathcal{B}_1$ is in the forgery game, and when $b=1$, forger $\mathcal{B}_1$ is in the random message attack game.
		
		Therefore, Hybrids $\textsf{H}_1$ and $\textsf{H}_2$  are computationally indistinguishable.		
	\end{proof}


	\textsf{Claim C.3.}
		Hybrids $\textsf{H}_2$ and $\textsf{H}_3$ are computationally indistinguishable when the hash function $H$ is a random oracle.
	
	\begin{proof}
		Since in $\textsf{H}_2$ we replaced $Z^{x_2}$ with a random value $h^{\gamma_2}$ from $G_2$, the probability that the adversary can make a hash query $H(Y^{x_1},h^{\gamma_2})$ is negligible. When $(Y^{x_1},h^{\gamma_2})$ is not queried, its hash value (i.e., the session key) is an unknown random value to the adversary, same as in $\textsf{H}_3$. The claim follows.
	\end{proof}

	\textbf{Case 2: $\mathcal{A}$ has compromised $z$.}
	
	In this case, we rely on the ephemeral DH share $y$ of the server to ensure confidentiality of the session key. The security proof is quite similar to that of Case 1. The hybrid experiments are described below.
	
	$\bullet$ \textbf{Hybrid} $\textsf{H}_0$: This game is the same as a real interaction with the $\textsf{PriSrv}$ protocol. 
	
	$\bullet$ \textbf{Hybrid} $\textsf{H}_1$: This game is the same as in Case 1, except that $Y^{x_1}=X_1^y$ is replaced by a random value in group $G_1$.

	$\bullet$ \textbf{Hybrid} $\textsf{H}_2$: This game is the same as in Case 1. 
	
	$\bullet$ \textbf{Hybrid} $\textsf{H}_3$: This game is the same as in Case 1.\vspace{2mm}
	
	It suffices to prove that the hybrids $\textsf{H}_0$ and $\textsf{H}_1$ are computationally indistinguishable.

	\textsf{Claim C.4.}
		Hybrids $\textsf{H}_0$ and $\textsf{H}_1$ are computationally indistinguishable if the DDH assumption holds in group $G_1$.
	\begin{proof}
		The proof follows the same arguments as in Case 1 except that the DDH tuple $(h^z, h^{x_2}, h^{\gamma_2})$ on group $G_2$ used in the proof of the Case 1 is replaced by the DDH tuple $(g^y, g^{x_1}, g^{\gamma_1})$ on group $G_1$.
	\end{proof}

	Integrating the above proofs for the two cases, we conclude that PriSrv realizes secure service discovery with bilateral control.
\end{proof}

	\textsf{Lemma C.2.}	
	\label{lemma:PriSrvAnon-C}
	(\textbf{Client anonymity}) $\textsf{PriSrv}$ protocol satisfies client anonymity assuming the ACME scheme is private.

\begin{proof}
	We define a simulator $\mathcal{S}$ that simulates the challenger for the adversary $\mathcal{A}$ in the client anonymity security game.     
	
	At the beginning of the simulation, $\mathcal{A}$ submits a set of identities (with attributes and policies) $\{(uid_{j},\vec{x}_{j},f_{j})\}_{j=1}^n$ for parties $(P_1,\cdots,P_n)$, and two challenge tuples
	$(uid_{i_0}^*,\vec{x}_{i_0}^*,f_{i_0}^*)$ and  $(uid_{i_1}^*,\vec{x}_{i_1}^*,f_{i_1}^*)$, which are distinct from $\{(uid_{j},\vec{x}_{j},f_{j})\}_{j=1}^n$.
	
	We define a series of hybrid experiments.
	During the protocol execution, the simulator responds to the adversary's queries according to these hybrid experiments. 
	
	$\bullet$ Hybrid $\textsf{H}_1$: This is the real experiment $\textsf{Exp}_0$, where the simulator responds to adversary's queries as described in $\textsf{Exp}_0$.
	
	$\bullet$ Hybrid $\textsf{H}_2$:  This is the real experiment   $\textsf{Exp}_1$. \vspace{2mm} 
	
	Next, we prove that Hybrids $\textsf{H}_1$ and $\textsf{H}_2$ are computationally indistinguishable if the underlying ACME is private.
	
	Let $q$ be an upper bound on the number of sessions, where $\mathcal{A}$ activates the test party $P_T$ as the responder (the client). We define a sequence of $q+1$ hybrid experiments $\textsf{H}_{1,0},\cdots,\textsf{H}_{1,q}$, where hybrid experiment $\textsf{H}_{1,i}$ is defined as follows: $\textsf{H}_{1,i}$ is same as $\textsf{H}_1$ except that for the first $i$ times when $P_T$ is activated as the responder (the client) of the broadcast ciphertext, $P_T$ is instantiated using the credential of  $(uid_{i_1}^*,\vec{x}_{i_1}^*,f_{i_1}^*)$ for generating the response message. In all subsequent times, $P_T$ $P_T$ is instantiated using the credential of  $(uid_{i_0}^*,\vec{x}_{i_0}^*,f_{i_0}^*)$.
	
	By construction, $H_1\equiv H_{1,0}$ and $H_2\equiv H_{1,q}$.		
	We prove that for all $i\in[q]$, hybrid $\textsf{H}_{1,i-1}$ and $\textsf{H}_{1,i}$ are computationally indistinguishable assuming that the ACME scheme is private.		
	Suppose $\mathcal{A}$ is able to distinguish $\textsf{H}_{1,i-1}$ from $\textsf{H}_{1,i}$, we use $\mathcal{A}$ to construct an adversary $\mathcal{B}$ against ACME in the security game.
	
	First, $\mathcal{B}$ is given the public parameters $\textsf{mpk}$ of the ACME scheme. Then, $\mathcal{B}$ begins running $\mathcal{A}$ and obtains set of identities (with attributes and policies) $\{(uid_{j},\vec{x}_{j},f_{j})\}_{j=1}^n$ for parties $(P_1,\cdots,P_n)$, and two challenge tuples
	$(uid_{i_0}^*,\vec{x}_{i_0}^*,f_{i_0}^*)$ and  $(uid_{i_1}^*,\vec{x}_{i_1}^*,f_{i_1}^*)$, which are distinct from $\{(uid_{j},\vec{x}_{j},f_{j})\}_{j=1}^n$.
	
	$\mathcal{B}$ simulates the setup procedure in $\textsf{H}_{1}$ by creating anonymous credentials, private attribute/policy keys for each party. Then, $\mathcal{B}$ sends $\textsf{mpk}$ to $\mathcal{A}$ and begins simulating the protocol execution experiment for $\mathcal{A}$.
	
	$\bullet$ \textbf{Server broadcast queries}. These are handled exactly as in $\textsf{H}_{1}$ and $\textsf{H}_{2}$.

	$\bullet$ \textbf{Client initialization queries}. When $\mathcal{A}$ activates a client $C$ to respond to a broadcast $(S,bid,\vec{x}_s,f_s)$, if $C\neq P_T$, algorithm $\mathcal{B}$ simulates the response as in the real scheme. If $C=P_T$, then let $\ell$ be the number of times $\mathcal{A}$ has activated $P_T$ to respond to a broadcast. $\mathcal{B}$ queries the ACME key generation or decryption oracle to obtain a decrypted broadcast message, and performs the checks on the decrypted broadcast message.	
	$\mathcal{B}$ then proceeds as below.
	
	- If $\ell<i-1$, $\mathcal{B}$ constructs the response message as described in $\textsf{H}_{2}$, that is using $(uid_{i_1}^*,\vec{x}_{i_1}^*,f_{i_1}^*)$ in client's response message.
	
	- If $\ell\geq i$, $\mathcal{B}$ constructs the response message as described in $\textsf{H}_{1}$, that is using $(uid_{i_0}^*,\vec{x}_{i_0}^*,f_{i_0}^*)$ in client's response message.
	
	- If $\ell= i-1$,  $\mathcal{B}$ selects random DH shares $x_1,x_2\xleftarrow{\$} \mathbb{Z}_p^*$, and generates a MAC key $K_s$. It submits  $(\textsf{cred}_{i_0},\vec{x}_{i_0}^*,f_{i_0}^*,MSG_c)$ and $(\textsf{cred}_{i_1},\vec{x}_{i_1}^*,f_{i_1}^*,MSG_c)$ to the ACME challenger, where $MSG_c=(K_s,M_c)$, $M_c=(``C\rightarrow S",bid,sid,X_1=g^{x_1},X_2=h^{x_2},Z)$, $\textsf{cred}_{i_b}$ is the anonymous credential for $uid_{i_b}^*$, $b\in\{0,1\}$. Then, it receives a ciphertext $\overline{CT}_c$ from the challenger. $\mathcal{B}$ runs MAC scheme to obtain $\sigma_c$ from $K_c$ and $M_c$. $\mathcal{B}$ outputs the response $(bid,sid,\sigma_c,\overline{CT}_c)$.
	
	$\bullet$ \textbf{Server response queries}. 	
	When $\mathcal{A}$ delivers a message to server, $\mathcal{B}$ responds as below.
	
	- $\mathcal{B}$ parses $\mathcal{A}$'s message as $(bid,sid,\sigma_c,CT_c)$.	
	
	- If $\mathcal{B}$ is in the pre-challenge phase, or if $\mathcal{B}$ is in the post-challenge phase, and either $CT_c \neq \overline{CT}_c$ or $B$ is allowed to obtain the decryption key for $CT_c$, $\mathcal{B}$ queries the ACME decryption or key generation oracle to decrypt $CT_c$.
	
	If $\mathcal{B}$ is in the post-challenge phase and $CT_c=\overline{CT}_c$, $\mathcal{B}$ simulates the response by using the $MSG_c=(K_s,M_c)$ it has chosen for generating $\overline{CT}_c$.
	
	$\bullet$ \textbf{Client finish queries}. These are handled exactly as in $\textsf{H}_{1}$ and $\textsf{H}_{2}$. They are independent of the ACME parameters.
	
	$\bullet$ $\textsf{RevealState}$ \textbf{and} $\textsf{RevealKey}$ \textbf{queries}. These are handled exactly as in $\textsf{H}_{1}$ and $\textsf{H}_{2}$. 
	
	$\bullet$ $\textsf{Corrupt}$ \textbf{queries}. If $\mathcal{A}$ corrupts a party $P\neq P_T$, $\mathcal{B}$ queries the ACME challenger for the private attribute/policy keys for $P$, and returns the keys as well as ephemeral secrets in the local storage of $P$ to $\mathcal{A}$.
	
	At the end of the game, $\mathcal{A}$ outputs a guess for whether it is in $\textsf{H}_{1}$ or $\textsf{H}_{2}$. $\mathcal{B}$ forwards the guess to its security game.
	
	To complete the proof, we show that $\mathcal{B}$ is an elegible ACME adversary in the privacy security game since $\mathcal{B}$ does not need to request its challenger to decrypt the challenge ciphertext or return secret keys that can decrypt the challenge ciphertext. 
	
	By construction, if $\mathcal{B}$ receives an encryption of $(uid_{i_0}^*,\vec{x}_{i_0}^*,f_{i_0}^*)$ from the ACME challenger, then it has correctly simulated the client's response queries according to the specification of hybrid $\textsf{H}_{1,i-1}$ for $\mathcal{A}$. If it receives an encryption of $(uid_{i_1}^*,\vec{x}_{i_1}^*,f_{i_1}^*)$ from the ACME challenger, then it has correctly simulated the client's response queries according to the specification of hybrid $\textsf{H}_{1,i}$ for $\mathcal{A}$.
	
	
	Hence, if the ACME is private, then $\textsf{H}_{1}$
	and
	$\textsf{H}_{2}$ are computationally indistinguishable.
	
	Therefore, we have proved that PriSrv satisfies client anonymity.	
\end{proof}

	\textsf{Lemma C.3.}
	\label{lemma:PriSrvAnon-S}
	(\textbf{Server anonymity}) $\textsf{PriSrv}$ protocol satisfies server anonymity assuming the ACME scheme is private.

\begin{proof}
	This proof is similar to the proof of client anonymity (in Lemma F.7) except that $P_T$ initiates broadcast as a server in the simulation. We omit the details of the proof for briefty.

\end{proof}


\end{document}